\documentclass[journal,onecolumn]{IEEEtran}
\usepackage{latexsym}
\usepackage{graphicx}
\usepackage{amsfonts,amssymb,amsmath}
\usepackage[table]{xcolor}

\usepackage[inline]{enumitem}

\usepackage{jabbrv}
\usepackage{cite}

\usepackage[hyphens]{url}
\usepackage{xspace}

\colorlet{lgray}{gray!40}
\usepackage{mathtools}
\usepackage{multirow}

\usepackage{scalerel}

\def\BibTeX{{\rm B\kern-.05em{\sc i\kern-.025em b}\kern-.08em T\kern-.1667em\lower.7ex\hbox{E}\kern-.125emX}}
\newcommand{\xmed}{\mu_{0.5}}
\newcommand{\vphi}{\varphi}

\newcommand{\expk}{\exp_{\kappa}}
\newcommand{\lnk}{\ln_{\kappa}}
\newcommand{\kp}{\kappa}
\newcommand{\hkp}{\hat{\kp}}
\newcommand{\hmu}{\hat{\mu}}
\newcommand{\hsi}{\hat{\sigma}}
\newcommand{\kpe}{$\kp$-exponential\xspace}
\newcommand{\kpl}{$\kp$-logarithm\xspace}
\DeclareMathOperator*{\dd}{d}

\newcommand{\PDF}{PDF\xspace}

\newcommand{\nll}{\mathrm{NLL}}

\newcommand{\om}{\vartheta}
\newcommand{\pa}{\partial}
\newcommand{\bfs}{\mathbf{s}} 
\newcommand{\rfsom}[1]{{{#1}}(\bfs;\om)}
\newcommand{\Xsom}{{X}(\bfs;\om)}
\newcommand{\N}{\mathcal N}

\renewcommand{\deg}{\ensuremath{^{\circ}}\xspace}

\newcommand{\bfx}{\mathbf{x}}
\newcommand{\bfy}{\mathbf{y}}

\newcommand{\bfr}{\mathbf{r}}

\newcommand{\bfC}{\mathbf{C}}
\newcommand{\Do}{\mathcal{D}}

\newcommand{\Nsim}{N_{\mathrm{sim}}}

\newcommand{\EE}{\mathbb{E}} 
\newcommand{\E}{\mathrm{e}} 

\newcommand{\R}{\mathbb{R}}
\newcommand{\Rd}{\mathbb{R}^{d}}

\newcommand{\Or}{\mathcal{O}}
\newcommand{\al}{\alpha}

\newcommand{\bfso}{{\bfs}_\ast}

\newcommand{\bmthe}{{\boldsymbol{\theta}}}  
\newcommand{\bmphi}{{\boldsymbol{\beta}}}
\newcommand{\bmze}{{\boldsymbol{\zeta}}}  

\newcommand{\bXo}{\mathbf{x}}

\newcommand{\xsam}{\bXo_{\mathrm S}}
\newcommand{\ysam}{\mathbf{y}_{\mathrm S}}
\newcommand{\haysam}{\mathbf{\hat{y}}_{\mathrm S}}

\newcommand{\Samp}{{\mathbb{S}_N}}
\newcommand{\bfrho}{\boldsymbol{\rho}}
\newcommand{\bfI}{\mathbf{I}}

\newcommand{\mx}{\mu_X}
\newcommand{\varx}{\sigma^{2}_{X}}

\newcommand{\noise}{\epsilon}

\newcommand{\la}{\lambda}
\newcommand{\hyperg}{{}_{1}F_{1}}

\newcommand{\beq}{\begin{equation}}
\newcommand{\eeq}{\end{equation}}

\newcommand{\bit}{\begin{itemize}}
\newcommand{\eit}{\end{itemize}}
\newcommand{\erf}{\mathrm{erf}}

\newcommand{\argmax}{{\mathrm{argmax}}}

\newtheorem{definition}{Definition} 
\newtheorem{propo}{Proposition}
\newtheorem{rem}{Remark}

\newtheorem{theorem}{Theorem}
\newtheorem{lemma}{Lemma}
\usepackage{pdfpages} 
\usepackage{pgffor} 
\usepackage{xr} 

\makeatletter
\AtBeginDocument{\let\LS@rot\@undefined}
\makeatother

\newif\ifarXiv
\arXivtrue

\begin{document}
	
\title{Stochastic Processes with Modified Lognormal Distribution Featuring  Flexible Upper Tail}
	
\author{Dionissios T. Hristopulos
\IEEEmembership{Senior Member, IEEE}
and Anastassia Baxevani and Giorgio Kaniadakis
\thanks{D. T. Hristopulos is with the School of Electrical and Computer Engineering, Technical University of Crete, Chania 73100, Greece (e-mail: dchristopoulos@tuc.gr).}
\thanks{Anastassia Baxevani is with the Department of Mathematics and Statistics, University of Cyprus, Nicosia, Cyprus (e-mail: baxevani@ucy.ac.cy)}
\thanks{Giorgio Kaniadakis is with the Department of Applied Science and Technology, Politecnico di Torino, 10129 Turin, Italy (e-mail: giorgio.kaniadakis@polito.it)}
}	
\markboth{}{Hristopulos, Baxevani, Kaniadakis: $\kp$-Lognormal  Stochastic Processes}

\maketitle

\IEEEpeerreviewmaketitle

\begin{abstract}
Asymmetric, non-Gaussian probability distributions are often observed in the analysis of natural and engineering datasets. The lognormal distribution is a standard model for data with skewed frequency histograms and fat tails. However, the lognormal law severely restricts the asymptotic dependence of the probability density and the hazard function for high values. Herein we present a family of three-parameter non-Gaussian probability density functions that are based on generalized \kpe and \kpl  functions and investigate its mathematical properties.  These $\kp$-lognormal
densities represent continuous deformations of the lognormal with lighter right tails,  controlled by the parameter $\kp$.  In addition, bimodal distributions are obtained for certain parameter combinations. We derive closed-form analytic expressions for the main statistical functions of the $\kp$-lognormal distribution. For the moments, we derive   bounds that are based on hypergeometric functions as well as series expansions.  Explicit expressions for the gradient and Hessian of the negative log-likelihood are obtained to facilitate numerical maximum-likelihood estimates of the $\kp$-lognormal parameters from data.  We also formulate a joint probability density function for $\kappa$-lognormal stochastic processes by applying Jacobi's multivariate theorem to a latent Gaussian process.  Estimation of the $\kp$-lognormal distribution based on synthetic and real data is explored. Furthermore, we investigate applications of $\kp$-lognormal processes with different covariance kernels in   time series forecasting and spatial interpolation using warped Gaussian process regression. Our results are of practical interest for modeling skewed distributions in various scientific and engineering fields.
\end{abstract}
		
	
	
\section{INTRODUCTION}
\label{sec:Intro}

\IEEEPARstart{T}{he} Gaussian (normal) distribution is one of the most popular models in science. It is used for point-like random variables, for stochastic processes that exhibit temporal dependence~\cite{Papoulis02}, and for spatially extended processes~\cite{Christakos92,Chiles12}.  It is also widely used in machine learning,  as for example  in the  Gaussian process regression framework~\cite{Rasmussen06,dth22}. Despite the dominant role of the Gaussian model, data often display skewed probability distributions with heavy tails~\cite{Andersson21}.  Such distributions can assign  considerably higher probabilities than the Gaussian law for large deviations from the mean. A paradigmatic heavy-tailed distribution  is the lognormal model~\cite{Sornette04a}.

The lognormal distribution is ubiquitous in natural and engineered processes. It is used   to model intermittent energy dissipation in fluid turbulence~\cite{Mccomb90,Ditlevsen10,Benzi23,Pearson18} and the distribution of local stiffness in living cells~\cite{Meng24}; it finds applications in  mining~\cite{Journel80,Krige81,Rivoirard13,Cressie06}, geology~\cite{Kondolf00}, hydrology~\cite{Adlouni08,Kedem87}, geochemistry~\cite{Zhang05}, lifetime analysis of electronic components~\cite{Zhang12} and
wireless communications~\cite{Mehta07,Heliot09,Enserink13}.
The lognormal is used to model the distribution of permeability~\cite{Dagan93,Wen96,dth03awr,Sudicky10,Zhou10} and fluid velocity~\cite{Kohlbecker06} in porous media,  particle size distributions in coagulation-free vapor growth processes~\cite{Soderlund98}, and the distribution of matter in the universe~\cite{Coles91,Xavier16,Tosone20}.
It is also a candidate distribution for earthquake recurrence times~\cite{Nadeau95,Abaimov07,Abaimov08,dth14}, and  it is  employed in pharmacokinetics due to its ability to model asymmetry and positive skewness~\cite{Elassaiss20}.
As emphasized in~\cite{Buzski14}, the lognormal distribution is crucial in the description of brain processes, reflecting the fact that in highly interconnected systems operations  tend to be multiplicative rather than additive.
Finally, the lognormal is used to  model heteroskedastic stochastic processes, because it captures   variance changes over time, as for example, in the well-known Black-Scholes-Merton model~\cite{Black73,Merton73}. Theoretical generative models for the lognormal  distribution are reviewed in~\cite{Koch66,Mitzenmacher04,Sornette04a,Mouri13}. The impact of the lognormal's upper (right) tail on extreme events is discussed in~\cite{Taleb20}.

The upper tail of the lognormal may be too ``heavy'' for certain applications. This happens if the tail assigns unrealistically high probabilities to large values of the process.
In such cases, truncated lognormal models are used~\cite{Johnson94,Verster12}.  For example, the Kolmogorov-Avrami-Mehl-Johnson crystallization model of random nucleation and growth exhibits a grain size distribution which evolves into a truncated lognormal form for specific nucleation and growth rates~\cite{Teran10}. A shorter-than-lognormal tail  was also observed in ecological population abundance data~\cite{Halley02,Sizling09}. The pore size distribution of hydrated cement paste  exhibits a lighter-than-lognormal   right tail,  due to an upper limit of the particle size~\cite{Diamond72}. Geochemical data also show evidence of
a lighter-than-lognormal tail~\cite{Zhang05}.  Finally, the lognormal model has an asymptotically declining hazard rate which is a potential shortcoming for failure time analysis.

The above considerations motivate the need for asymmetric, right-skewed probability distributions with a flexible right tail which is  lighter than the lognormal tail.  Such a distribution model optimally should continuously deform into the natural lognormal model by controlling a deformation parameter. This paper pursues this quest by exploiting the Kaniadakis deformation of the exponential and logarithmic functions to introduce the $\kp$-lognormal probability model. The latter is shown to incorporate the natural lognormal as well as asymmetric distributions with lighter tails;  it also closely approximates the normal distribution for certain parameter combinations, while in certain regions of the parametric space it leads to bimodal distributions. We discuss the mathematical properties of the $\kp$-lognormal marginal distribution (including the statistical moments and the hazard rate)   and their dependence on the parameter $\kp$.  We  formulate $\kp$-lognormal stochastic processes based on Jacobi's multivariate theorem and the \kpe transformation. Predictive equations in the framework of \emph{warped Gaussian processes}, suitable for $\kp$-lognormal processes, are also formulated; these are used for time series forecasting  and spatial interpolation.

The remainder of this paper has the following structure:
Section~\ref{sec:Methods} presents necessary notation and definitions, the main properties  of modified exponential and logarithm functions, as well as a short review of the lognormal distribution and associated stochastic process. In Section~\ref{sec:kappa-lognormal} we derive the $\kp$-lognormal distribution and its main statistical functions. We also present bounds and a power-series expansion for the statistical moments. In addition, we derive the hazard rate of the $\kp$-lognormal and investigate its asymptotic behavior. In Section~\ref{sec:kp-lognormal-process} we introduce the $\kp$-lognormal stochastic process in terms of the joint probability density function.  We also formulate predictive  equations by focusing on warped  Gaussian process regression. We formulate maximum likelihood estimation for the $\kp$-lognormal distribution in Section~\ref{sec:estim}. In Section~\ref{sec:applications} we study applications related to the estimation of the probability model, time series forecasting, and spatial regression using simulated and real datasets.   Finally, Section~\ref{sec:Conclusions} presents a brief discussion of the results and some open questions. Longer proofs are relegated to the Appendix, and additional information is presented in the Supplement.

\IEEEpubidadjcol
\section{Methodological Background} \label{sec:Methods}
In this section, we introduce notation and  review the main mathematical properties of lognormal stochastic processes, as well as the $\kappa$-exponential and $\kappa$-logarithmic functions.  We also present  new results regarding the dependence of the $\kp$-modified functions on $\kp$ (cf. Propositions~\ref{propo:bounds-kpe}, \ref{propo:kpl-increase}, and~\ref{propo:kpl-convexity}).  Certain useful properties of the \kpe and \kpl that are not used herein are given in the Supplement (Section~1).

\subsection{Notation}

\begin{enumerate}[wide,labelwidth=!,labelindent=0pt]\itemsep0.3em

\item Vector and matrix quantities will be denoted by boldface font, e.g., $\mathbf{A}$.  The superscript $\top$, e.g., $\mathbf{A}^\top$, denotes the  transpose of the vector or matrix $\mathbf{A}$.

\item $\R$ denotes the set of real numbers and $\mathbb{N}$ the set of positive integers; $\bfs \in \Do \subset \Rd$ denotes a vector in a $d$-dimensional feature space where $d \in \mathbb{N}$.

\item $\N(\mu, \sigma^2)$ denotes the normal distribution with mean $\mu$ and variance $\sigma^2$; $Y(\om) \overset{d}{=}\N(\mu, \sigma^2)$ means that the random variable $Y(\om)$ follows the normal distribution with mean $\mu$ and variance $\sigma^2$. $\phi(\cdot)$  and $\Phi(\cdot)$ denote the probability density and the cumulative  distribution functions of a standard normal distribution respectively.

\item ${\rm LN}(\mu,\sigma)$ denotes the lognormal distribution with parameters $\mu \in \R$ and  $\sigma>0$, and $\kp{\rm LN}(\mu,\sigma,\kp)$, where $\kp\ge 0$, the $\kp$-lognormal distribution.

\item The symbol $\triangleq$  is used to define  mathematical entities.

\item $\Xsom$, $\rfsom{Y}$ represent  random variables  defined over a probability space $(\Omega, \mathcal{F}, P)$ indexed by  $\bfs$,  where $\om \in \Omega$ is the state index in the sample space $\Omega$, $\mathcal{F}$ is a $\sigma-$algebra of events, and $P$ is the probability measure for these events~\cite{Yaglom87,Papoulis02,dth20}.

\item $\mathcal X=\{ \Xsom, \bfs \in \Do, \om \in \Omega \}$ and $\mathcal Y=\{\rfsom{Y}, \bfs \in \Do,\om \in \Omega \}$ represent stochastic processes with the index $\bfs \in \Do$. In the following, we will use for brevity $\Xsom$  and $\rfsom{Y}$ to refer to the stochastic processes $\mathcal X$ and $\mathcal Y$ respectively as well as the corresponding random variables.

\item The expectation operator over  the probability space is denoted
 by $\EE[\cdot]$.

\item The vector $\xsam \triangleq (x_{1}, \ldots, x_{N})^{\top}$ denotes sample values of $\Xsom$ for the respective sampled features  $\Samp=\{ \bfs_{1}, \ldots, \bfs_{N} \}$, where $\bfs_{i} \in \Do \subset \Rd, \, i=1, \ldots, N$. In real-world scenarios, the sampled values involve a noise term.  For simplicity, we will assume Gaussian white noise $\noise_i \overset{d}{=}\N(0, \sigma_{\noise}^2)$.

\item The symbol $\ln(\cdot)$ denotes the natural logarithm, while $\erf(\cdot)$ is the error function defined  as $\erf(z)=\frac{2}{\sqrt{\pi}}\int_{0}^z e^{-t^2}\dd{t}$.

\item The symbols $\expk(\cdot)$ and $\lnk(\cdot)$ will be used for the modified exponential and logarithmic functions respectively. If the deformation factor $\kp$ is viewed as a parameter, the total derivative $\dd \expk(x)/\dd x$ will be used. To study the dependence of $\kp$-modified functions on $\kp$,  partial derivatives, namely, $\pa \expk(x)/\pa x$ and $\pa \expk(x) /\pa \kp$ will be used.

\item Sample-based averages will be denoted by the overline symbol, for example,  $\overline{x}$ is the sample mean. Sample-based estimates of a distribution parameter $\theta$ will be denoted by  $\hat{\theta}$.

\end{enumerate}


The following two sections focus on the definitions and most relevant properties of the \kpe and the \kpl.  The mathematical properties of the \kpe and the \kpl are thoroughly studied in~\cite{Kaniadakis01a}.

\subsection{The $\kp$-Exponential Function}
\label{ssec:kappa-exp}

The base of the exponential function is the Euler number $e$ which can be defined as the  limit\[
e = \lim_{n \to \infty} \left( 1 + \frac{1}{n}\right)^{n}\,.
\]
Therefore, the natural exponential function $\exp(y), \, y \in \R\,,$ corresponds to the limit
\beq
\label{eq:exponential}
\exp(y) = \lim_{n \to \infty}\left( 1 + \frac{y}{n}\right)^{n}, \; \mbox{where}\; n \in {\mathbb N}\,.
\eeq

\noindent The Kaniadakis $\kappa$-exponential is defined by means of the expression~\cite{Kaniadakis01,Kaniadakis05}
\begin{equation}
    \label{eq:exp-kappa}
    \expk(y) \triangleq \left( \sqrt{1 +  {\kappa^2}{y^2} }  +
    {y}{\kappa}\right)^{1/\kappa},
\end{equation}
with $0 \leq \kappa <1$ and $y \in \R$. The function~\eqref{eq:exp-kappa} emerges naturally within the framework of special relativity, where the deformation
factor $\kappa$ is proportional to the reciprocal light speed
~\cite{Kaniadakis09}. In that context, $\expk(y)$ is the
relativistic generalization of the natural exponential $\exp(y)$.  Herein we take an  agnostic viewpoint and consider~\eqref{eq:exp-kappa} as a mathematical generalization of $\exp(y)$.

\begin{figure}
\centering
\includegraphics[width=0.5\linewidth]{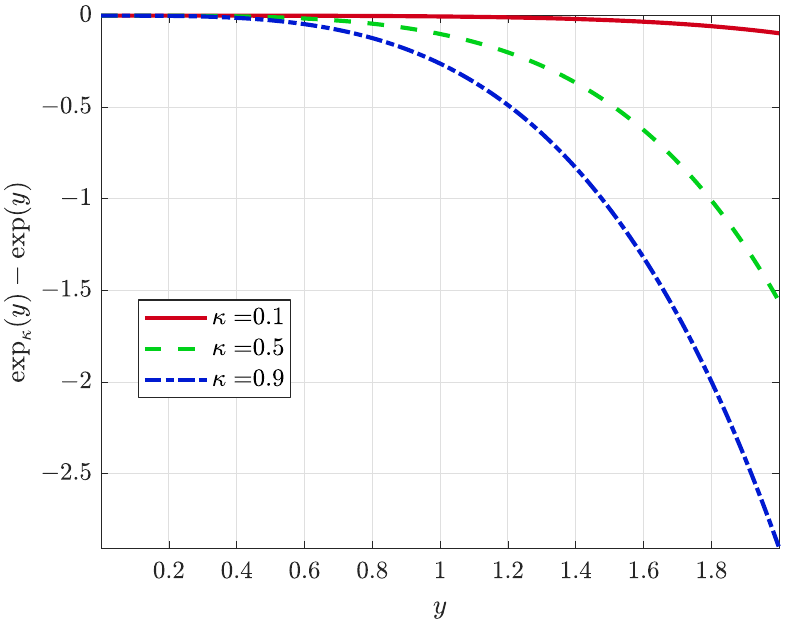}
\caption{Plots of the difference $\expk(x)-\exp(x)$ for  $ \kappa=0.1, 0.5, 0.9$ based on the \kpe definition~\eqref{eq:exp-kappa}  for $x \ge 0$, which confirm that the natural exponential is the upper bound of the \kpe, namely $\exp(x) \ge \expk(x)$, for $x \ge 0$ (cf. Proposition~\ref{propo:bounds-kpe}).  For $x<0$ (not shown) the sign of the difference is reversed, that is, $\expk(x) \ge \exp(x)$, and the natural exponential is a lower bound of the \kpe. }
\label{fig:expk}
\end{figure}

By replacing $1/\kappa$ with $n$ and realizing that at the limit $\kappa \to 0$ ($n \to \infty$), $1+{\kappa^2}{y^2} \approx 1$, it follows that the $\kappa \to 0$  limit of the modified exponential~\eqref{eq:exp-kappa} is the same as the $n \to \infty$ limit in~\eqref{eq:exponential}.  The \kpe function is visually compared to the natural exponential in Fig.~\ref{fig:expk}.  While in the relativistic theory the constraint $0\le \kp < 1$ makes sense (strictly speaking,   $-1 < \kp <1$), mathematically the \kpe~\eqref{eq:exp-kappa} and other $\kappa$-modified functions can be defined for $\kp>1$. Herein, we thus focus on $\kp \ge 0$.
A Taylor series expansion of $\expk(y)$ around $y=0$ is available~\cite{Kaniadakis13}.  We elaborate on this expansion in the Supplement (Section~1).

\medskip \paragraph{Asymptotic behavior}
The function $\expk(y)$ exhibits
\emph{power-law} asymptotic behavior as $y\to \infty$~\cite{Kaniadakis05, Kaniadakis13}:
\beq
\expk(y) \sim  \big|\,2\kp y\big|^{\,\pm1/\kp}, \; \mathrm{as} \; y\to \pm\infty.
\eeq

\noindent In light of the above,  the \kpe  $\expk(-y)$ exhibits a \emph{heavy right tail} for $y \to +\infty$,
i.e.
\beq
\label{eq:power-law-kpe}
\expk(-y) \sim (2\kp y)^{\,-1/\kp}, \; \mathrm{as} \; y\to \infty.
\eeq
Hence, $\expk(-y)$ can be used to model probability distributions with high probability density (power law)  in the  tail.

\medskip \paragraph{Monotonicity and convexity} The \kpe is continuously differentiable with respect to $y$. Its derivatives are proportional to
$\expk(y)$, but they also involve a modifying factor that depends on $\kp$ and $y$ and vanishes at $\kp=0$.
The first two derivatives are given by the following expressions
\beq
\label{eq:expk-first-deriv}
\frac{\dd \expk(y)}{\dd y} = \frac{\expk(y)}{ \sqrt{1 + (\kp\,y)^2} }\,,
\eeq
\beq
\frac{\dd^2 \expk(y)}{\dd y^2} = \expk(y) \,\frac{\sqrt{1 + \kp^{2}\,y^2} - \kp^2\,y}{ \left( \,1 + \kp^{2}\,y^2 \,\right)^{3/2} }\,.
\label{eq:expk-second-deriv}
\eeq
It follows from~\eqref{eq:expk-first-deriv} that the first derivative is nonnegative for all $y \in \R$. Hence, the
\kpe is a monotonically increasing function of $y$ for fixed $\kp$.  Since $\expk(y)$ and the denominator in~\eqref{eq:expk-second-deriv} are nonnegative for all $y \in \R$, it is easy to see that the \kpe is a convex function of $y$ for all $\kp \in [\,0,1\,]$. The numerator $\sqrt{1 + \kp^{2}\,y^2} - \kp^2\,y$ is clearly positive for $y<0$, while for  $y>0$ it is enough to observe that $1 + (\kp\,y)^2 - \kp^4\,y^2=1+ \kp ^2 y^2(1-\kp^2)>0$ for all $\kp \in[0,1]$.

\medskip
\begin{propo}[Bounds of \kpe]
\label{propo:bounds-kpe}
The \kpe  decreases monotonically as a function of $\kp \ge 0$ for $y \ge 0$ and  increases monotonically  as  function of  $\kp \ge 0$ for $y <0$.  Thus, $\exp(y) \le \expk(y)$ for $y<0$  and $\exp(y) \ge \expk(y)$  for $y \ge 0$ for all $\kp\ge 0$.
\end{propo}

\smallskip

\begin{IEEEproof}
Viewing $\expk(y)$  as a function of both $y$ and $\kp$,  its partial derivative  with respect to $\kp$ is given by
\begin{align}
\label{eq:expk-deriv-kappa}
\frac{\pa \expk(y)}{\pa \kp }= & \frac{g(\kp\,y)\, \expk(y)}{\kp^{2}\sqrt{1 +  {\kappa^2}{y^2} }}\,,
\\[1ex]
g(\kp\,y)=  & -\sqrt{1 +  {\kappa^2}{y^2} }\, \ln\left( \sqrt{1 +  {\kappa^2}{y^2} }+ y\kp\right) + \kp\,y\,.
\end{align}
The denominator of the \kpe's partial derivative is positive for all $y \in \R$ and $\kp > 0$.  The sign of the partial derivative  $\pa \expk(y)/\pa \kp$ is determined by the the sign of  $g(\kp y)$. To find the latter, consider that   $\pa g(\kp y)/\pa y$ is given by
\[
\frac{\pa g(\kp y)}{\pa y}= -\frac{\ln(\sqrt{1+{\kappa^2}{y^2}} + \kp y)\, y\kappa^2}{\sqrt{1+ {\kappa^2}{y^2}}}\,.
\]
For $y>0$ it holds that $\sqrt{1+{\kappa^2}{y^2}} + \kp y>1$, the logarithm of the above is positive, and thus $\pa g(\kp y)/\pa y$ is negative for $y>0$.
For $y<0$, since $\sqrt{1+{\kappa^2}{y^2}} < 1 + \lvert y \rvert \kp$, it also holds that $y\, \ln(\sqrt{1+{\kappa^2}{y^2}} + \kp y) >0$ since the argument of the logarithm is less than one.  Hence,   $\frac{\pa g(\kp y)}{\pa y}  < 0$ for $y < 0$ as well. Finally, for $y=0$ it holds that $\frac{\pa g(\kp y)}{\pa y}=0$.  Therefore, $g(\kp y)$ is  a monotonically decreasing function for all $y \in \R$ and $\kp>0$.

The monotonic decline of $g(\cdot)$, together with the fact that $g(0)=0$, implies that $g(\cdot)$ takes negative (positive) values for $y>0$ ($y<0$).
Since the sign of ${\pa \expk(y)}/{\pa \kp }$ is controlled by $g(\kp y)$ according to~\eqref{eq:expk-deriv-kappa}, it holds that $\pa \expk(y)/\pa \kp < 0$ for any $y \ge 0$ and $\kp > 0$ while  $\pa \expk(y)/\pa \kp > 0$ for any $y < 0$ and $\kp > 0$.

At $\kp=0$, the derivative~\eqref{eq:expk-deriv-kappa} vanishes---as can be shown by applying de l'Hospital's rule and noticing that the leading-order term of the Taylor expansion for $g(\kp y)$ around $\kp=0$ is proportional to $\kp^{3}$. Thus, $\exp(y)$ provides an upper bound of the $\kp$-exponential for $y \ge 0$. For $y<0$ the fact that $g(\kp y)>0$  implies that $\expk(y)$ is an increasing function of $\kp$; thus, $\exp(y) \le \expk(y)$ for $y<0$ and the exponential provides a lower bound of $\expk(y)$.
\end{IEEEproof}

\medskip

The exponential upper bound of the \kpe is illustrated in Fig.~\ref{fig:expk}.

\subsection{The $\kp$-Logarithm Function}
\label{ssec:kappa-logarithm}

\begin{figure}
\centering
\includegraphics[width=0.5\linewidth]{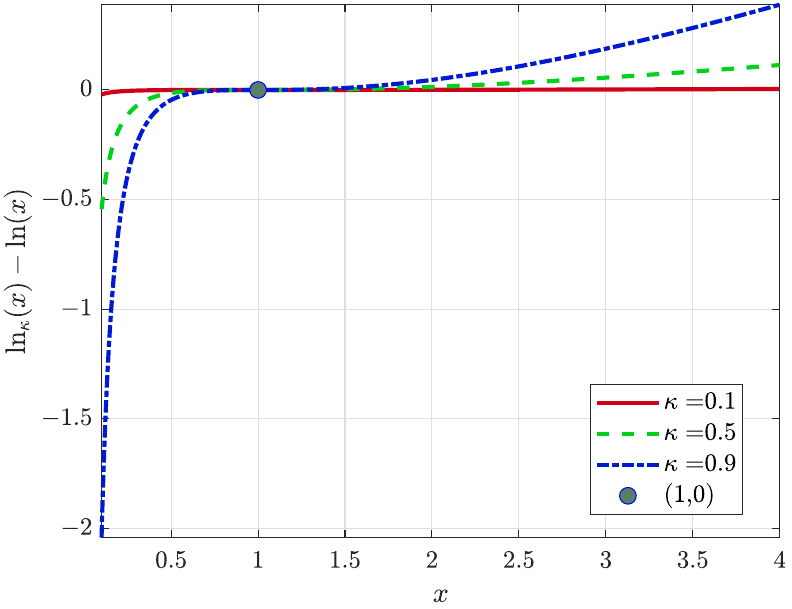}
\caption{Plots of the difference $\expk(y)-\exp(y)$ for  $ \kappa=0.1, 0.5, 0.9$ based on the \kpe definition~\eqref{eq:exp-kappa}  for $y \ge 0$, which confirm that the natural exponential is the upper bound of the \kpe, namely $\exp(y) \ge \expk(y)$, for $y \ge 0$ (cf. Proposition~\ref{propo:bounds-kpe}).  For $y<0$ (not shown) the sign of the difference is reversed, that is, $\expk(y) \ge \exp(y)$, and the natural exponential is a lower bound of the \kpe.}
\label{fig:lnk}
\end{figure}

The inverse of the $\kappa$-exponential is the
\emph{$\kappa$-logarithm}~\cite{Kaniadakis13}, defined by the following function for $x>0$:
\begin{equation}
    \ln_{\kappa}(x) \triangleq \frac{x^{\kappa}-x^{-\kappa}}{2\kappa}\,, \quad \kappa >0\,.
    \label{eq:kappa_log}
\end{equation}
The \kpl function is compared to the natural logarithm in Fig.~\ref{fig:lnk}.

The $\kappa$-logarithm can be used to normalize  skewed, non-negative data by means of the transformation $x \mapsto y \triangleq \lnk(x)$. The \kpl transform is similar to the well-known Box-Cox transform $x \mapsto y \triangleq (x^\lambda -1)/\lambda$~\cite{Box64}.  Both the Box-Cox and the $\kp$ logarithm yield the logarithmic transformation at the respective limits $\lambda \to 0$ and $\kp \to 0$. However, the inverse of the $\kappa$-logarithm is the \kpe $y \mapsto x \triangleq \expk(y)$ which is stable even for $y<0$ as evidenced in~\eqref{eq:exp-kappa}. In contrast, the inverse Box-Cox transform $y \mapsto x \triangleq (\lambda y +1)^{1/\lambda}$  fails for $\la y + 1<0$~\cite{dth22_kan}.

\medskip \paragraph{Taylor expansion} It follows from~\eqref{eq:kappa_log} that as $x \to 0$, $\lnk(x) \to -\infty$ just as the natural logarithm.  It is also straightforward to see---by writing $x^\kp = \exp(\kp\ln (x))$ and using the Taylor series expansion of the exponential---that $\lim_{\kp \to 0} \lnk(x) = \ln (x)$. In particular, the small-$\kp$ expansion of~\eqref{eq:kappa_log} yields

\beq
\label{eq:lnk-expansion}
\lnk(x) = \ln(x) + \frac{1}{6} \,\kp^2 \,\ln(x)^3 + \frac{1}{120} \,\kp^4 \,\ln(x)^5 + \Or(\kp^6) \,.
\eeq
Hence, the above expansion recovers the natural logarithm for $\kp=0$.  According to expansion~\eqref{eq:lnk-expansion}, $\lnk(x) \le \ln(x)$ for $x \le 1$ and $\lnk(x) > \ln(x)$ for $x > 1$. Hence, the difference $\lnk(x)-\ln(x)$ changes sign at $x=1$ as evidenced in Fig.~\ref{fig:lnk}.

\medskip \paragraph{Asymptotic behavior}
For $\kp >0 $, it follows that
\beq
\label{eq:lnk-asympt}
\lnk(x) \sim x^{\kp}/(2\kp), \; \mathrm{as} \; x\to \pm\infty \,.
\eeq

\medskip \paragraph{Monotonicity of \kpl with respect to $\kp$} Partial derivatives are introduced here  to consider variations of $\lnk(x)$ with respect to both $x$ and $\kp$. The derivative of $\lnk(x)$ with respect to $\kp$ is given by
\beq
\label{eq:lnk-deriv-kappa}
\frac{\pa\lnk(x)}{\pa\kp}= \ln(x)\, \frac{x^{k}+ x^{-k}}{2\kp}-\frac{\lnk(x)}{\kp}\,.
\eeq

\beq
\label{eq:lnk-first-deriv}
\frac{\pa \lnk(x)}{\pa x} = \frac{1}{2x} \left(x^{\kp} + x^{-\kp} \right),
\eeq
\beq
\label{eq:lnk-second-deriv}
\frac{\pa^2 \lnk(x)}{\pa x^2} =\frac{(\kp-1)x^{\kp}-(\kp+1)x^{-\kp}}{2x^2} \,.
\eeq

Note that $\pa \lnk(x)/\pa x >0$ for all $x \ge 0$ and $\kp > 0$,
and thus the \kpl  is a monotonically increasing function.

\medskip

\begin{propo}[\kpl increases faster than the natural logarithm]
\label{propo:kpl-increase}
(i) The \kpl defined by~\eqref{eq:kappa_log} increases faster than the natural logarithm for all $x>0$. (ii) The rate difference $d_{\ln}(x,\kp) \triangleq \frac{\pa \lnk(x)}{\pa x}- \frac{\dd \ln(x)}{\dd x}$ increases with increasing $\kp$, indicating a faster increase of the \kpl with $x$ for higher $\kp$.
\end{propo}

\smallskip

\begin{IEEEproof}
(i) We need to show that $\frac{\pa \lnk(x)}{\pa x}> \frac{\dd \ln(x)}{\dd x}$ for all $x>0$. Since $\dd \ln(x)/\dd x=1/x$, the above is equivalent to $x^\kp + x^{-\kp} >2$ for $x > 0$.  Defining $z \triangleq x^\kp$, it suffices to show that $z +1/z>2$; since $z>0$, this is equivalent to $z^2 -2z +1 = (z-1)^2>0$ which is true for all $z \in \R$ (and thus for $z>0$).

\smallskip
(ii)   The $\kp$-derivative of the rate difference,  $\frac{\pa d_{\ln}(x,\kp)}{\pa \kp} =\frac{\pa^{2} \lnk(x)}{\pa x \pa \kp}=\frac{\ln(x)}{2x}(x^{\kp}-x^{-\kp})$, is nonnegative  for all $x>0$: for $0<x<1$ both  $\ln(x)$ and $x^{\kp}-x^{-\kp}$ are negative, while  both functions are positive for $x >1$.
\end{IEEEproof}

\medskip

\begin{propo}[\kpl convexity]
\label{propo:kpl-convexity}
(i) For  $0 \le  \kp \le 1$, the \kpl  is concave for all $x>0$. (ii) For  $\kp>1$ the \kpl is concave for $0<x<x_{+}$ and convex for $x>x_{+}$ where $x_{+} \triangleq \left( \frac{\kp +1}{\kp -1}\right)^{1/2\kp}$ is a $\kp$-dependent \emph{inflection point}.
\end{propo}

\begin{IEEEproof}
(i) By inspection of~\eqref{eq:lnk-second-deriv}, it follows that the second derivative $\frac{\pa^2 \lnk(x)}{\pa x^2}$ is  negative  for $\kp \in [0,1]$ and $x>0$,  which implies that $\lnk(x)$ is \emph{concave}.

(ii) For  $\kp >1$, the zeros of the second-order derivative~\eqref{eq:lnk-second-deriv}  are determined by the zeros of the function  $g(z) \triangleq (\kp -1)z-(\kp +1)z^{-1}$ which represents the numerator of the fraction, where $z\triangleq x^{\kp} >0$. The zeros of $g(z)$ are the same as the two non-zero roots of $z\,g(z)=0$, namely $ z_{\pm } = \pm x_{+}^{\kp}$.  In addition, the sign of $z\,g(z)$ is the same as that of $g(z)$ for $z>0$. Hence, since $z\,g(z)= (\kp-1)(z-z_{+})(z- z_{-})$, the sign of $z\,g(z)$ is negative for $0<z<z_{+}$ and positive for $z>z_{+}$; consequently, the sign of the second-order derivative of $\lnk(x)$  is negative for $0\le x< x_{+}$ and positive for $x>x_{+}$ where $x_{+} \triangleq \left( \frac{\kp +1}{\kp -1}\right)^{1/2\kp}.$ Therefore, $x_{+}$ is an inflection point, where the behavior of $\lnk(x)$ changes from concave to convex.
\end{IEEEproof}

\medskip

\smallskip

\medskip

\subsection{Review of Lognormal Stochastic Processes}

\begin{definition}[Lognormal stochastic process]
\label{defi:lognormal-stochastic-process}
Given a probability space $(\Omega, \mathcal{F}, \textsl{P})$,  the collection of real-valued, scalar random variables  $ \{ Y(\bfs,\om): \bfs \in \Do, \, \om \in \Omega \}$ is a scalar, real-valued \emph{stochastic process} $Y: \Do \times  \Omega \mapsto \R$.
Let $\{Y(\bfs;\om)\}$  represent a \emph{stationary Gaussian stochastic process} with mean $\mu$, variance $\sigma^2$ and covariance kernel $C_{Y}(\bfr)$, where $\bfr$ is the  lag between any two points  $\bfs, \bfs' \in \Do$. Then, the stochastic process defined by $X(\bfs;\om)=\exp\left[Y(\bfs;\om)\right]$ $:\Do \times  \Omega \mapsto \R_{+}$
is a \emph{stationary lognormal stochastic process}.
\end{definition}

\medskip
The main statistics and probability functions of a lognormal stochastic process are given below.

\begin{enumerate}[wide,labelwidth=!,labelindent=0pt]\itemsep0.4em
\item Marginal probability density function (PDF)
\beq
\label{eq:lognormal}
f_{X}(x)=\frac{1}{\sqrt{2\pi}\sigma\, x}\E^{-\left(\ln (x) -\mu\right)^2 / 2\sigma^2}=\frac{1}{x\sigma}\phi\left( \frac{\ln (x) - \mu}{\sigma}\right)\;, \; x \in (0, +\infty)\,,
\eeq
where $\phi(\cdot)$ is the PDF of the standard normal distribution.

\item Median ($\xmed$), mode ($x_{\ast}$), and mean ($\mu_{X}$):
\beq
\label{eq:ln-mean-mode-median}
\xmed= \E^{\mu}\;, \; x_{\ast}= \E^{\mu - \sigma^2}\;, \; \mu_{X}= \E^{\mu + \sigma^2/2}\,.
\eeq
The median of the lognormal is also known as the \emph{geometric mean}  of $X(\bfs;\om)$.

\item Variance ($\sigma^{2}_{X}$) and skewness ($s_{X}$):

\beq
\label{eq:logn-var-skew}
\sigma^{2}_{X}= \mu_{X}^{2}\, \left[ \exp(\sigma^2)-1\right]\, , \; s_{X}=\left[ \exp(\sigma^2) + 2\right]\, \sqrt{\exp(\sigma^2)-1}\, .
\eeq

    \item The marginal cumulative distribution function (CDF) is defined as $F_{X}(x) = P(X(\om)\le x)$, where $x>0$:
    \beq
    F_{X}(x) = \frac{1}{2}\left[1 + \erf\left( \frac{\ln (x) - \mu}{\sigma\sqrt{2}}\right)  \right]= \Phi\left(\frac{\ln (x)-\mu}{\sigma}\right) \,,
    \eeq
    where $\Phi(\cdot)$ is the  CDF of the standard normal distribution.

    \item The quantile function $x_{p}=Q_X(p)$ is defined so that  $P\left(X (\om)\le x_{p}\right)=p$, where $0 \le p \le 1$:
    \beq
    \label{eq:qf-ln}
    Q_X(p) = \exp\left[ \mu + \sqrt{2\sigma^2}\, \erf^{-1}(2p-1)\right]\,.
    \eeq

    \item Covariance kernel
    \beq
    \label{eq:logn-cova}
    C_{X}(\bfr)= \mu_{X}^{2}\,\left( \E^{C_{Y}(\bfr)}-1 \right)\,.
    \eeq

\end{enumerate}

\medskip

\section{Modified Lognormal Distribution}
\label{sec:kappa-lognormal}

\begin{definition}[$\kp$-lognormal distribution]
\label{defi:kln-rv}
If $ Y(\om): \om \in \Omega$ represents a scalar, real-valued  Gaussian random variable in the probability space $(\Omega, \mathcal{F}, \textsl{P})$ with mean $\mu$, and variance $\sigma^2$,
the  random variable $X(\om)$  defined by means of the \kpe transformation
\begin{subequations}
\label{eq:kp-transforms}
\beq
\label{eq:kln-rv}
X(\om) \triangleq \expk\left[Y(\om)\right] = \left[ \sqrt{  1 +  Y^{2}(\om){\kp^2} }  + Y(\om) {\kp}\right]^{1/\kp}\,, \; \kp \ge 0\,,
\eeq
follows the $\kappa$-lognormal distribution, $\kp{\rm LN}(\mu,\sigma,\kp)$.
The inverse transformation  is given by the \kpl, that is,
\beq
\label{eq:kln-rv-inv}
Y(\om) \triangleq \lnk \left[ X(\om)\right] = \frac{X^{\kp}(\om) - X^{-\kp}(\om)}{2\kp}.
\eeq
\end{subequations}
\end{definition}

\smallskip

\noindent The marginal probability functions for the $\kp$-lognormal distribution are defined below.


\subsection{Probability Functions of $\kp$-Lognormal Distribution}
\label{ssec:kp-logn-probab}

\begin{enumerate}[wide,labelwidth=!,labelindent=0pt]\itemsep0.4em

\item Marginal probability density function (PDF)

By  defining the nonlinear  monotone transformation $x= g(y)$, where $g(\cdot) = \expk(\cdot)$, and
$ y = g^{-1}(x)$ where $g^{-1}(\cdot) = \lnk(\cdot)$, the
 random variable $X=g(Y)$
follows the \emph{$\kp$-lognormal distribution}, if $Y$ is normally distributed. The respective PDF is given by the following function

\beq
\label{eq:klogn-pdf}
f_{X}(x) = \frac{1}{2\, \sqrt{2\pi}\sigma} \, \E^{- \left(\lnk (x) - \mu\right)^2/2\sigma^2} \,
\left( x^{\kp-1} + x^{-\kp-1}  \right), \; x>0\,, \, \kp \ge 0\,.
\eeq


\noindent The $\kp$-lognormal PDF is obtained using the conservation of probability under the monotone variable transformation  by means of the PDF of a normal $\N(\mu, \sigma^2)$ and the derivative of \kpl which is given by \eqref{eq:lnk-first-deriv}.
This PDF decays faster (has a shorter right tail) than the lognormal, because the function $\lnk(y)$ increases with $y$ faster than the natural logarithm $\ln(y)$.

A comparison of the $\kp$-lognormal with the lognormal PDF for $\kp <1$ is shown in the left frame of Fig.~\ref{fig:lognormk-PDF}. In this case the $\kp$-lognormal  is similar to the lognormal PDF.
For all values of
$\kp$, the right tail of the $\kp$-lognormal density is shorter than that of the lognormal, while it approaches the lognormal tail in the limit $\kp \to 0$. On the other hand, the $\kp$-lognormal density has more weight than the lognormal over an intermediate range of values that depends on $\kp$ (as well as $\mu$ and $\sigma$) a property which is useful in applications.

\begin{figure}
\centering
\includegraphics[width=0.49\linewidth]{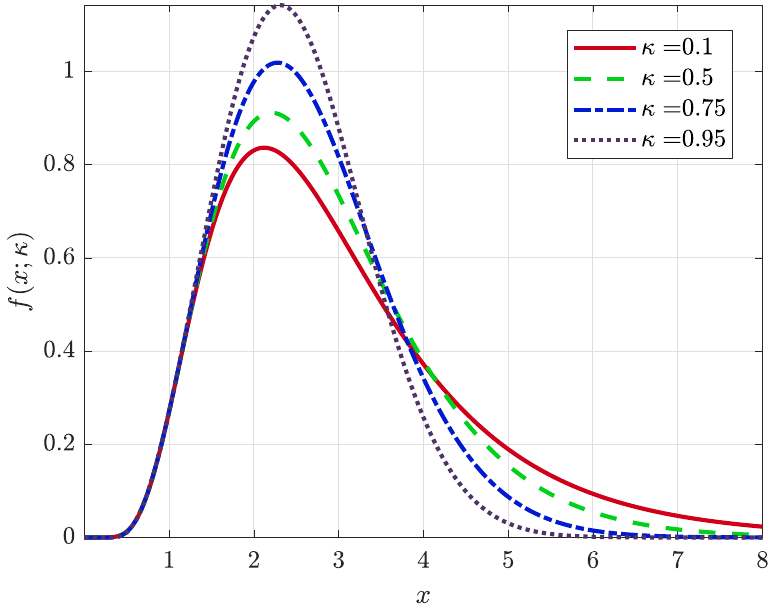}
\includegraphics[width=0.49\linewidth]{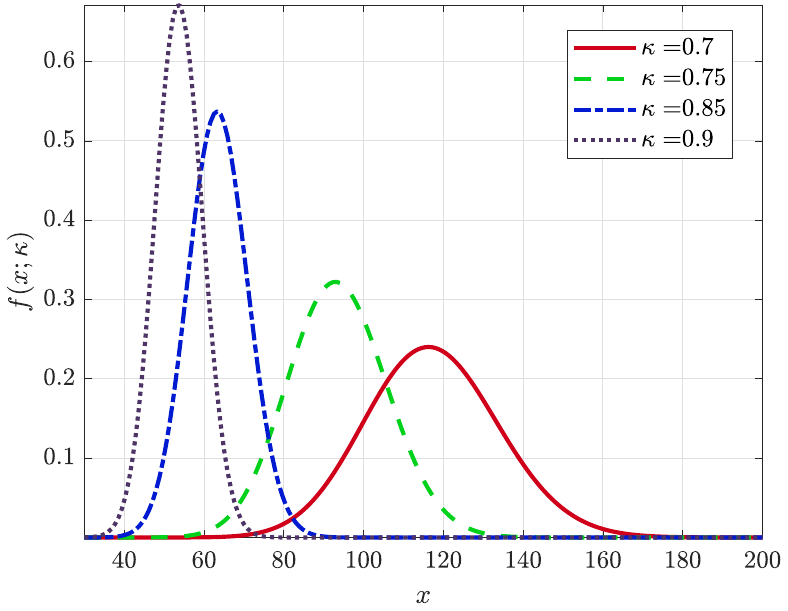}
\caption{{\bf Left:} Unimodal probability density functions of $\kp$-lognormal distributions defined by \eqref{eq:klogn-pdf} with parameters $\mu=1$ and $\sigma=0.5$. The lognormal probability density is recaptured at the limit $\kp=0$. {\bf Right:} Approximately Gaussian $\kp$-lognormal PDFs with parameters  $\mu=20$ and $\sigma=2$ for $0.7 \le \kp \le 0.9$.}
\label{fig:lognormk-PDF}
\end{figure}

\begin{figure}
\centering
\includegraphics[width=0.49\linewidth]{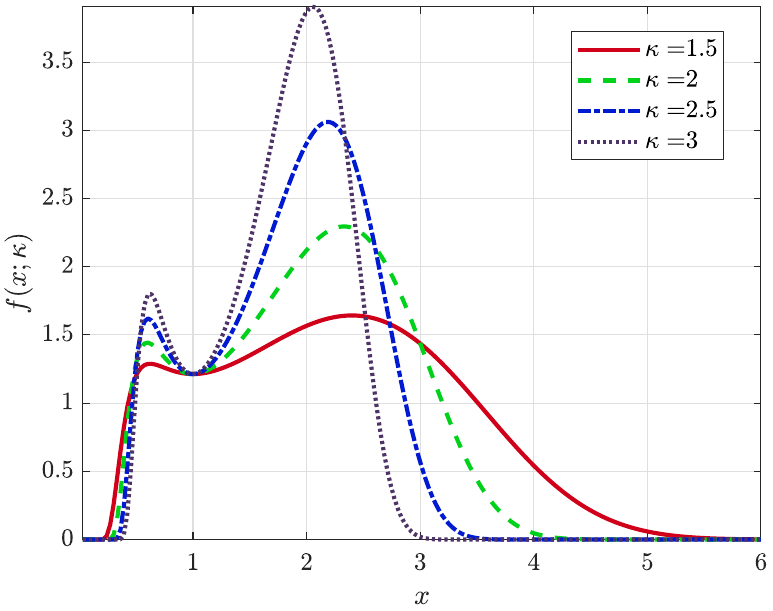}
\includegraphics[width=0.49\linewidth]{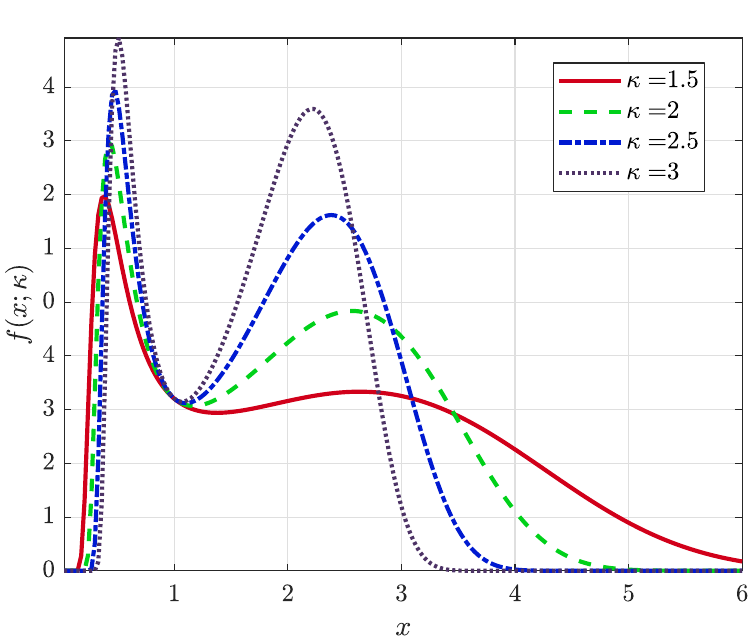}
\caption{{\bf Left:} Bimodal probability density functions of $\kp$-lognormal distributions defined by \eqref{eq:klogn-pdf} with parameters $\mu=1$ and $\sigma=1$ for $1.5 \le \kp \le 3$. {\bf Right:} Bimodal $\kp$-lognormal PDFs with parameters  $\mu=1$ and $\sigma=1.5$ for $1.5 \le \kp \le 3$.}
\label{fig:lognormk-PDF-kappa-gt-1}
\end{figure}

\begin{rem}[On the asymmetry of $\kp$-lognormal PDF]
\label{rem:gaussian-limit}
The asymmetry of the $\kp$-lognormal PDF is controlled by the combination of $\mu$, $\sigma$ and $\kp$.  For example, assume $\kp=1$,  then $\lnk(x)=(x- x^{-1})/2$ so that for $x\gg 1$ it holds $\lnk(x) \approx x/2$.  At the same time, $x^{\kp-1}+ x^{-\kp-1} = 1+ x^{-2} \approx 1$  for $x\gg 1$. Hence, the right tail of the PDF~\eqref{eq:klogn-pdf} is approximately normal. Moreover, if $\mu \gg 1$, the PDF~\eqref{eq:klogn-pdf} is approximately a normal PDF with mean $2\mu$ and standard deviation equal to $2\sigma$---because most of the mass is concentrated around $2\mu$ where the condition $x \gg 1$ holds.  This behavior is illustrated by the PDFs obtained for $\mu=20$ and $\sigma=2$  shown in Fig.~\ref{fig:lognormk-PDF} (right frame).
\end{rem}

Figure~\ref{fig:lognormk-PDF-kappa-gt-1} presents PDF plots  that exhibit bimodality for $\kp>1$. For the cases shown, PDFs with higher $\kp$ display more pronounced peaks.  For the same values of $\kp \in [1.5, 3]$ and $\mu=1$, the relative position of the highest peak depends on $\sigma$. Hence, for $\sigma=1$ (left frame of Fig.~\ref{fig:lognormk-PDF-kappa-gt-1}) the most prominent peaks occur for larger $x$, while for $\sigma=1.5$ (right frame), the tallest peaks appear at the lower peak locations.

\begin{rem}[On bimodality]
The plots shown in Figs.~\ref{fig:lognormk-PDF}-\ref{fig:lognormk-PDF-kappa-gt-1} should not lead readers to the erroneous conclusion that values $0<\kp<1$ give unimodal and $\kp \ge 1$ bimodal distributions. In fact,  for the $\kp$ values shown in Fig.~\ref{fig:lognormk-PDF-kappa-gt-1}, unimodal distributions are obtained for $\mu=-1$ and $\sigma=0.5$ (not shown).  Vice versa, $\kp<1$ leads to bimodal distributions for certain combinations of $\mu$ and $\sigma$ (e.g., for $\mu=4, \sigma=3, \kp=0.8)$. The shape of the $\kp$-lognormal PDF implicitly depends on the triplet $\mu, \sigma, \kp$  and the boundaries between different regimes are nonlinear (as per the arguments  in the proof of Theorem~\ref{theorem:peaks} in Appendix~\ref{app:modes}).
\end{rem}

\item Median
\beq
\label{eq:klogn-median}
\xmed= \expk(\mu)\,.
\eeq
The principle of \emph{quantile invariance} states that the quantiles of a probability distribution  remain invariant under a monotonic transformation.  Hence, since $\mu$ is the median of $Y$, quantile invariance requires that $\expk(\mu)$ be the median of $X=\expk(Y)$.

\item Marginal cumulative distribution function (CDF)

\noindent Using the same justification as for \eqref{eq:klogn-pdf}, the conservation of probability under the transform implies that $F_{X}(x)= \Phi(g^{-1}(x))$, leading to the following CDF for $x>0$:
\beq
\label{eq:klogn-cdf}
F_{X}(x) = \frac{1}{2}\left[1 + \erf\left( \frac{\lnk (x) - \mu}{\sigma\sqrt{2}}\right)  \right] = \Phi\left(\frac{\lnk(x)-\mu}{\sigma}\right) \,.
\eeq

\item Quantile function (defined for $0 \le p \le 1$)
\beq
\label{eq:qf-kappaln}
Q_X(p) = \expk\left[ \mu + \sqrt{2\sigma^2}\, \erf^{-1}(2p-1)\right]\,.
\eeq


\end{enumerate}
\medskip

\smallskip


\begin{theorem}[Modes of $\kp$-lognormal PDF]
\label{theorem:peaks}
The $\kp$-lognormal PDF defined  by~\eqref{eq:klogn-pdf} is unimodal or bimodal depending on the values of $\mu, \sigma \ge 0$ and $\kp \ge 0$. For $\kp=0$ (lognormal distribution) the PDF has a single mode at $x_{{\rm mode}}=\exp(\mu-\sigma^2)$. For $\kp>0$ the PDF can sustain up to three modes. For $0< \kp \le 1$ the modes (at least one) are in the interval $(0, \xmed)$. For $\kp>1$, if $\xmed< x_{+}$, where $x_{+}= \left[(\kp+1)/(\kp-1) \right]^{1/2\kp}$ as defined in Proposition~\ref{propo:kpl-convexity}, there is at least one mode in $(0, \xmed)$ while a mode can potentially emerge  in $(x_{+}, \infty)$.  For $\kp>1$ and $\xmed> x_{+}$, the mode(s) are located inside the intervals $(0,  x_{+})$ or $(\xmed, \infty)$.

For $\kp>0$ the locations of the mode(s)   are determined from the zeros of the \emph{characteristic polynomial}
\begin{align}
\label{eq:polynom-df-dx}
p_{1}(z) = z^6 - a \,z^5 + b\, z^{4} -2a\, z^3  + c \, z^{2} -a \,z -1\,,
\end{align}
where $a=2\mu\kp$,  $b=1 -4\kp\sigma^{2}(\kp-1)$ and $c=4\kp\sigma^{2} (\kp+1) -1$. If we denote by $z_r$  the positive real root(s) of $p_{1}(z)$, then the stationary points of the PDF are given by $x_{{\ast};r}=z_{r}^{1/\kappa}$,  $r=1, \ldots, R$, with $R \in \{1,  3\}$.
\begin{enumerate}
\item For $R=1$, there is a unique mode  given by $x_{{\rm mode}}=x_{{\ast};1}$.

\item For $R=3$   there are two modes  located at $x_{{\rm mode};1}=x_{{\ast};1}$ and $x_{{\rm mode};2}=x_{{\ast};3}$ (the ordering $x_{{\ast};1} < x_{{\ast};2} <x_{{\ast};3}$ is assumed).

\item If at most one of the coefficients $a, b, c$ is non-zero, then $R=1$ and the PDF is unimodal. The converse is generally not true, unless  one of the following conditions holds:
$\mu=0$,  $\kp = 0$,  $\sigma^2 = \frac{1}{4\kp(\kp-1)}$ or  $\sigma^2 = \frac{1}{4\kp(\kp+1)}$.

\item Finally, while $R=5$ (implying three modes) is not excluded by Descartes' rule if $a,b,c>0$,
five positive roots have not been detected in numerical investigations.

\end{enumerate}
\end{theorem}

\smallskip


\begin{IEEEproof}
The proof of Theorem~\ref{theorem:peaks} is given in Appendix~\ref{app:modes}.
\end{IEEEproof}

\medskip

\begin{rem}[Number of Modes of $\kp$-Lognormal PDF]
Descartes' rule of signs allows but does not require multiple (3 or 5) positive roots of the polynomial $p_{1}(z)$~\eqref{eq:polynom-df-dx}. Numerical evaluation of roots is necessary for a given combination of $\mu, \sigma, \kp$ (except for the combinations that lead to single positive roots, see Table~\ref{tab:number-of-roots} in Appendix~\ref{app:modes}). We conducted a small numerical study of the real positive roots of $p_{1}(z)$ for a rectangular grid comprising $10^6$ parameters with 100 nodes per side and parameters in the range $\mu \in [-5, 5]$, $\sigma \in [0, 3]$ and $\kp \in [0, 5]$. About 39\% of the parameter combinations admit three positive roots (corresponding to a bimodal PDF) with the remaining 61\% admitting a single positive root (corresponding to a unimodal PDF).  The numerical investigation did not identify a combination of $(\mu, \sigma, \kp)$ that admits five roots, even though  parameter vectors $(a,b,c) \in (0, \infty)^{3}$---which according to Descartes' rule is the only admissible sign combination that allows up to five positive roots---were included among the cases studied. The numerical calculation of the roots of $p_{1}(z)$ was performed using the  Matlab\circledR\xspace function \texttt{roots}.
\end{rem}

\subsection{Marginal Moments of $\kp$-Lognormal Distribution}
\label{ssec:kp-logn-moments}

\begin{theorem}[Statistical moments of integer order]
\label{theorem:scaling}
Let $m_{X;\ell}(\kp;\mu,\sigma)= \EE[X^{\ell}(\om)]$, where $\ell \in \mathbb{N}$,  denote the moment of order $\ell$ of a $\kp$-lognormal variable $X(\om)$ such that $Y(\om)=\lnk \left[ X(\om)\right] \overset{d}{=} \N(\mu, \sigma^2)$.   The $\kp$-lognormal moments can be obtained from the following integral
\beq
\label{eq:moments-integral}
m_{X;\ell}(\kp;\mu,\sigma) = \int_{-\infty}^\infty \exp_{\kp/\ell}(\ell y)\,f_{Y}(y;\mu,\sigma^2) \, \dd y\,,
\eeq
where $f_{Y}(y;\mu,\sigma^2)=\frac{1}{\sigma}\,\phi\left(\frac{y-\mu}{\sigma}\right)$  is the  PDF of a $\N(\mu, \sigma^2)$.
Moreover, the moments satisfy the following scaling relation
\beq
\label{eq:moment-scaling}
m_{X;\ell}(\kp;\mu,\sigma)=  m_{X;1}(\kp/\ell; \ell\mu, \ell\sigma)\,.
\eeq
\end{theorem}

\begin{IEEEproof}
The proof is given in Appendix~\ref{app:scaling}.
\end{IEEEproof}

\medskip

\begin{rem}[On moment scaling]
The scaling relation~\eqref{eq:moment-scaling} implies that if a closed-form expression is available for the first-order moment  of the $\kp$-lognormal (for all possible values of  $\mu, \sigma$, and $\kp$), this expression can be used to calculate the moments of  order $\ell \in \mathbb{N}$. Another implication of~\eqref{eq:moment-scaling} is that the $\kp$-lognormal moments of any order  $\ell \gg 1$ (so that $\kp/\ell \to 0$) are essentially equal to the expectation of the lognormal  with parameters $\ell\mu$ and $\ell\sigma$.
\end{rem}

\medskip

\begin{figure}
\centering
\includegraphics[width=0.75\linewidth]{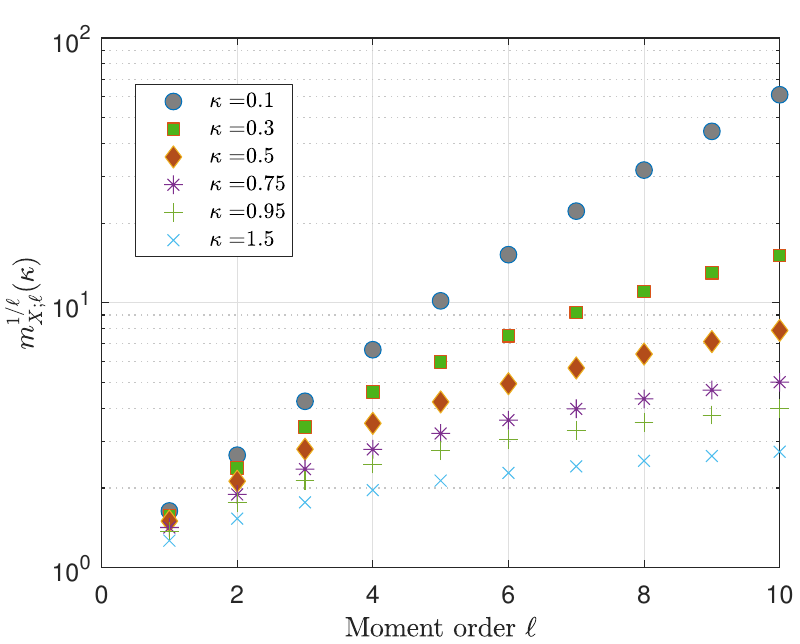}
\caption{Moments of order $\ell \in \{1, 2, \ldots, 10\}$ for the $\kp$-lognormal distribution with different $\kappa$ values.  All curves are obtained by numerical evaluation of the integral~\eqref{eq:moments-integral} for $\mu=0$ and $\sigma=1$.  The horizontal axis shows the moment order $\ell$. The vertical axis uses a logarithmic scale to display  $m_{X;\ell}^{1/\ell}(\kp)$ where $m_{X;\ell}(\kp) \triangleq m_{X;\ell}(\kp;0,1)$; the $\ell$-th root of the order-$\ell$ moments is used in order to maintain a common scale for different $\ell$. }
\label{fig:moments}
\end{figure}

In the following, if $\mu=0$ and $\sigma=1$ we use  the shorthand notation $m_{X;\ell}(\kp) \triangleq m_{X;\ell}(\kp;0,1)$ for the moments of order $\ell$.
We evaluate the moments $m_{X;\ell}(\kp)$ for $\ell=1, 2, \ldots 10$  by numerical integration of~\eqref{eq:moments-integral}.
The functions $m_{X;\ell}^{1/\ell}(\kp)$ that  are  plotted in Fig.~\ref{fig:moments} represent the $\ell$-th roots  of the $\ell$-order moments---thus all $m_{X;\ell}^{1/\ell}(\kp)$ have the same units as $X$.  As evidenced in Fig.~\ref{fig:moments}, the $\ell$-th roots of the moments  increase faster with $\ell$ for  $\kp$ values closer to zero.  This pattern is due to the increased weight in the tail of the $\kp$-lognormal PDF  as $\kp \to 0$.

\medskip

\subsection{Moment Bounds}
\label{ssec:moment-bounds}

The moments $m_{X;\ell}(\kp;\mu,\sigma)$ in~\eqref{eq:moments-integral} are bounded from below by zero, since $\expk(y) \ge 0$. The upper bounds depend on the values of $\mu$ and $\sigma$: if the latter are such that  $f_{Y}(y;\mu,\sigma^2)$
is practically zero for $y<0$, then it becomes evident from the defining integral,
that a moment of order $\ell$ for $\kp>0$ is bounded from above by the same-order lognormal moment (with the same $\mu$ and $\sigma$ as the $\kp$-lognormal), because $\left[\expk(y) \right]^\ell \le \left[ \exp(y)\right]^\ell$ for $y>0$. In fact, given the monotonic dependence of \kpe on $\kp$ the ordering
$m_{X;\ell}(\kp';\mu,\sigma) < m_{X;\ell}(\kp;\mu,\sigma)$ if $\kp' > \kp$ holds. On the other hand, the above argument cannot be extended if $f_{Y}(y;\mu,\sigma^2)$
is non-negligible for $y<0$, because $\expk(y) > \exp(y)$ for $y<0$.
In the following, we derive stricter bounds on the moments of the $\kp$-lognormal distribution.

\medskip
\begin{theorem}[Moment bounds]
\label{theorem:bounds}
Let $X$ denote a $\kp$-lognormal process so that $\lnk \left[ X(\om)\right] \overset{d}{=}\N(\mu, \sigma^2)$ with $\mu>0$. Then, the moments   $m_{X;\ell}(\kp;\mu,\sigma)$ of order $\ell$ respect the following inequalities
\begin{subequations}
 \beq
 \label{eq:moments-lb-ub}
 UB_{X;\ell}(\kp;\mu,\sigma) \ge m_{X;\ell}(\kp;\mu,\sigma) \ge LB_{X;\ell}(\kp;\mu,\sigma)\,,
 \eeq
where $LB_{X;\ell}(\kp;\mu,\sigma)$ and $UB_{X;\ell}(\kp;\mu,\sigma)$ represent, respectively, lower and upper bounds of $m_{X;\ell}(\kp;\mu,\sigma)$ given by
 \begin{align}
\label{eq:moments-lb}
LB_{X;\ell}(\kp;\mu,\sigma)=  &
\frac{(2\kp)^{\ell/\kp} \, 2^{\frac{\ell-\kp}{2\kp}} \, \sigma^{\ell/\kp} \, }{\sqrt{2\pi}}\, \E^{-\frac{\mu^2}{2\sigma^2}}\, \left[ \Gamma\left(\frac{\ell+ \kp}{2\kp}\right) \hyperg\left(\frac{\ell+\kp}{2\kp}, \, \frac{1}{2} \,; \frac{\mu^2}{2\sigma^2}\right) \right.
\nonumber \\
& \left. \quad \quad  + \,
\Gamma\left(\frac{\ell+ 2\kp}{2\kp}\right)\frac{\mu\sqrt{2}}{\sigma}\,  \hyperg\left(\frac{\ell+ 2\kp}{2\kp}, \, \frac{3}{2} \,; \frac{\mu^2}{2\sigma^2}\right) \right] +  \frac{\E^{\ell\mu +\ell^{2}\sigma^{2}/2}}{2}
\left[ 1 - \erf\left( \frac{\mu + \ell \sigma^2}{\sqrt{2}\sigma}\right) \right]\,,
\end{align}
\begin{align}
\label{eq:moments-ub}
UB_{X;\ell}(\kp;\mu,\sigma)= & \frac{1}{2}\left[ 1 - \erf\left( \frac{\mu}{\sqrt{2}\sigma}\right) \right] + \frac{2^{\ell/\kp-1}\, \E^{-\mu^2/2\sigma^2}}{\sqrt{\pi}}\, \, \sum_{m=0}^{n_\ast}\,\,  \frac{n_\ast ! \,}{m! \, (n_\ast - m)!}\left(2\kp^{2}\sigma^{2} \right)^{m}
\nonumber \\
& \times \left[\Gamma\left(m+\frac{1}{2}\right)\, \hyperg\left( m+\frac{1}{2}\,, \frac{1}{2}\,; \frac{\mu^2}{2\sigma^2}\right) + \Gamma(m+1)\, \frac{\mu\sqrt{2}}{\sigma}\,\hyperg\left( m+1\,, \frac{3}{2}\,; \frac{\mu^2}{2\sigma^2}\right)  \right]\,.
\end{align}
\end{subequations}

\noindent $\Gamma(\cdot)$ is the Gamma function,  $\hyperg(a, b; z)$  is  the \emph{confluent hypergeometric function}~\cite[9.210]{Gradshteyn07},  $\erf(\cdot)$ is the error function, and $n_\ast = \lceil \frac{\ell}{2\kp} \rceil$ is the nearest integer that is not smaller than $\frac{\ell}{2\kp}$.
\end{theorem}
\smallskip

\begin{IEEEproof}
The proof of Theorem~\ref{theorem:bounds} is given in Appendix~\ref{app:bounds}.
\end{IEEEproof}

\medskip

\begin{rem}[Scaling relation for moment bounds]
The moment bounds~\eqref{eq:moments-lb} and~\eqref{eq:moments-ub} respect  scaling relations analogous to~\eqref{eq:moment-scaling} for the moments, i.e.,
\begin{subequations}
\begin{align}
\label{eq:scaling-ub}
{UB}_{X;\ell}(\kp;\mu,\sigma) = &
{UB}_{X;1}\left(\frac{\kp}{\ell};\ell\mu,\ell\sigma\right)\,,
\\
\label{eq:scaling-lb}
{LB}_{X;\ell}(\kp;\mu,\sigma) = &
{LB}_{X;1}\left(\frac{\kp}{\ell};\ell\mu,\ell\sigma\right)\,.
\end{align}
 \end{subequations}

\noindent The scaling relations follow directly from~\eqref{eq:moments-lb} and~\eqref{eq:moments-ub} because the bounds depend on the ratios $\ell/\kp$ and $\mu/\sigma$; such terms are invariant under the transformation $\ell\mapsto 1$, $\kp \mapsto \kp/\ell$, $\mu \mapsto \ell\mu$, and $\sigma \mapsto \ell \sigma$. The lower bound also includes the term $\exp\left( \ell\mu+\ell^2\sigma^2/2\right)$ which is invariant under the transformations $\ell \mapsto 1$, $\mu \mapsto \ell\mu$, $\sigma \mapsto \ell \sigma$.  The upper bound contains  terms $\kp^{2m}\sigma^{2m}$ which also remain invariant under $\ell\mapsto 1$, $\kp \mapsto \kp/\ell$ and $\sigma \mapsto \ell \sigma$.
\end{rem}

\medskip

\begin{rem}[Tightness of the bounds]
Tightness of both the lower and upper bounds requires $\sqrt{1+\kp^{2}y^{2}} \approx \kp y$, which is easier to achieve over a wide range of $y$ as $\kp$ moves away from zero. For the upper bound, it is also necessary that $\ell/2\kp$ be equal to an integer value.
Both the lower and upper bounds contain terms that involve the error function. These terms become important only if the arguments of the error functions are considerably less than one, namely
$\mu/\sigma \ll \sqrt{2}$ (for the upper bound) and $\mu/\sigma + \ell\sigma \ll \sqrt{2}$ (for the lower bound).
\end{rem}

\medskip

\begin{figure}
\centering
\includegraphics[width=0.49\linewidth]{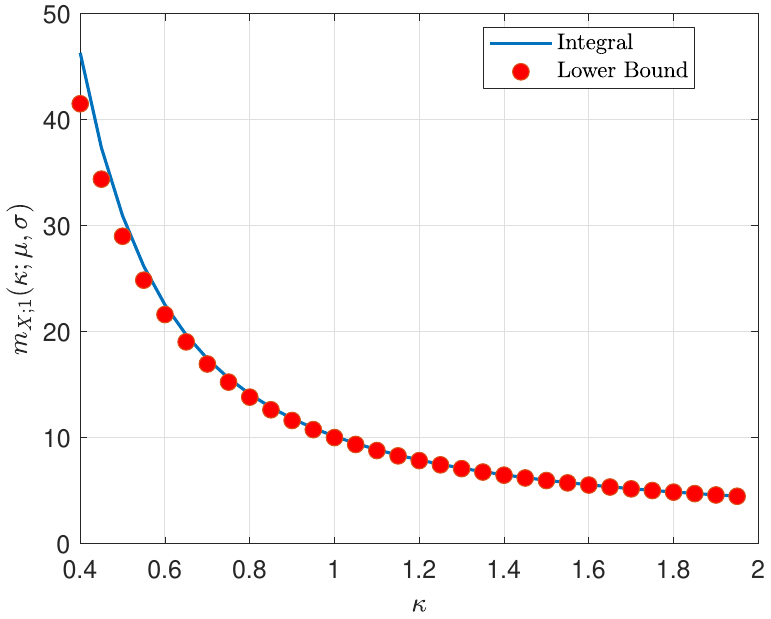}
\includegraphics[width=0.49\linewidth]{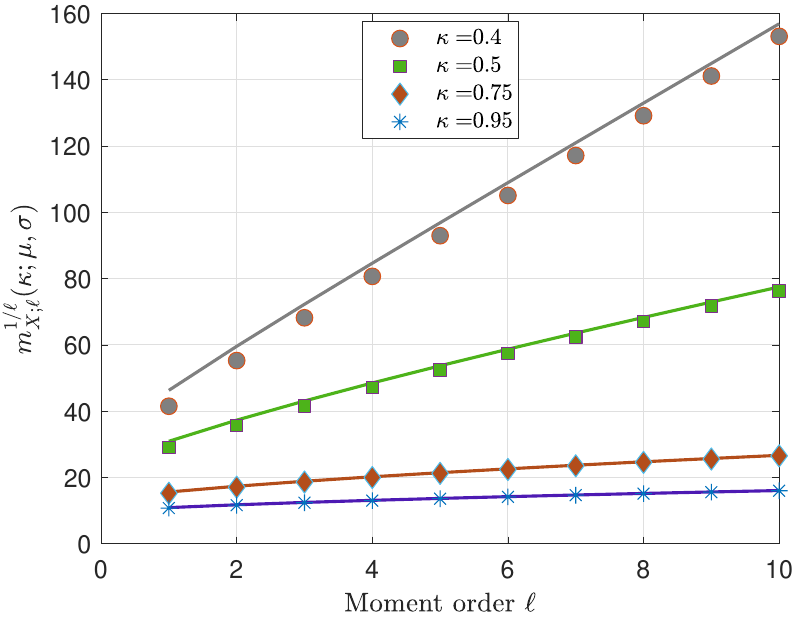}
\caption{\textbf{Left:} Mean of the lognormal distribution (continuous line, blue online) with $\mu=5$ and $\sigma=2$ calculated by means of numerical integration (cf.~\eqref{eq:moments-integral}) and lower bound of the first-order moment based on~\eqref{eq:moments-lb} (circle markers, red online). \textbf{Right:} Root of order $\ell$ of the expectation $m_{X;\ell}(\kp;\mu,\sigma)$ versus $\ell$ for $\mu=5$ and $\sigma=2$. Continuous lines represent the  numerical integral~\eqref{eq:moments-integral} while markers denote the lower bounds~\eqref{eq:moments-lb}. Four different values of $\kp$ are used: $\kp \in \{0.4, 0.5, 0.75, 0.95 \}$.  }
\label{fig:moments-lb}
\end{figure}

Figure~\ref{fig:moments-lb} compares the moments obtained by means of  numerical integration of~\eqref{eq:moments-integral} with the lower bounds~\eqref{eq:moments-lb}  for a $\kp$-lognormal model  with $\mu=5$, $\sigma=2$ and varying $\kp \ge 0.4$.   The mean $m_{X;1}(\kp;\mu, \sigma)$ shows excellent agreement for $\kp \in [0.4, 2]$, while the moments $m_{X;\ell}(\kp;\mu,\sigma)$ for $\ell \in \{2, \ldots, 10\}$ also show very good agreement for $\kp=0.4, 0.5, 0.75, 0.95$.  Higher  $\kp \in [1.40, 1.50, 1.75, 1.95]$ (not shown) yield even better agreement.

\begin{figure}
\centering
\includegraphics[width=0.49\linewidth]{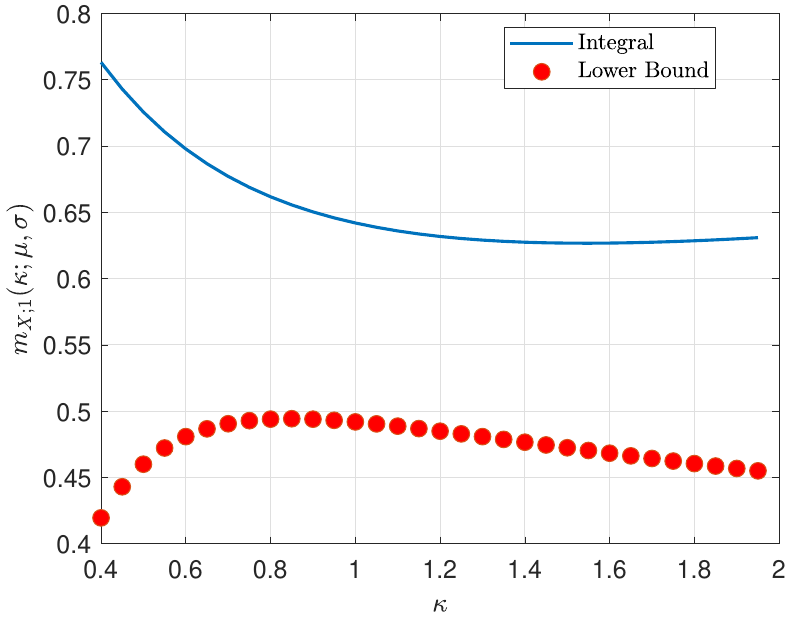}
\includegraphics[width=0.49\linewidth]{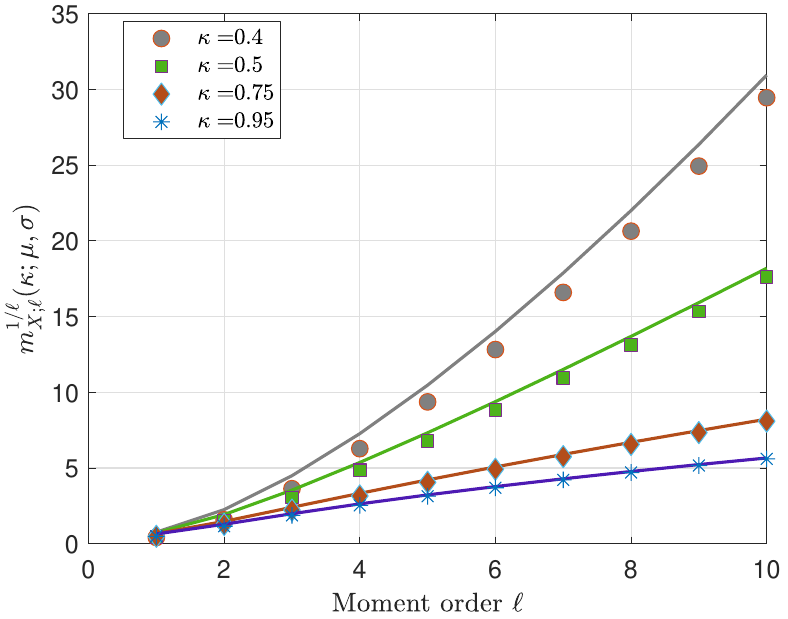}
\caption{\textbf{Left:} Mean of the $\kp$-lognormal distribution (continuous line) with $\mu=-2$ and $\sigma=2$ versus $\kp$ (continuous line, blue online).  The expectation is calculated by means of numerical integration (cf.~\eqref{eq:moments-integral}). The lower bound of the mean  as a function of $\kp$, based on~\eqref{eq:moments-lb},  is shown using circle markers (red online). \textbf{Right:} Root of order $\ell$ of the moment $m_{X;\ell}(\kp;\mu,\sigma)$ versus $\ell$ for $\mu=-2$ and $\sigma=2$. Continuous lines represent the  numerical integral~\eqref{eq:moments-integral} while markers denote the lower bounds~\eqref{eq:moments-lb}. Four different values of $\kp$ are used: $\kp \in \{0.4, 0.5, 0.75, 0.95 \}$.  }
\label{fig:moments-lb-neg-mu}
\end{figure}

Table~\ref{tab:moments_bounds} compares the  $\ell$-th root moments $m_{X;\ell}^{1/\ell}(\kp;\mu,\sigma)$ evaluated by numerical integration, cf.~\eqref{eq:moments-integral}, with both the lower and upper bounds derived from~\eqref{eq:moments-lb} and~\eqref{eq:moments-ub}. It is evident that the lower bound provides a very accurate approximation (in agreement with Fig.~\ref{fig:moments-lb}), especially for higher values of $\kp$.  The upper bound is tight for $\kp \ge 0.5$ and values such that $\ell/2\kp \in {\mathbb N}$.
This is not surprising since the upper bound is constructed based on $n_\ast = \lceil \frac{\ell}{2\kp} \rceil$ (cf. Theorem~\ref{theorem:bounds}). For $\kp=0.4$,  the upper bound significantly exceeds the ``true'' value of the moments except for integer values of $\ell/2\kp$. For smaller $\kp$, both bounds become less accurate.  Figure~\ref{fig:moments-lb-neg-mu} shows the same plots now obtained for $\mu=-2$ and $\sigma=2$.
We notice that due to the negative $\mu$ the value of the mean is in the range $[0.6, 0.8]$ for all $\kp$. The lower bound, on the other hand, is approximately limited to values between  0.4 and 0.5; hence, it  underestimates the  mean but correctly captures its  order of magnitude. For $\ell \in \{1, 2, \ldots, 10\}$, the lower bound provides an accurate approximation for $\kp \in [0.5, 0.95]$.



\begin{table}[!ht]
\caption{Estimates of the $\kp$-lognormal moments $m_{X;\ell}^{1/\ell}(\kp;\mu,\sigma)$ for $\kp=0.4, 0.5, 0.75, 0.95$ and $\ell \in \{1, 2, \ldots, 10\}$. The ratio $\ell/2\kp$ of the moment order $\ell$ over $2\kp$ is also shown. All results are obtained for $\mu=5$ and $\sigma=2$.  The following abbreviations are used:  LB: Lower bound based on~\eqref{eq:moments-lb}. MOM: Numerical moment estimates based on the integral~\eqref{eq:moments-integral}. UB: Upper bound based on~\eqref{eq:moments-ub}. The bounds provide tight approximation of the moments for integer $\ell/2\kp$.}
\label{tab:moments_bounds}
\centering
\begin{tabular}{c|c|cccccccccc}
\hline
 & $\ell$  & 1 & 2 & 3 & 4& 5 & 6 & 7 & 8 & 9 & 10  \\
 \hline
\multirow{4}{*}{$\kp=0.4$} & LB & 41.49  & 55.28 &  68.19 &  80.69 &   92.96 & 105.10 &  117.16 & 129.17 & 141.15 & 153.12
\\
 & MOM & 46.29 &  59.58 & 72.27 &  84.66 &  96.86 & 108.96 &  120.99 & 132.98 &  144.95 & 156.91 \\
& UB & 242.46 &  114.25 &   94.31 &   88.86 &  154.15 &  144.21 &  139.21 &  136.92& 194.66 & 189.55
  \\
& $\ell/2\kp$  & 1.25 &   2.5 &    3.75 &  5 &  6.25 &   7.5 &  8.75&   10 &  11.25 &  12.5 \\
\hline
\multirow{4}{*}{$\kp=0.5$} & LB & 29.00  & 35.68 &  41.62 &  47.14 &   52.39 & 57.44 &  62.33 & 67.10 & 71.76 & 76.34
\\
 & MOM &    30.92 &   37.30 &  43.09  & 48.52 &  53.71 &   58.71  & 63.56 &  68.30 &  72.93 &  77.49 \\
  & UB &    32.98 &   39.00 &  44.62 &   49.95 &  55.06 &   60.00 &  64.82 &   69.52 &   74.13 &   78.66
  \\
  & $\ell/2\kp$  &  1  &  2 & 3 &  4  & 5 &  6 &  7 & 8  & 9 &   10\\
\hline
\multirow{4}{*}{$\kp=0.75$} & LB & 15.24  & 17.07 &  18.61   & 19.98 &   21.24 &  22.41  & 23.51 &  24.56  &  25.56 &   26.54
\\
 & MOM &    15.62  & 17.36  & 18.86  & 20.21  & 21.44  & 22.60    & 23.70 &  24.73  & 25.73 &  26.69 \\
  & UB &    43.61 &  52.64  & 19.11  & 27.45  & 35.41 &  22.80   & 28.70  & 34.65 &  25.90  & 30.77
  \\
  & $\ell/2\kp$  &   0.67   & 1.33  &  2  &  2.67  &  3.33  &  4  &  4.67  & 5.33 & 6 &  6.67 \\
\hline
\multirow{4}{*}{$\kp=0.95$} & LB &    10.76  & 11.64  & 12.38   & 13.02  & 13.60 &  14.14  & 14.65  & 15.12  & 15.57  & 16.00
\\
 & MOM &    10.91 & 11.75 &  12.47 &  13.10  & 13.68 &  14.21   & 14.71 & 15.18 &  15.63 &  16.06 \\
& UB &   56.35  & 68.49  & 21.35  & 31.94  & 18.48  & 25.55   & 17.85  & 23.30  & 17.81  & 22.33
 \\
 & $\ell/2\kp$  &  0.53  &  1.06  &  1.58  & 2.11  &  2.63    & 3.16  &  3.69  &  4.21  &  4.74  & 5.27\\
\hline
\end{tabular}
\end{table}

\subsection{Power Series Expansions of the Moments}
Based on the moment integral~\eqref{eq:moments-integral}, we can develop a power series expansion for the moment of order $\ell$ by expanding the $\kp$-exponential. The expansion can be developed around different points. Below,  we focus on the Taylor series expansion of the moments around $\mu$.

\medskip
\begin{theorem}[Power series of $\kp$-lognormal moments]
\label{theorem:power-series-moments}
Let $X$ be a random variable that follows the $\kp$-lognormal distribution with parameters $\mu \ge 0, \sigma \ge 0, \kp \ge 0$ and $\ell \in \mathbb{N}$. Then,

\begin{subequations}
\label{eq:moments-lnk-taylor}
\beq
\label{eq:moments-approx}
m_{X;\ell}(\kp;\mu,\sigma) =\exp_{\kp/\ell}(\ell \,\mu) \, \left[ 1 + \ell\,
\sum_{q=1}^{\infty}  \frac{\sigma^{2q}}{q!\, 2^{q}}\,\frac{g_{2q}(\mu;\kp,\ell)}{\left({1+ \kp^{2} \mu^{2}}\right)^{2q-1/2}}\,  \right] \,,
\eeq
where the functions $g_{n}(\cdot)$ satisfy the recursive relation
\beq
\label{eq:gn}
g_{n+1}(\mu;\kp,\ell)= \left(1+  \kp^{2} \mu^{2}\right)\, \frac{\dd g_{n}(\mu;\kp,\ell)}{\dd \mu} - \left[ (2n-1)\kp^{2} \mu - \ell \sqrt{1+\kp^{2} \mu^{2}}\right] \, g_{n}(\mu;\kp,\ell)\,,
\eeq
with $g_{1}(\mu;\kp,\ell)=1$.
\end{subequations}

\end{theorem}

\begin{IEEEproof}
The proof of Theorem~\ref{theorem:power-series-moments} is given in Appendix~\ref{app:power-series}.
\end{IEEEproof}

\medskip


\begin{rem}
The functions $g_{n}(\mu;\kp,\ell)$ for $n=1, 2, 3, 4$ are given by
\begin{subequations}
\begin{align}
g_{1}(\mu;\kp,\ell) = & 1\,,
\\
g_{2}(\mu;\kp,\ell) = & \ell \sqrt{1+ \kp^2\mu^2} - \mu\kp^2   \,,
\\
g_{3}(\mu;\kp,\ell) = & -3  \kp^2 \, \ell \mu\,\sqrt{1+ \kp^2 \mu^2} - \kp^2+2  \kp^4 \mu^2+ \ell^2 \, \left( 1 +  \kp^2\mu^2 \right)\,,
\\
g_{4}(\mu;\kp,\ell) = &  \ell   \, \kp^2 \, \sqrt{1+ \kp^2  \mu^2} \left( 11  \kp^2 \, \mu^2  -4\right) + 3\kp^2  \mu \left(  3 \kp^2  - 2 \kp^4  \mu^2 - 2  \ell^2   - 2  \kp^2  \mu^2  \ell^2 \right)
\nonumber \\
& + \ell^3 \, \sqrt{1+ \kp^2  \mu^2}  \, \left( 1 + \kp^2  \mu^2\, \right)\,.
\end{align}

\end{subequations}

\noindent From the above, only the even-order $g_{2}(\mu;\kp,\ell)$ and $g_{4}(\mu;\kp,\ell)$ contribute to the moment  expansion~\eqref{eq:moments-approx}.
\end{rem}

At the limit of the lognormal distribution, that is for $\kp=0$, the moment series~\eqref{eq:moments-approx} for fixed $\ell$ is expressed as
\begin{subequations}
\beq
\label{eq:moments-approx-zero-kappa}
m_{X;\ell}(\kp=0;\mu,\sigma) =\exp(\ell \,\mu) \, \left[ 1 + \ell\,
\sum_{q=1}^{\infty}  \frac{\sigma^{2q}}{q!\, 2^{q}}\,g_{2q}(\mu;0,\ell)\,  \right] \,,
\eeq
where the functions $g_{n}(\cdot)$ satisfy the recursive equation
\beq
\label{eq:gn-zero-kappa}
g_{n+1}(\mu;0,\ell)=  \frac{\dd g_{n}(\mu;0,\ell)}{\dd \mu} + \ell  \, g_{n}(\mu;0,\ell)\,, \;\mbox{for}\; n>1\,,
\eeq
with $g_{1}(\mu;0,\ell)=1$.
The above recursive equation is solved by
$g_{n+1}(\mu;0,\ell)= \ell^{n}$.
\end{subequations}
Hence, the power series~\eqref{eq:moments-approx-zero-kappa}  can be explicitly summed yielding  the lognormal moments, i.e., $m_{X;\ell}(\kp=0;\mu,\sigma) =\exp\left(\ell \,\mu+ \frac{1}{2}\ell^2\sigma^2\right)$.

\smallskip
\begin{figure}
\centering
\includegraphics[width=0.49\linewidth]{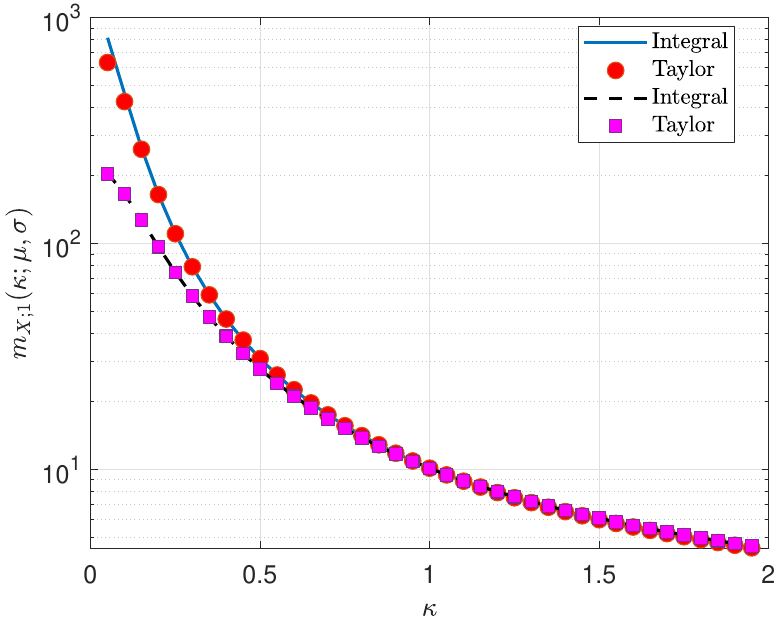}
\includegraphics[width=0.49\linewidth]{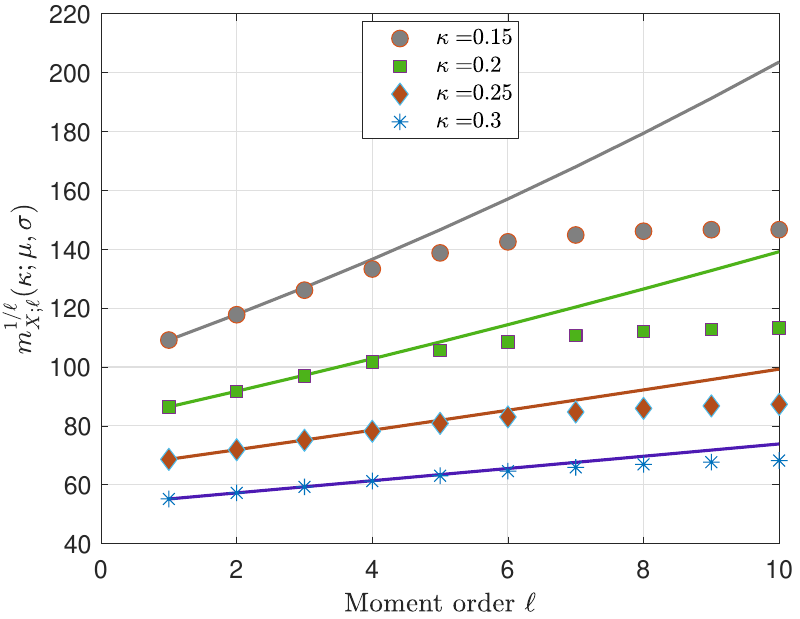}
\caption{\textbf{Left:} Mean of the $\kp$-lognormal distribution with $\mu=5$ and $\sigma=2$ (continuous line) and $\sigma=0.9$ (dashed line) calculated by means of numerical integration (cf.~\eqref{eq:moments-integral}) for $\kp \in [0.05, \, 1.95]$.  The Taylor-series moment approximation~\eqref{eq:moments-lnk-taylor} is also shown (circles for $\sigma=2$ and square markers for $\sigma=0.9$). A
logarithmic (base 10) scale is used for the vertical axis. \textbf{Right:} Root of order $\ell$ of the moment $m_{X;\ell}(\kp;\mu,\sigma)$ versus $\ell$ for $\mu=5$ and $\sigma=0.5$. Continuous lines represent the  numerical integral~\eqref{eq:moments-integral} while markers denote the Taylor-based expression~\eqref{eq:moments-lnk-taylor}. Four different values of $\kp$ are used: $\kp \in \{0.15, 0.2, 0.25, 0.30 \}$.  The Taylor series moment approximations include only the low-order terms $g_{2}(\mu;\kp,\ell)$ and $g_{4}(\mu;\kp,\ell)$.}
\label{fig:moments-taylor}
\end{figure}

Figure~\ref{fig:moments-taylor} tests the accuracy of the Taylor series moment expansion~\eqref{eq:moments-lnk-taylor}.  In the left frame, the expectation of the $\kp$-lognormal distribution, obtained from the numerical integral~\eqref{eq:moments-integral}, is compared with the mean obtained by means of the Taylor series~\eqref{eq:moments-lnk-taylor} for $\ell=1$, truncated at second order, that is for $q_{\ast}=2$. The  parameters of the $\kp$-lognormal distribution are $\mu=5$ and $\sigma \in \{0.9, 2\}$.  For all $\kp \in [0.05, 1.95]$ there is very good agreement between the true mean and the approximation; the agreement is better for the lower normal variance, that is for $\sigma=0.9$, as expected. Furthermore, the agreement between the true mean and the truncated series improves as $\kp$ increases.   The right frame compares the true order-$\ell$ moments (continuous lines) with the respective truncated series approximations (markers) for $\kp \in \{0.15, 0.20, 0.25, 0.30 \}$.  Excellent agreement is obtained for $\ell=1, 2, 3, 4$ for all $\kp$ examined. The agreement deteriorates for moments of order $\ell \ge 5$ for lower values of $\kp$. The results shown are obtained for $\mu=5$ and $\sigma=0.5$.   The accuracy of the approximation is reduced as $\sigma$ increases.

\subsection{Behavior of $\kp$-Lognormal  Tail versus the Lognormal Tail}
\label{ssec:tail}
As stated in the Introduction,  the tail of the lognormal distribution is too long for certain applications---a comparison presented in the Supplement (Section~2) between the normal and lognormal distributions  confirms this behavior.

To  intuitively explore this observation,  we consider the behavior of lognormal quantiles for probability values $p_{\ast} \triangleq 1 - 1/N$ that are close to one. The quantiles $x_{\max} \triangleq Q(p_{\ast})$   can be viewed as typical extremes  obtained from $N=2^L$, where $L \in \mathbb{N}$, independent  samples of a lognormal variable.   We focus on the \emph{dimensionless ratio} $q_{\ast}(\kp) \triangleq x_{\max}/\xmed$ of the ``typical extreme'' value over the median. This ratio represents the \emph{normalized typical  extreme}. For the lognormal distribution it is given by 
\begin{align}
\label{eq:q-ratio-max}
q_{\ast}(\kp=0) & \triangleq \frac{Q_{\mathrm{LN}}\left(p_{\ast} \right)}{Q_{\mathrm{LN}}\left(0.5\right)} = \exp\left[  \sqrt{2\sigma^2}\, \erf^{-1}(1-2/N)\right]\,,
\end{align}
where $Q_{\mathrm{LN}}$ is  the quantile function of the lognormal distribution. 
For  stochastic processes,  normalized typical  extremes can be smaller---for given $N$---because the values at  neighboring nodes are not independent.  
This implies an effective system size $N_{\textrm{eff}}\approx N/\xi^d$ (instead of $N$), where $\xi$ is the correlation length of the stochastic process~\cite[Chap.~13]{dth20}.

\begin{figure}[!ht]
\centering
\includegraphics[width=0.49\linewidth]{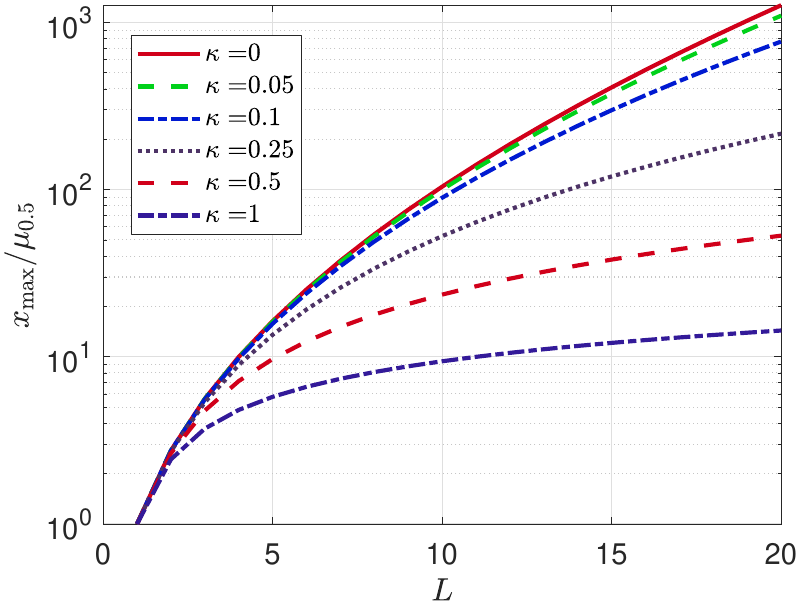}
\includegraphics[width=0.49\linewidth]{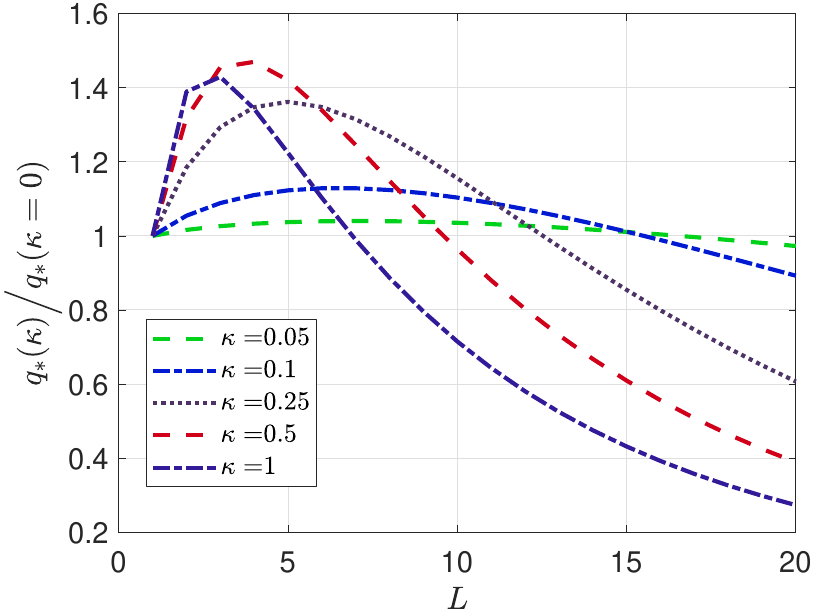}
\caption{\textbf{Left:} Ratio $q_{\ast}(\kp) \triangleq x_{\max}/\mu_{0.5}$ for the $\kappa$-lognormal distribution with different $\kappa$ including the lognormal limit $\kp=0$ based on~\eqref{eq:q-ratio-max}
; $x_{\max}=Q(1-1/N)$ where $N=2^L$ and $\xmed=Q(0.5)$. The quantile function~\eqref{eq:qf-kappaln} is used to calculate $x_{\max}$. Different $L$ values correspond to  probability levels between $p_{\ast}=0.5$ for $L=2$  and $p_{\ast} \approx 0.999999046$ for $L=20$. \textbf{Right:} Ratio of $q_{\ast}(\kp)$ for the $\kappa$-lognormal with $\kp>0$ over  $q_{\ast}(\kp=0)$ for the lognormal distribution. All curves are generated using $\mu=0$ and $\sigma=1.5$.}
\label{fig:lognormk-ratios}
\end{figure}

To highlight the tail difference between the lognormal and $\kp$-lognormal, we calculate the normalized typical extreme as a function of $N$ assuming that both distributions have the same mean, $\mx$, and variance, $\sigma^{2}_X$.  To calculate 
$q_{\ast}(\kp)$ we  use the $\kp$-quantile function given by~\eqref{eq:qf-kappaln} in~\eqref{eq:q-ratio-max}.  
We select  $\mu=0$, $\sigma=1.5$, and estimate the $\kp$-lognormal mean, $\mu_X$, and variance, $\sigma^{2}_X$, for different $\kp$ based on the integral~\eqref{eq:moments-integral}.  Then, we obtain the lognormal mean, $\mu'$, and variance, ${\sigma'}^{2}$, by inverting~\eqref{eq:ln-mean-mode-median} for the mean and~\eqref{eq:logn-var-skew} for the variance leading to 
\beq
\label{eq:ln-invert-mean-variance}
\sigma'^2 = \ln\left( 1 + \frac{\sigma^{2}_X}{\mu_{X}^{2}}\right)\,, \quad \mu' = \ln \mu_X - \ln \left(\sqrt{1 + \frac{\sigma^{2}_X}{\mu_{X}^{2}}}\,\right)\,. \quad
\eeq
Equations~\eqref{eq:ln-invert-mean-variance} ensure that the lognormal and $\kp$-lognormal distributions share the same mean and variance. 
The behavior of the $\kp$-lognormal typical extreme values is shown in the parametric plots (left frame of Fig.~\ref{fig:lognormk-ratios}). 
For $\kappa =0.05$ the $\kp$-lognormal typical extremes are close to those of the lognormal. However, the values of normalized typical extremes are considerably reduced for $\kp>0.25$.  
Figure~\ref{fig:lognormk-ratios} (right frame) also focuses on the ratio 
$q_{\ast}(\kappa)\big/q_{\ast}(\kappa=0)$.  This ratio is close to one for $\kp=0.05$ since the two distributions have the same mean and variance and $\kp=0.05$ is close to the lognormal.  As  $L$ increases beyond a $\kp$-dependent threshold, the ratio $q_{\ast}(\kappa)\big/q_{\ast}(\kappa=0)$ decreases monotonically. This behavior is more pronounced (smaller threshold and steeper decline) for higher $\kp$.

\subsection{Hazard Rate of the $\kp$-Lognormal}
\label{ssec:hazard}

The \emph{hazard rate} (also known as \emph{hazard function} and \emph{failure rate}), $h(x)$, is a key concept in reliability engineering and survival analysis~\cite{Trivedi82,Oconnor12}.  Its definition involves the survival function, $S_{X}(x)$, also known as complementary CDF (CCDF): $S_{X}(x)=1-F_{X}(x)$. The hazard  function, $h(x)$,  is then defined as 
\beq\label{eq:hazard}
h(x)=\frac{f_{X}(x)}{S_{X}(x)}.
\eeq

\paragraph{Significance} For an ensemble of identical systems under  ``stress, where the latter is denoted by the random variable $X(\om)$,  $h(x)$ is  the  conditional probability that a system fails at stress level  $X(\om)=x$ 
considering that identical systems have survived for $X(\om) \le x$. Alternatively,  $X(\om)$ could  represent the time to failure.  For example, in  lifetime analysis it represents the time to catastrophic failure of a mechanical or electrical component.  For systems that can sustain multiple failure events, such as geological faults, it represents the recurrence time to the next event (e.g., an  earthquake  above a certain threshold magnitude).   For random failure events,  lifetime and recurrence times  follow the exponential distribution with rate parameter $\lambda$.  The exponential hazard rate equals the constant $\lambda$, a result which is anticipated due to the randomness of the failure events. 

\paragraph{Asymptotic behavior} The dependence of $h(x)$ for large $x$ is a matter of interest. If $h(x)$ increases  as $x \to \infty$,  the probability of the next failure event  increases with the
time elapsed since the last event~\cite{Sornette97}; for non-repairable systems, an increasing $h(x)$ implies that the failure probability   increases with elapsed time. Such  behavior represents an aging effect~\cite{Oconnor12}.
On the other hand,  decreasing hazard rates indicate that the probability of  new occurrences (or a single failure) is
reduced  with increasing waiting time; this behavior is often observed in electronic parts and is attributed to a ``burn-in'' effect~\cite{Oconnor12}.

\paragraph{Lognormal hazard rate} The hazard function of the lognormal distribution is given by
\beq
\label{eq:logn-hazard}
h(x) = \frac{\phi\left( \frac{\ln (x) - \mu}{\sigma}\right)}{x\,\sigma\left[ 1-\Phi\left( \frac{\ln (x) - \mu}{\sigma}\right)\right]}\,,\; \mbox{for}\; x \ge 0.
\eeq
The hazard rate~\eqref{eq:logn-hazard} starts at zero for $x=0$, rises to a maximum, and then slowly decreases to zero~\cite{Sweet90}. Hence, the lognormal distribution is not suitable for survival or failure-time data. 
To determine the asymptotic dependence of $h(x)$ we use $z(x) \triangleq (\ln (x)-\mu)/\sigma$ and subsequently (dropping the dependence of $z$ on $x$ for convenience), $h(z)=\phi(z)\E^{-\mu-\sigma z}/\sigma [1-\Phi(z)]$. As $z\to \infty$,  $\phi(z)$, $\E^{-\mu-\sigma z}$, and $1-\Phi(z)$ tend to zero.  To lift the indeterminacy, we use L'H\^{o}spital's rule, 
that is, $\phi(z)/\left[ 1-\Phi(z)\right] \sim \phi'(z)/\left[-\phi(z)\right] = -[\ln \phi(z)]'\,,$ where $\phi(z) = \exp(-z^{2}/2)/\sqrt{2\pi}$ and the prime denotes differentiation with respect to $z$. Since  $[\ln \phi(z)]'=-z$, it follows that $h(x) \sim z(x)/x\,\sigma$, that is,  
\beq
\label{eq:hazard-logn-asym}
h(x) \sim \frac{\ln (x) -\mu}{x\sigma^{2}}\,, \quad x \to \infty.
\eeq

\paragraph{$\kp$-lognormal hazard rate} Based on~\eqref{eq:klogn-pdf} and~\eqref{eq:klogn-cdf}, the $\kp$-lognormal hazard function is given by
\beq
\label{eq:logkn-hazard}
h_{\kp}(x)= \frac{\phi\left( \frac{\lnk(x) - \mu}{\sigma}\right)\left(x^{\kp-1} + x^{-\kp-1} \right)}{2\,\sigma\left[ 1-\Phi\left( \frac{\lnk(x) - \mu}{\sigma}\right)\right]}\,.
\eeq
We investigate the asymptotic behavior of $h_{\kp}(x)$ following the same procedure, now setting  $z=(\lnk(x)-\mu)/\sigma$ and using $h_{\kp}(x) \sim z\left(x^{\kp-1} + x^{-\kp-1} \right)/(2\sigma)$.  Since, according to~\eqref{eq:lnk-asympt}, as $x \to \infty$  $\lnk(x) \sim x^{\kp}/(2\kp)$ and $\left(x^{\kp-1} + x^{-\kp-1} \right) \approx x^{\kp-1}$, it follows that
\beq
\label{eq:hazard-klogn-asym}
h_{\kp}(x) \sim \frac{ x^{2\kp-1} -2\kp\mu x^{\kp-1} }{4\kp\sigma^{2}}\,, \quad x \to \infty \,,
\eeq
which implies that the dominant term is $x^{2\kp-1}/(4\kp\,\sigma^2)$.

\begin{figure}[!ht]
\centering
\includegraphics[width=0.5\linewidth]{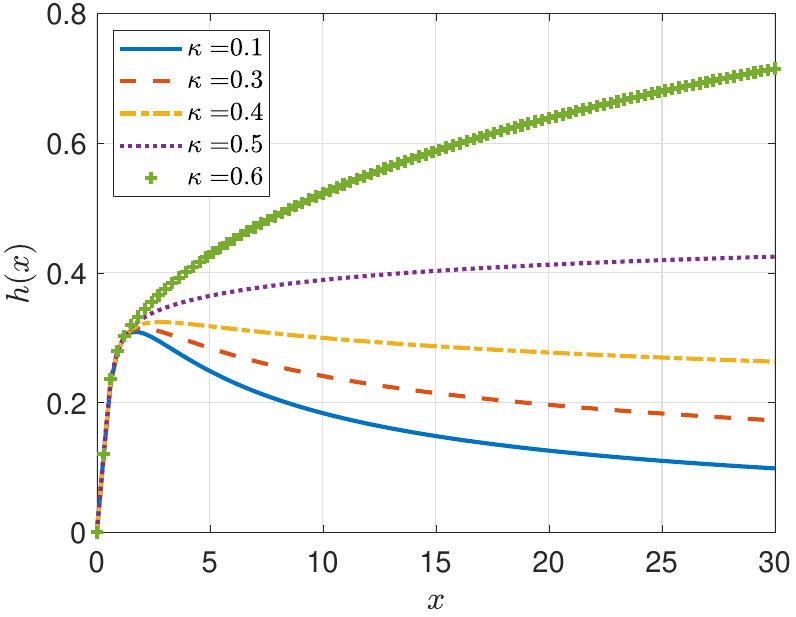}
\caption{Hazard function $h_{\kp}(x)$ for the $\kp$-lognormal distribution based on~\eqref{eq:logkn-hazard} for five different values of $\kp$. All curves are obtained with $\mu=\sigma=1$. The transition from asymptotically declining to increasing $h_{\kp}(x)$ at $\kp=0.5$ is evident.}
\label{fig:hazard_klognormal}
\end{figure}

It follows from the above analysis that $\kp=0.5$ corresponds to a critical point where the hazard rate asymptotically tends to the constant $1/2\sigma^2$. Hence, for $\kp=0.5$ the $\kp$-lognormal  behaves asymptotically as the exponential hazard rate with $\lambda =1/2\sigma^2$. On the other hand, $h_{\kp}(x)$ is a decreasing  function of $x$ for $\kp<0.5$ (thus exhibiting ``burn-in'' effect) and an increasing function for $\kp>0.5$ (demonstrating the aging effect). In all three cases, however, $h_{\kp}(x=0)=0$, and thus the $\kp$-lognormal distribution does not support the ``infant mortality'' (early failures) effect, unlike the Weibull distribution with shape modulus less than one~\cite{Oconnor12}. Nonetheless, the $\kp$-lognormal with $\kp>0.5$ is, unlike the lognormal, more suitable for failure-time analysis.
Plots of $h_{\kp}(x)$ for different values of $\kp$ confirming the near-zero and the asymptotic behavior are shown in  Fig.~\ref{fig:hazard_klognormal}.

\subsection{Failure of Simple Scaling Invariance}
\label{ssec:simple-scaling}
The lognornal distribution has a scaling property which ensures the invariance of the  CDF under a scaling operation: If $X(\om) \overset{d}{=}{\rm LN}(\mu, \sigma)$ and  $\la >0$ then for  $X'(\om)=\la X(\om)$  it follows that $X'(\om) \overset{d}{=}{\rm LN}(\mu', \sigma)$, where $\mu'=\mu+\ln \la$.
It can easily be shown  by replacing $\ln (x)$ in~\eqref{eq:klogn-cdf} with $\ln(\la x)$ and  translating $\mu \mapsto \mu'=\mu+\ln \la$ that the CDF remains invariant.
The $\kp$-lognormal distribution lacks this simple scaling property for $\kp >0$.

To study the CDF transformation of  $X'(\om)=\la X(\om)$, for $\la>0$ we need the \kpl composition law~\cite{Kaniadakis05}
\[
\lnk(\la x)=\lnk(\la)\sqrt{1+ \kp^{2}\lnk^{2}(x)}+\lnk(x)\sqrt{1+ \kp^{2}\lnk^{2}(\la)}\,
\]
which shows that the mapping $\left(\lnk(\la x)-\mu\right)/\sigma \mapsto \left(\lnk(x)-\mu'\right)/\sigma'$ is not possible in general for $\mu'$, $\sigma'$ independent of $x$.  There are two cases when such a mapping is approximately possible.
\begin{enumerate}
\item If $\kp\,\lvert\,  \lnk(x) \rvert \ll 1$ for all $x$ such that the values of $f_{X}(x)$ in~\eqref{eq:klogn-pdf} exceed some arbitrary threshold (e.g., $10^{-6}$), then the CDF is invariant under the mapping $\mu \mapsto \mu' =\mu\sqrt{1+ \kp^{2}\lnk^{2}(\la)}+\lnk(\la)$ and $\sigma \mapsto \sigma'=\sigma \sqrt{1+ \kp^{2}\lnk^{2}(\la)}$.  The condition $\kp\,\lvert\,  \lnk(x) \rvert \ll 1$ is equivalent to
$1< x \ll \expk(1/\kp)$ or $1>x \gg \expk(-1/\kp)$.
\item If $\kp\,\lvert\,  \lnk(x) \rvert \gg 1$ for all $x$ such that $f_{X}(x)$ is not negligible, then the CDF is invariant under the mapping $\mu \mapsto \mu' =\mu\left[\sqrt{1+ \kp^{2}\lnk^{2}(\la)}+\kp\lnk(\la)\right] +\lnk(\la)$ and $\sigma \mapsto \sigma'=\sigma \left[\sqrt{1+ \kp^{2}\lnk^{2}(\la)}+\kp\lnk(\la)\right]$.  The condition $\kp\,\lvert\,  \lnk(x) \rvert \gg 1$ is equivalent to
$x \gg \expk(1/\kp)$ or $x \ll \expk(-1/\kp)$.
\end{enumerate}


\section{$\kp$-lognormal stochastic process}
\label{sec:kp-lognormal-process}

\begin{definition}[$\kp$-lognormal process]
\label{defi:kln-ssp} 
The $\kappa$-lognormal stochastic process $\{X(\bfs;\om): \bfs \in \Rd\,, \om \in \Omega\}$ in the probability space $(\Omega, \mathcal{F}, \textsl{P})$ is defined as the process obtained by applying the \kpe transformation on a scalar, real-valued  Gaussian stochastic process  $\{ Y(\bfs;\om)\}$ as follows
\beq
\label{eq:kln-ssp}
X(\bfs;\om) \triangleq \expk\left[Y(\bfs;\om)\right] = \left( \sqrt{  1 +  {\kp^2}\,Y^{2}(\bfs;\om) }  + {\kp}\,Y(\bfs;\om) \right)^{1/\kp}\,, \; \kp \ge 0\,,
\eeq
The finite-dimensional distributions of the $\kp$-lognormal process are given below in~\eqref{eq:klogn-joint}. 
\noindent The inverse transformation that recovers the latent Gaussian stochastic process is given by the \kpl, that is,
\beq
\label{eq:kln-ssp-inv}
Y(\bfs;\om) \triangleq \lnk \left[ X(\bfs;\om)\right] = \frac{X^{\kp}(\bfs;\om) - X^{-\kp}(\bfs;\om)}{2\kp}.
\eeq
\end{definition}
\smallskip

We focus on \emph{weakly stationary} (also known as second-order stationary) stochastic processes with  constant mean $\mu$ and covariance kernel $C_{Y}(\bfs, \bfs'\,;\,\bmthe)$ that depends  on the input vectors purely via the lag $\bfs-\bfs'$, where $\bfs, \bfs' \in \Rd$ and   $\bmthe \in \R^{m}$  is the kernel parameter vector. 

\subsection{Joint Density of $\kp$-Lognormal  Processes}
In order to derive the joint density of the $\kp$-lognormal process we use Jacobi's multivariate theorem~\cite{Papoulis02}.  

\begin{theorem}[Jabobi's multivariate theorem]
 \label{theorem:Jacobi-mult}
Let $\mathbf{X}(\om)=({X}_1(\om), \ldots, {X}_N(\om))^\top$ and $\mathbf{Y}(\om)=(Y_1(\om), \ldots, Y_N(\om))^\top$ denote two $N$-dimensional  continuously-valued random vectors; the index $i=1, \ldots, N$ refers to the input vector $\bfs_i$. The vector variable $\mathbf{Y}(\om)$ is obtained by means of the transformation $\mathbf{Y}(\om) = \mathbf{g}\, \left(\mathbf{X}(\om)\right)$, where $\mathbf{g}(\cdot)$ is a vector function  with elements $Y_l = g_l(\mathbf{X}),\ l=1,\dots, N$.
The functions $g_l(\cdot)$ are assumed to be continuous, possess continuous partial derivatives with  respect to all inputs, and define one-to-one mappings. Then, there exist unique inverse functions  $g_{l}^{-1}(\cdot)$ such that ${X}_{l}(\om) = g_{l}^{-1}\big(\mathbf{Y}(\om)\big)\,.$

In addition, the joint \PDF $f_{\mathbf{Y}}(\mathbf{y})$ is given by means of the following equation
\beq
\label{eq:klogn-joint}
f_{\mathbf{X}}(\mathbf{x}) = f_{\mathbf{Y}}\left(\mathbf{g}(\mathbf{x})\right)\, \big| \det(\mathbb{J}) \big|\,,
\eeq
where $\mathbb{J}$ is the Jacobian of the transformation $\mathbf{y} \mapsto \mathbf{x}$. $\mathbb{J}$ is given by the following matrix of
partial derivatives
\begin{subequations}
\label{eq:Jacobian}
\beq
\label{eq:Jacobian-1}
\mathbb{J} = \left[ {J}_{i,j}\right]_{i,j=1}^{N}\,,
\eeq
where 
\beq
\label{eq:Jacobian-2}
{J}_{i,j} = \frac{\partial g_{i}(\mathbf{x})}{\partial x_j}, \; i,j=1, \ldots, N.
\eeq
\end{subequations}
\end{theorem}

\noindent For the $\kp$-lognormal  process, we apply Jacobi's multivariate theorem with the transformation $y_{i}=g_{i}(x_i)$, where $g_{i} (x_i)\triangleq \lnk(x_i)$. Therefore, ${J}_{i,j}=  \delta_{i,j} \, \frac{\pa \lnk(x_{i}) }{ \pa x_{j}}$, where $\delta_{i,j}$ is the Kronecker delta, leading to ${J}_{i,i}= \frac{1}{2}\left(x_{i}^{\kp-1} + x_{i}^{-\kp-1} \right)$, for $i=1, \ldots, N$ and $J_{i,j}=0$ for $i \neq j$. Hence, assuming that $\mathbf{Y} \overset{d}{=} \N\left(\boldsymbol{\mu}, \, \mathbf{C}_{Y} \right)$, the $\kp$-lognormal joint PDF is given by 
\beq
\label{eq:klogn-pdf-joint}
f_{\mathbf{X}}(\mathbf{x}) =\frac{1}{\left( 2\sqrt{2\pi}\right)^{N}\, \big| \det(\mathbf{C}_{\mathbf{Y}}) \big|^{1/2}}\, \E^{-\frac{1}{2}\left( \lnk(\bfx) -\boldsymbol{\mu}\right)^\top \,\mathbf{C}_{\mathbf{Y}}^{-1} \left( \lnk(\bfx) -\boldsymbol{\mu}\right)}\, \prod_{i=1}^{N}\left(x_{i}^{\kp-1} + x_{i}^{-\kp-1} \right)\,,
\eeq
where $\lnk(\bfx)$ and $\boldsymbol{\mu}$ are  $N\times 1$ vectors with $i$-th elements $\lnk (x_{i})$ and $\mu$ respectively for all $i=1, \ldots, N$. Furthermore, the PDF~\eqref{eq:klogn-pdf-joint} can be extended to non-stationary models by considering proper covariance kernels or by introducing temporal dependence in $\boldsymbol{\mu}$. 

\subsection{Predictive Distribution}
\label{ssec:warped-GPR}
In this section, we formulate warped Gaussian process regression for the $\kp$-lognormal stochastic process.

\paragraph{Warping} Let $\xsam \triangleq (x_{1}, \ldots, x_{N})^{\top}$ represent the sample values of the $\kp$-lognormal process $\Xsom$ for respective  features  $\Samp=\{ \bfs_{1}, \ldots, \bfs_{N} \}$. The sampled values may contain a noise term $\{ \noise_i \}_{i=1}^{N}$. To keep notation simple, we do not use a different symbol for ``noise contaminated'' data.  However, the presence of noise is accounted for by including a diagonal noise variance term in the covariance kernel. Assume that we aim to estimate $\Xsom$ for a new input  $\bfso$.  Using the framework of Gaussian process regression~\cite{Rasmussen06},  the answer involves the marginal conditional distribution of  $X(\bfso;\om)$.  For processes with asymmetric probability distributions, the concept of warped Gaussian processes (w-GPs) is used~\cite{Snelson03,dth22_warp}.  In w-GPs, the observation space is warped by means of a nonlinear monotonic transformation, so that learning and prediction are performed with a \emph{latent Gaussian process}. In spatial statistics this approach is known as Gaussian anamorphosis~\cite{Chiles12,dth20}.

\paragraph{\kpl warping} Herein, we study a warping transform  implemented by means of the \kpl, assuming that the resulting latent process $\rfsom{Y}$ is a stationary Gaussian process with mean $\mu$, variance $\sigma^2$, and covariance kernel $C_{Y}(\bfs-\bfs')$ as in Definition~\ref{defi:kln-ssp}.   Furthermore, we allow for contamination of the sample by Gaussian white noise with variance $\sigma^{2}_\noise$.


\paragraph{Prediction in latent space}  The latent-space sample vector $\ysam \triangleq \left(y_{1}, \ldots, y_{N}\right)^\top$ is obtained by applying the \kpl to each element of $\xsam$. For the latent Gaussian process $Y(\bfs;\om)$ the marginal predictive distribution function  $f_{\ast}(y_\ast \,\vert\, \bfso; \ysam,\Samp)$ becomes the conditional Gaussian $\mathcal{N}\left(\mu_{\ast}, \sigma^{2}_{\ast}\right)$. The posterior mean, $\mu_{\ast}$, which represents the optimal forecast, $\hat{y}_\ast$,  is given by
\begin{align}
\mu_{\ast} = \mu + \bfC_{Y}(\bfs_{\ast},\Samp)\, \left[\,\bfC_{Y}(\Samp,\Samp) + \sigma_{\noise}^2 \, \mathbf{I}\, \right]^{-1}\,\left(\ysam - \mu\, \boldsymbol{1} \,\right)\,,
\label{eq:predictive-mean}
\end{align}
and the posterior variance by
\begin{align}
\sigma^{2}_{\ast} = \sigma^2 - \bfC_{Y}(\bfs_{\ast},\Samp)\,  \left[\,\bfC_{Y}(\Samp,\Samp) + \sigma_{\noise}^2 \,\mathbf{I}\, \right]^{-1}\,\bfC_{Y}^\top(\bfs_{\ast},\Samp)\,.
\label{eq:predictive_variance}
\end{align}

\noindent In the above, the following notation is used:
\bit\itemsep0.3em
\item $\bfC_{Y}(\bfs_{\ast},\Samp)$ is the $1\times N$ kernel vector with elements $C_{Y}(\bfs_{\ast},\bfs_i)$, $i=1, \ldots, N$,
\item $\bfC_{Y}(\Samp,\Samp)$ is the $N\times N$ kernel matrix with elements  $C_{Y}(\bfs_i,\bfs_j)$, $i,j=1, \ldots, N$,
\item $\bfI$ is the $N\times N$ identity matrix,
\item $\boldsymbol{1}$ is the $N\times 1$ vector of ones.
\eit


\paragraph{Prediction in observation space} Since the marginal predictive distribution   $f(y_\ast \,\vert\, \bfso; \ysam,\Samp)$ is Gaussian, $\mu_\ast$ represents the mean, median, and mode of $f(y_\ast \,\vert\, \bfso; \ysam,\Samp)$ and  the unequivocal optimal prediction in warped space. The predictive marginal PDF in the observation space is the following  $\kp$-lognormal, where $x_{\ast} \triangleq \expk(y_{\ast})$
\beq
\label{eq:predictive-marginal}
f_{\ast}(x_\ast\,\vert\, \bfso; \xsam,\Samp)  = \frac{1}{2\, \sqrt{2\pi}\sigma_{\ast}} \, \E^{- \left(\lnk (x)_{\ast} - \mu_\ast\right)^2/2\sigma_{\ast}^2} \,
\left( x_{\ast}^{\kp-1} + x_{\ast}^{-\kp-1}  \right)\,.
\eeq

The principle of \emph{quantile invariance} affirms that the quantiles of a probability distribution  remain invariant under a monotonic transformation.  Hence, if we assume that (i) $Y(\om)=g\left(X(\om)\right) \overset{d}{=} \N(0,1)$, (ii) $y_\alpha$ is the $\alpha \in [0, 1]$ quantile, that is, $\Phi(y_\alpha)=\alpha$, then $F_{\!_X}(x_{\alpha})=\alpha$ where $x_{\alpha}= g^{-1}(y_{\alpha})$~\cite{dth20}.   For $\kp$-lognormal processes, \( g(\cdot)=\lnk(\cdot)\) and $g^{-1}(\cdot)= \expk(\cdot)$.  Since $\Phi(\mu_\ast)=0.5$, quantile invariance implies that the predictor $\hat{x}_{\ast} \triangleq \expk(\mu_{\ast})$ corresponds to the median of the predictive distribution~\eqref{eq:predictive-marginal}.
In warped space, the \emph{prediction interval} at confidence level $(1-\alpha) \times 100 \%$ for $0 < \alpha < 1$ is \(\left[\,\mu_{\ast} - z_{1-\alpha/2}\,\sigma_{\ast}, \; \mu_{\ast} + z_{1 - \alpha/2}\, \sigma_{\ast}\,\right] \)
where  $z_{1 - \alpha/2} = \Phi^{-1}(1 - \alpha/2)$ is the  $(1 - \alpha/2) \times 100 \%$ quantile of the standard normal distribution. For example, if $\alpha=0.05$,  $z_{1-\alpha/2} = \Phi^{-1}(0.975)=1.96$. Then, based on the principle of quantile invariance, the respective \emph{prediction interval in observation space} is given by
\beq
\label{eq:prediction-interval}
\left[\,g^{-1}\left(\mu_{\ast} -  z_{1-\alpha/2}\,\sigma_{\ast}\right), \; g^{-1}\left(\mu_{\ast} + z_{1 - \alpha/2}\, \sigma_{\ast}\right)\,\right]\,.
\eeq

\begin{rem}[Mode predictor]
Even if $X(\bfs;\om) \overset{d}{=} \kp{\rm LN}(\mu,\sigma,\kp)$, the predictive distribution could be approximately Gaussian, namely,
$f_{\ast}(x_\ast\,\vert\, \bfso; \xsam,\Samp)  \overset{d}{=} \N(x_\ast\,, \sigma_{\ast}^2)$. As discussed in Section~\ref{ssec:kp-logn-probab} (see Fig.~\ref{fig:lognormk-PDF}), certain combinations of $(\mu,\sigma,\kp)$ lead to nearly Gaussian PDF. In this case, the median predictor $X$ at $\bfso$, that is, $\hat{x}_{\ast} \triangleq \expk(\mu_{\ast})$ is adequate.
If, on the other hand, $f_{\ast}(x_\ast\,\vert\, \bfso; \xsam,\Samp)$ is significantly skewed, the mode  predictor could be of interest
\beq
\label{eq:mode-predictor}
\hat{x}_{\rm mode}(\bfs_\ast)=\argmax_{x_\ast \in \R} \,f_{\ast}(x_\ast\,\vert\, \bfso; \xsam,\Samp)=\argmax_{ \{x_{\ast;r}\}_{r=1}^{R}} f_{\ast}(x_{\ast;r}\,\vert\, \bfso; \xsam,\Samp) \,.
\eeq
where $x_{\ast;r}$ for $r=1, \ldots, R$, represent the $\kp$-th root of the zeros of the characteristic polynomial $p_{1}(z)$ (cf. Theorem~\ref{theorem:peaks}).
\end{rem}


\section{Estimation of $\kp$-lognormal Parameters}
\label{sec:estim}

\subsection{Estimating the Marginal Distribution}
\label{ssec:mle-marginal}

\paragraph{Marginal likelihood} Maximizing the likelihood of the marginal $\kp$-lognormal distribution with respect to the parameters given $\xsam$ is equivalent to minimizing the respective negative log-likelihood (NLL).
Based on~\eqref{eq:klogn-pdf}, the NLL (normalized by the sample size $N$) for the marginal $\kp$-lognormal  distribution is given by
\beq
\label{eq:klogn-nll}
\nll(\bmphi) = \ln\left(2\, \sqrt{2\pi}\sigma \right)
+ \frac{1}{2N\,\sigma^2}\sum_{i=1}^{N }\left(\lnk (x_{i}) - \mu\right)^2 - \frac{1}{N}\sum_{i=1}^{N}\, \ln \left( x^{\kp-1}_{i} + x^{-\kp-1}_{i}  \right)\,,
\eeq
where $\bmphi=(\mu,\sigma,\kp)^\top$.
Respectively, the estimates of the optimal parameters $\hat{\mu}, \hat{\sigma}, \hat{\kp}$ are given by
\beq
\label{eq:estim-kln}
\left( \hat{\mu}, \hat{\sigma}, \hat{\kp}\right) = \arg\min_{\bmphi} \, \nll(\bmphi)\,.
\eeq

\paragraph{Gradient of NLL} The gradient  $\nabla_{\bmphi}\nll(\bmphi) = \left(\frac{\partial \nll(\bmphi)}{\partial \mu}\,, \frac{\partial \nll(\bmphi)}{\partial \sigma} \,,   \frac{\partial \nll(\bmphi)}{\partial \kappa}\right)^\top$  of the NLL~\eqref{eq:klogn-nll} involves the following partial derivatives
\begin{subequations}
\label{eq:nll-gradient}
\begin{align}
\label{eq:partial-nll-mu}
\frac{\partial \nll(\bmphi)}{\partial \mu} = & -\frac{1}{N\sigma^2}\sum_{i=1}^{N} \left( \lnk (x_{i}) - \mu\right)\,,
\\
\label{eq:partial-nll-sigma}
\frac{\partial \nll(\bmphi)}{\partial \sigma} = & \frac{1}{\sigma} - \frac{1}{N\,\sigma^3}\sum_{i=1}^{N} \left( \lnk (x_{i})-\mu \right)^{2}\,,
\\
\label{eq:partial-nll-kappa}
\frac{\partial \nll(\bmphi)}{\partial \kappa} = &
\frac{1}{N\,\sigma^2}\sum_{i=1}^{N}\left( \lnk (x_{i})-\mu \right)\left[ \frac{\ln (x_{i})\,\left( x_{i}^\kappa + x_{i}^{-\kappa}\right) }{2\kappa} - \frac{\lnk (x_{i})}{\kappa}\right] - \frac{1}{N}\sum_{i=1}^{N}\frac{\ln (x_{i}) \left( x_{i}^{\kappa-1} - x_{i}^{-\kappa-1}\right)}{\left( x_{i}^{\kappa-1} + x_{i}^{-\kappa-1}\right)}\,.
\end{align}
\end{subequations}

The system of equations~\eqref{eq:nll-gradient} that determines NLL stationary points  is not in general  analytically solvable. Hence, we are forced to seek numerical solutions by means of optimization methods; the latter rely on initial guesses for the parameters. Finally, the NLL Hessian matrix is evaluated in Appendix~\ref{app:hessian}.

\medskip

\begin{rem}[Initialization of parameter vector]
Note that the point  $\bmphi_0 \triangleq ( \mu_0, \sigma_{0}, \kappa_0\,)^{\top}$ with $\kappa_0=0$, $\mu_0=\frac{1}{N}\sum_{i=1}^{N}\ln (x_{i})$, and $\sigma^{2}_0 =\frac{1}{N}\sum_{i=1}^{N}(\ln (x_{i}) - \mu)^2$ is a stationary point of the NLL. It is straightforward to verify that the NLL partial derivatives with respect to $\mu$ and $\sigma$ vanish at this point according to~\eqref{eq:partial-nll-mu}-\eqref{eq:partial-nll-sigma}. It also follows from~\eqref{eq:partial-nll-kappa} that  the derivative ${\partial \nll(\bmphi)}/{\partial \kappa}$ also vanishes at $\kp=0$: since  $\lim_{\kappa \to 0}x_{i}^\kappa = \lim_{\kappa \to 0}x_{i}^{-\kappa}=1$, it follows that $\lim_{\kappa \to 0} \frac{1}{2\kp}\left[ \,\ln (x_{i})\,\left( x_{i}^\kappa + x_{i}^{-\kappa}\right)   - 2 \lnk (x_{i})\right]=0$ and $ \lim_{\kappa \to 0} \left( x_{i}^{\kappa-1} - x_{i}^{-\kappa-1}\right)=0$; thus, both summands participating in ${\partial \nll(\bmphi)}/{\partial \kappa}$ vanish.   Therefore, using $\bmphi_0$ as initial guess in a gradient descent or a naive Newton's algorithm will lead to the lognormal distribution.
\end{rem}

\subsection{Estimating Joint Dependence}
\label{ssec:mle-joint}

A $\kp$-lognormal process $\Xsom$, as specified in Definition~\ref{defi:kln-ssp}, involves the  parameter triplet $\bmphi=(\mu, \sigma, \kp)^\top$ of the marginal PDF, and the parameters $\bmthe$ of the covariance kernel $C_{Y}(\bfr;\bmthe)$ associated with $Y(\bfs;\om)=\lnk \left[ X(\bfs;\om)\right]$.  MLE requires optimization of the following  NLL function
\begin{align}
\label{eq:nll-joint-klogn}
\nll(\bmphi,\bmthe) = & \frac{1}{2} \ln \left[\sigma^{2\,N}\det\bfrho(\bmthe)\right] + \frac{1}{2\sigma^{2\,N}}\left( \ysam -\mu\,\bfI \right)^\top \bfrho^{-1}(\bmthe)\,\left( \ysam -\mu\,\bfI \right)
\nonumber \\
 & + \frac{N}{2} \ln2\pi  - \sum_{i=1}^{N} \ln \left(x_{i}^{\kp-1} + x_{i}^{-\kp-1} \right)\,,
\end{align}
where $\ysam = \lnk(\xsam)$, $\bfI$ is an $N\times 1$ vector of ones and $\bfrho(\bmthe)$ is the $N\times N$ correlation matrix such that $\rho_{i,j} \triangleq C_{Y}(\bfs_i - \bfs_j)/ \sigma^{2}$.  Based on numerical experiments, we observed that optimization of the parameter vector $\bmze=(\bmthe^\top , \bmphi^\top)^\top$ by means of local optimization methods can lead to sub-optimal results. For certain problems (e.g., when estimating parameters for an ensemble of many realizations), global optimization methods may be prohibitively expensive.  The estimation procedure that we propose employs profile likelihood and involves the following steps:
\begin{enumerate}
\item
Determine the parameters $\bmphi \triangleq \left(\mu, \sigma, \kp\right)^\top$  using maximum likelihood estimation (MLE) to fit the data to the marginal $\kp$-lognormal distribution (cf. Section~\ref{ssec:mle-marginal}). This step is sufficient if only the marginal dependence is of interest.

\item Generate the transformed dataset   ${\haysam} \triangleq \left( \lnk (x_{1}),  \ldots, \lnk (x_{N})\right)^\top$ by means of the \kpl transformation with $\kp=\hat{\kp}$.

\item Finally, keep $\hat{\kp}$ fixed and apply MLE to estimate optimal parameters $\mu, \sigma, {\bmthe}$  using $\haysam$.


\end{enumerate}

\section{Applications}
\label{sec:applications}
In this section we apply the formalism developed above to simulated and real data. In the Supplement (Section~3), we provide additional  comparisons between the lognormal and $\kp$-lognormal distributions.

\subsection{Estimation of Marginal Distribution from Simulated Data}
\label{ssec:estim-kappa-simul}

To generate independent $\kp{\rm LN}(\mu, \sigma, \kappa)$ variates, we generate a $N \times 1$ random vector $\bfy$ drawn from $\N(0,1)$ and then apply the \kpe as follows to generate synthetic data
\[
\bfx = \expk(\mu + \sigma\bfy)\,.
\]
We then estimate the $\kp{\rm LN}(\mu, \sigma, \kappa)$  parameters from the samples using MLE and the simpler  quantile fitting  (described below).

\paragraph{MLE} We propose the following strategy to determine initial guesses for the parameters: (i) First, we select an initial $\kappa_0$; if the histogram shows signs of bimodality, we set $\kappa_0 >1$. (ii) Then, we calculate the sample-based median, $\hat{\mu}_{0.5}$, and estimate $\mu_0$ by inverting~\eqref{eq:klogn-median}, which yields $\mu_0=\lnk(\hat{\mu}_{0.5},\kp)$. (iii) Finally, we estimate $\sigma_0$ by taking advantage of~\eqref{eq:qf-kappaln} which connects $\sigma$ to the quantiles of the distribution, assuming $\mu$ and $\kp$ are known. The initial guess becomes
\beq
\label{eq:sigma-qf}
{\sigma}_0 = \frac{1}{M}\sum_{m=1}^{M}
\frac{\lnk\left(\hat{Q}(p_m), \kp_0 \right) - {\mu}_{0}}{\sqrt{2}\, \erf^{-1}(2p_m -1)}\,,
\eeq
where $\hat{Q}(p_m)$ are sample-based quantiles for $M$ arbitrarily selected probability levels $p_m$.
Herein, we use $N=10^3$ and $p_m=0.01\,m$, where $m=1, \ldots, M=99$.

\begin{figure}
\centering
\includegraphics[width=0.49\linewidth]{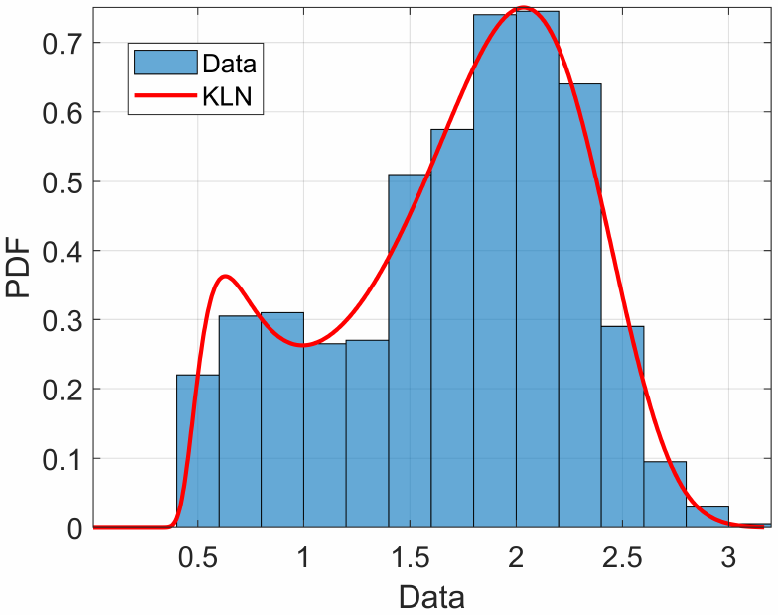}
\includegraphics[width=0.49\linewidth]{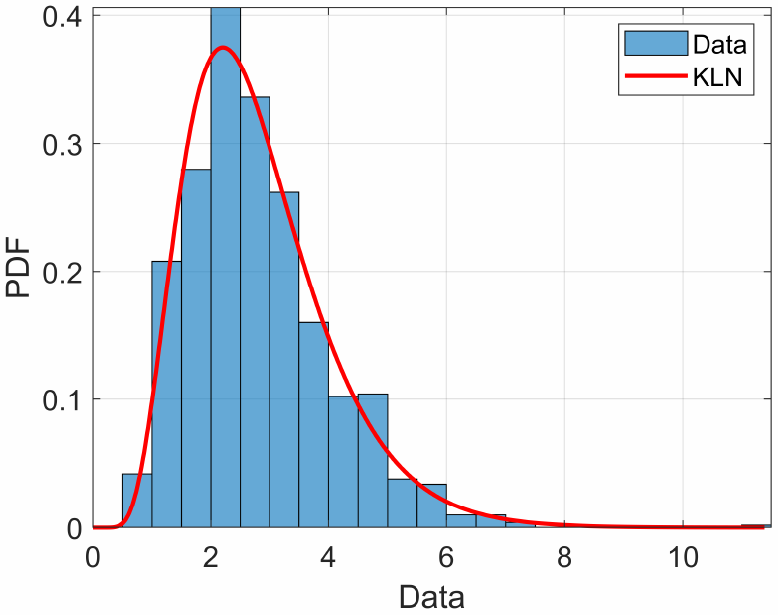}
\caption{Histograms and ML-estimated PDFs obtained from  samples comprising 1000 random numbers from the $\kp{\rm LN}(\mu, \sigma, \kappa)$ distribution. {\bf Left.}  Model A: $\mu=1$, $\sigma=1$, $\kappa=3$. The MLE parameter vector for Model A is $(0.92, 0.94, 2.89)$. {\bf Right.} Model B: $\kp{\rm LN}(\mu, \sigma, \kappa)$ with  $\mu=1$, $\sigma=0.5$, $\kappa=0.5$. The MLE parameter vector for Model B is $(1.00, 0.51, 0.52)$.}
\label{fig:MLE-fits-simulated}
\end{figure}

The MLE solution is derived using the multistart global optimization method~\cite{Ugray07} (in the Matlab \texttt{GlobalSearch} function). In deriving the results shown in Fig.~\ref{fig:MLE-fits-simulated} and Table~\ref{tab:mle_simulated_data}, we use as initial guess $\kappa_0=1.5$, while $\mu_0$ and $\sigma_0$ are calculated as described above.

\paragraph{Quantile Fitting} This procedure aims to estimate the $\kp{\rm LN}(\mu, \sigma, \kappa)$  parameters by matching the empirical and model-based quantiles.  It provides a computationally easier optimization problem than MLE.  We estimate the optimal parameters by minimizing the following univariate cost function
\beq
\label{eq:cost-qf}
u(\kappa) = \sqrt{\sum_{m=1}^{M} \left[\hat{Q}(p_m)-  Q\left(p_m; \hat{\mu}(\kp),\hat{\sigma}(\kp),\kappa \right)\right]^2}\,,\, \kp \in [0, \kp_{\max}]\,,
\eeq
where $\hat{\mu}(\kp)=\lnk(\hat{\mu}_{0.5},\kp)$ and $\hat{\sigma}(\kp)$ is obtained  by means of
$\hat{\sigma}(\kp)=\left[\lnk(\hat{Q}(0.95), \kp ) - \hat{\mu}(\kp)\right]/{\sqrt{2}\, \erf^{-1}(0.9)}$, that is from~\eqref{eq:qf-kappaln} using the empirical 95\% quantile. The optimal parameter vector is then given by
$\hat{\kp}= \arg\min_{\kp \in [0, \kp_{\max}]} u(k)$, while $\hat{\mu}$, $\hat{\sigma}$ are obtained by means of $\hat{\mu}=\lnk(\hat{\mu}_{0.5},\hat{\kp})$ and by replacing $\kp_0$ with $\hat{\kp}$ and $\mu_0$ with $\hat{\mu}$ in~\eqref{eq:sigma-qf} to obtain $\hat{\sigma}$.

Table~\ref{tab:mle_simulated_data} provides statistics (averages and standard deviations) of the $(\hat{\mu}, \hat{\sigma}, \hat{\kappa})$ estimates based on an ensemble of $\Nsim=100$ realizations for the $\kp$-lognormal distribution models A and B.  The estimates are obtained by means of MLE and quantile fitting as described above.  The ensemble-averaged parameter estimates obtained by means of both methods are quite similar and very close to the population parameters.  On the other hand, the  MLE-derived parameters have smaller standard deviations than those obtained by quantile fitting.

Figure~\ref{fig:MLE-fits-simulated} shows the histograms obtained from single realizations of two $\kp{\rm LN}(\mu, \sigma, \kappa)$ distributions: Model A with parameters $\mu=1$, $\sigma=1$, $\kappa=3$ and Model B with $\mu=1$, $\sigma=0.5$, $\kappa=0.5$. The plots also show the PDF curves obtained from~\eqref{eq:klogn-pdf} using ML estimates $(\hat{\mu}, \hat{\sigma}, \hat{\kappa})$ of the distribution parameters. Excellent agreement is observed between the sample histograms and the PDF curves, including faithful representation of the bimodality for Model A.

\begin{table}[!ht]
\caption{Statistics of estimates $(\hat{\mu}, \hat{\sigma}, \hat{\kappa})$ derived by maximum likelihood estimation (MLE) and quantile fitting (QF).  The averages, $\langle \cdot \rangle$ and standard deviations, ${\rm std}(\cdot)$ are based on ensembles
of $\Nsim=100$ realizations each of which comprises 1000 random numbers simulated from $\kp{\rm LN}(\mu, \sigma, \kappa)$. Two different experiments are conducted that simulate the  PDFs shown in Fig.~\ref{fig:MLE-fits-simulated}.  }
\centering
\begin{tabular}{c|cccccc}
Parameter estimates & $\langle \hat{\mu} \rangle$ & $\langle \hat{\sigma} \rangle$ & $\langle \hat{\kappa} \rangle$ & std$(\hat{\mu})$  & std$(\hat{\sigma})$  & std$(\hat{\kappa})$ \\
\hline
\multicolumn{7}{c}{\textbf{First simulation experiment:} $\mu=1, \, \sigma=1, \, \kappa=3$} \\ \hline
  MLE  & 1.0010  &  1.0011  & 2.9998 & 0.0561  &  0.0567 &   0.0976 \\\hline
QF  & 1.0007 & 1.0006 & 2.9990 & 0.0748 & 0.0814 & 0.1382\\
\hline
\hline
\multicolumn{7}{c}{\textbf{Second simulation experiment:} $\mu=1, \, \sigma=0.5, \, \kappa=0.5$} \\ \hline
  MLE  & 1.0013  &  0.4975 & 0.4933 & 0.0208  &  0.0206 &   0.0668 \\\hline
QF  & 1.0013 & 0.4980 & 0.4880 & 0.0274 & 0.0266 & 0.0913\\
\hline
\end{tabular}
\label{tab:mle_simulated_data}
\end{table}



\subsection{Estimation of Marginal Distribution for Berea Sandstone Permeability}
\label{ssec:Berea}

We analyze a Berea sandstone permeability data set that contains 1600 mini-permeameter measurements sampled on a 40cm$\times$40cm square grid with a 1-cm grid step. The summary statistics of the dataset are given in Table~\ref{Tab:Data Stat}.

\begin{table}[!ht]
\caption{Statistics of Berea sandstone permeability measurements. The units are in millidarcies (md).}
\centering
\renewcommand\tabcolsep{5pt} 
\renewcommand\arraystretch{1.2} 
\begin{tabular}{ccccccc}
\hline\hline
Mean (md)  &  Median (md)  &  Min (md)  &  Max (md)  &  Standard deviation (md)  &  Skewness & Kurtosis\\
\hline
52.53  &  55.00     & 19.50  &  111.50  &  15.78 & 0.38  &  3.13 \\
\hline
\end{tabular}
\label{Tab:Data Stat}
\end{table}

\medskip

\begin{figure}[!ht]
\centering  \includegraphics[width=0.5\linewidth]{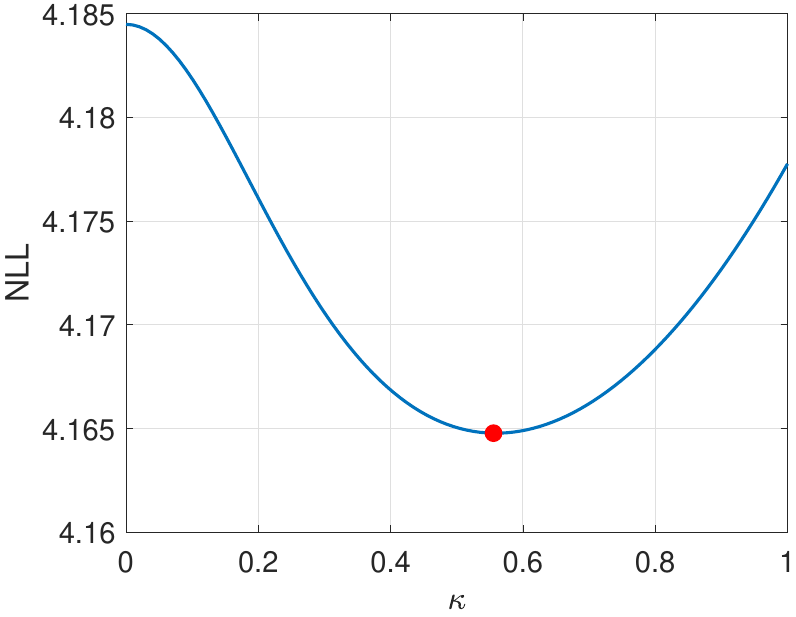}
\caption{Profile NLL curve (NLL per site) as a function of $\kp$ for the Berea permeability data, obtained from~\eqref{eq:klogn-nll}.  The parameters $\hat{\mu}_{i}$ and $\hat{\sigma}_{i}$  for a given $\kp_{i}$ are  determined by maximizing the likelihood of the normal distribution for the  $\kp$-deformed dataset $\ysam=\lnk(\xsam)$. }
\label{fig:profile-nll}
\end{figure}

\begin{figure}[!ht]
\centering  \includegraphics[width=0.5\linewidth]{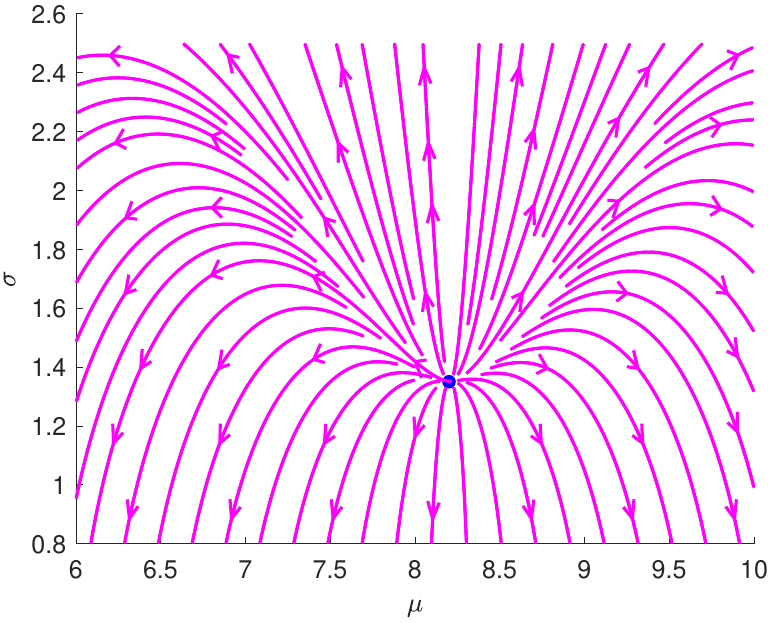}
\caption{NLL Streamlines  for the Berea permeability data on the $(\mu, \sigma)$ plane  obtained using the NLL gradient~\eqref{eq:nll-gradient}   for $\kappa=0.556$.  }
\label{fig:streamlines-nll-berea}
\end{figure}

\paragraph{Profile likelihood} We study the dependence of the optimized NLL obtained for selected values of $\kappa \in \{ \kappa_{1}, \ldots, \kappa_{\ell}\}$. For each $\kp_{i}, i=1, \ldots, \ell$, we calculate the $\kp$-deformed dataset $\ysam^{(i)}=\left(\lnk(x_1), \ldots, \lnk (x_N) \right)$, where $\kappa=\kappa_i$. We then  determine  $\hat{\mu}_{i}$ and $\hat{\sigma}_{i}$ by maximizing the likelihood of the normal distribution for $\ysam^{(i)}$. The $\hat{\mu}_{i}$ and $\hat{\sigma}_{i}$ are inserted in~\eqref{eq:klogn-nll} to obtain the respective   ${\rm{NLL}}(\hat{\mu}_{i}, \hat{\sigma}_{i} \mid \kp_{i})$. This leads to the profile NLL curve shown in Fig.~\ref{fig:profile-nll}, which features a minimum of NLL at $\hkp\approx 0.556$, that is,
$\hkp=\min_{\kp} \left\{\min_{\mu,\sigma}{\mathrm{NLL}} (\mu,\sigma \mid \kp) \right\}$.  ML estimation for $\kappa=0$ (lognormal distribution) returns $\hat{\mu}_0 \approx 3.97$ and $\hat{\sigma}_0 \approx 0.30$. For $\hkp\approx 0.556$ it follows that $\hat{\mu} \approx 8.2$ and $\hat{\sigma} \approx 1.35$.  The $\kp$-lognormal distribution corresponding to $(\hmu, \hsi, \hkp)$ is unimodal with its mode located at $x_{\rm{mode}} \approx 54.33$.
Figure~\ref{fig:streamlines-nll-berea} presents  NLL streamlines (lines that are tangential to the local gradient) for the Berea data in the $(\mu, \sigma)$ plane for $\kp=0.556$. The stationary point in this plane is $(8.20, 1.35)$, in agreement with the profile likelihood result; the streamlines radiate away from the stationary point indicating increasing gradient.

Table~\ref{tab:Berea_fits} lists the optimal parameters based on ML estimates of the $\kp$-lognormal and lognormal distributions to the Berea dataset.  MLE was conducted in the full parameter space $(\mu,\sigma,\kp)$ using global optimization to avoid potential local traps.  The table also contains the  optimal NLL per site and values of Akaike  and  Bayesian information criteria (AIC and BIC).  Both criteria favor the $\kp$-lognormal distribution.  The globally-optimized MLE parameter estimates are very close to those obtained by means of profile likelihood.

\begin{table}[!ht]
\centering
\caption{Statistics of the  Berea permeability dataset. Columns 2-4 contain the optimal $\kp$-lognormal parameters derived by MLE. Columns 5-7 display respectively the negative log-likelihood (NLL) per sampling location, as well as the AIC and BIC selection criteria. Columns 8-9 contain the optimal lognormal parameters, and columns 10-12 the NLL per site, the AIC and BIC respectively. Bold font indicates the model with the lowest AIC and BIC.}
\begin{tabular}{|cccccc||ccccc|}
\hline\hline
\multicolumn{6}{|c||}{$\kp$-Lognormal} &  \multicolumn{5}{c|}{Lognormal}
\\ \hline
{$\mu$} & {$\sigma$} & {$\kappa$} & {NLL} & {AIC} & {BIC} & {$\mu_0$} & {$\sigma_0$} & {NLL$_0$} & {AIC$_0$} & {BIC$_0$}\\ \hline
8.26 & 1.37 & 0.56 & 4.16 & \textbf{13333.28} & \textbf{13349.41} & 3.97 & 0.30 & 4.18 & 13394.34 & 13405.09\\[1ex]
\hline
\end{tabular}
\label{tab:Berea_fits}
\end{table}

\medskip

\begin{figure}
\centering
\includegraphics[width=0.51\linewidth]{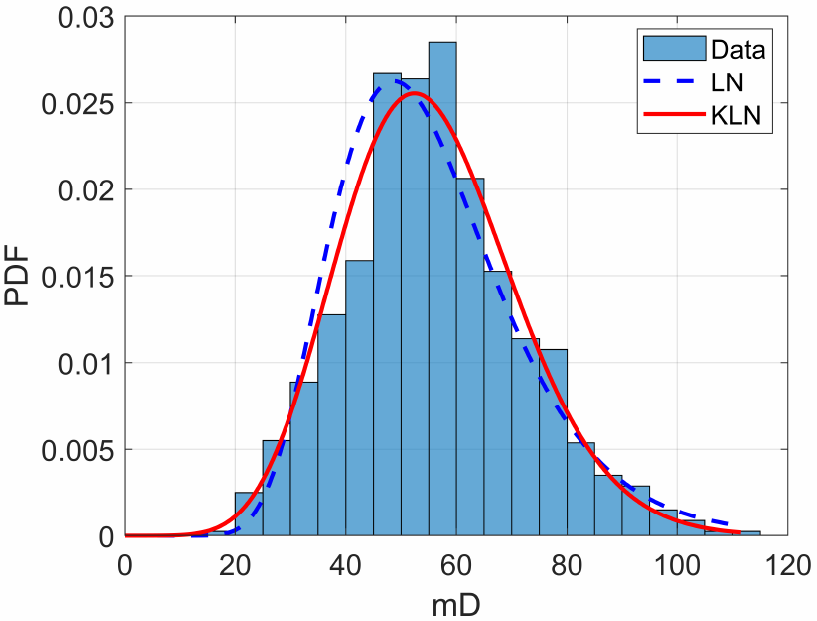}
\includegraphics[width=0.47\linewidth]{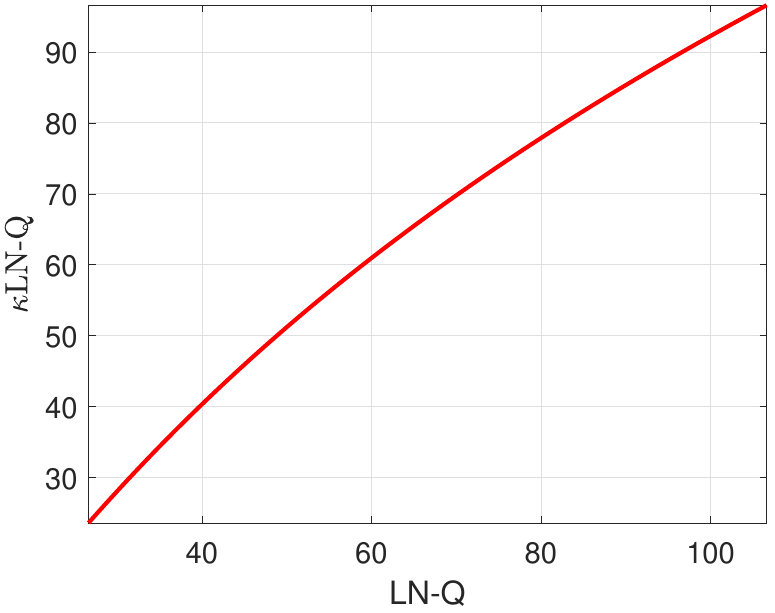}
\caption{\textbf{Left:} Histogram of Berea permeability data and MLE fits of the lognormal PDF (blue broken line) and the $\kp$-lognormal PDF (continuous line). The horizontal axis measures permeability and the units used are millidarcy (mD). \textbf{Right:} Q-Q plot of the lognormal quantiles (horizontal axis) versus the respective $\kp$-lognormal quantiles (vertical axis). The quantiles are obtained from the best-fit distributions for the Berea data.  The units of both axes are in milliDarcy.}
\label{fig:prob_fun_Berea}
\end{figure}

\paragraph{Comparison of lognormal and $\kp$-lognormal models} The histogram of  Berea permeability  is shown in Fig.~\ref{fig:prob_fun_Berea} (left frame) with superimposed PDF curves for the optimal lognormal and $\kp$-lognormal distributions.  As evidenced in the plot, the mode of the $\kp$-lognormal model is closer to the peak of the histogram than that of the lognormal; in addition, the $\kp$-lognormal right tail decays faster than the lognormal, also in agreement with the histogram.
The comparison of the PDF tails  is more compelling in terms of the quantile-quantile (Q-Q) plot (right frame of Fig.~\ref{fig:prob_fun_Berea}). The Q-Q plots are generated by the set of $N_Q$ points $\Big( Q_{\mathrm{LN}}(p_{i}), \;  Q_{\mathrm{KLN}}(p_{i}) \Big)$, $i=0, \ldots, N_Q -1$, where $p_{i} =0.01 + 0.98i/(N_{Q}-1)$ for $i=0, 1, \ldots, N_{Q}-1$. We use $N_Q=500$ herein leading to 500 probability levels in the interval $[0.1, 0.99]$.  The quantiles $Q_{\mathrm{LN}}(p_{i})$, $Q_{\mathrm{KLN}}(p_{i})$ are given by~\eqref{eq:qf-ln} and~\eqref{eq:qf-kappaln} respectively using the ML estimates for the respective $\mu$ and $\sigma$.  This plot highlights the fact that the  right-tail quantiles  of the lognormal  exceed those of the  $\kp$-lognormal distribution.

\medskip

\subsection{Time series forecasting using $\kp$-lognormal processes}
\label{ssec:forecasting}

This section investigates forecasting for  stochastic processes $\{ X(t;\om)\}$ defined by means of $X(t;\om)=\expk\left[ Y(t;\om)\right]$,
where $\{ Y(t;\om)\}$ is a scalar, real-valued  Gaussian stochastic process with mean $\mu$, variance $\sigma^2$ and correlation kernel $\rho_{Y}(\tau\,;\,\bmthe)$;  $t$ is the time index and $\tau=t-t'$ is the time lag.  The covariance kernel is given by $C_{Y}= \sigma^{2}\,\rho_{Y}$. Then, according to Definition~\ref{defi:kln-ssp},  $\{ X(t;\om)\}$  is a $\kp$-lognormal stationary stochastic process. We conduct forecasting in the framework of warped Gaussian process regression (cf. Section~\ref{ssec:warped-GPR}).

\paragraph{Simulation method} We use the multivariate normal (MVN) simulation  method~\cite[p. 50]{Johnson87}  to generate  samples of $\{ X(t;\om)\}$, namely time series comprising $N$   times $t_{k} \in [0, T]$, for $k=1,\ldots, N$.  The temporal dependence is determined by the $N\times N$  covariance
matrix  ${\bfC}_{Y}$ with elements $[\bfC_{Y}]_{k,l}={C_{Y}}(t_{k}-t_{l})$, for  $k,l=1, \ldots, N$, where  $C_{Y}(\tau)$ is the covariance kernel.
If  ${\bfC}_{Y} = {\bf A}{\bf A}^{T}$ is a factorization of ${\bf C}_{Y}$,  and  ${\bf z}$  is an $N \times 1$ vector of  independent  random numbers  from    $\mathcal{N}(0, 1)$,   the $N \times 1$ vector  ${\bf y} = {\bf A}\,{\bf z}$  is a realization of a zero-mean normal stochastic process with covariance kernel $C_{Y}(\cdot)$. Then, $\bfx = \expk(\mu + \sigma\bfy)$ provides a realization of the $\kp{\rm LN}(\mu,\bfC_{Y},\kappa)$ stochastic process with joint PDF given by~\eqref{eq:klogn-pdf-joint}. We use the
principal square root factorization of $\bfC_{Y}$ for numerical stability~\cite{Higham87}.

\paragraph{Study design} We follow the procedure described below.
\begin{enumerate}
\item  $\Nsim=500$ realizations are generated using the MVN method. Each realization represents a correlated sequence sampled at times $t_n=n$ where $n=1, \ldots N$ and $N=2^{10}$. The $\kp$-lognormal marginal is determined by the parameter vector $\bmphi=(\mu,\sigma,\kp)$, where $\mu=1$, $\sigma=1$, $\kp=3$. As shown in Figs.~\ref{fig:lognormk-PDF-kappa-gt-1} and~\ref{fig:MLE-fits-simulated}, this combination of parameters generates a bimodal distribution, which is clearly non-Gaussian.  $C_{Y}(\tau;\bmthe,\sigma)$  is given by the linear damped harmonic oscillator (LDHO) kernel~\cite{dth24} in the underdamped regime
\beq
\label{eq:ldho-kernel}
C_{Y}(\tau;\bmthe,\sigma)= \sigma^2 \, \E^{-\lvert \tau \rvert/(2\tau_c)}\left[ \cos(\omega_d \tau)+ \frac{1}{2 \omega_d \tau_{c}} \sin(\omega_d \lvert\tau \rvert)\right]\,,
\eeq
where $\bmthe=(\tau_{c}, \omega_{d})^\top$, $\tau_c =30$ is the relaxation (damping) time and $\omega_d =2\pi/50 \approx 0.126$ is the natural frequency of the damped oscillations.

\item Each realization is split into a training set containing $N_{\rm tr}=973$  data points, $N_{\rm tr}=\lfloor 0.95\,N\rfloor$, and a test set that comprises the remaining $N_{\rm te}=51$ points.
\item The model parameter vector $\bmze \triangleq(\bmphi^\top, \bmthe^\top)^\top$ is estimated over the training set using MLE as outlined in Section~\ref{ssec:mle-joint}.
\item The optimal model, specified by  $\hat{\bmze}=(\hat{\mu}, \hat{\sigma}, \hat{\kp}, \hat{\omega}_d, \hat{\tau}_c)$ is used to forecast the series at the test-set times $t_{N_{\rm tr}+k}$, $k=1, \ldots, N_{\rm te}$. We employ the multi-step (multiple output) forecasting strategy (without recursive feedback of predictions into the training set). Both median-based and mode-based predictions are used, respectively, given by the median of the predictive marginal PDF~\eqref{eq:predictive-marginal} and the mode predictor~\eqref{eq:mode-predictor}.

\item We quantify the forecasting performance  using different cross validation measures calculated by comparing the forecasts to the true test values.  Finally, we average the cross-validation measures over the $\Nsim$ realizations.
\end{enumerate}

\begin{figure}[!ht]
\centering
\includegraphics[width=0.5\linewidth]{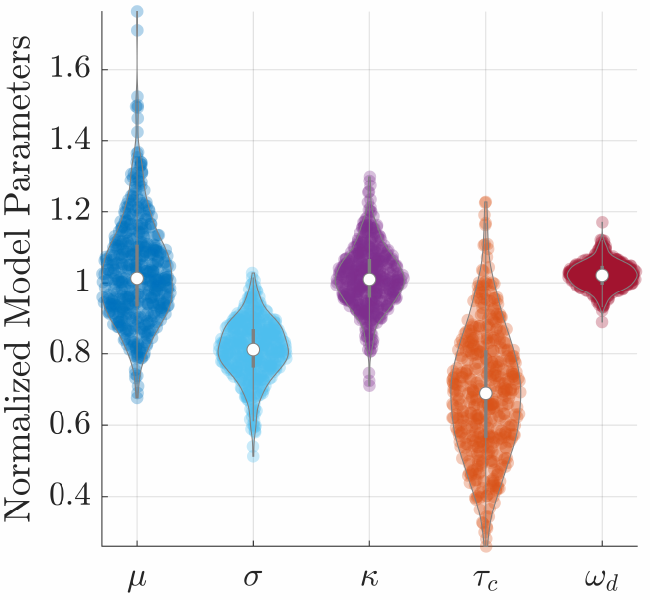}
\caption{Violin plots displaying the distribution of the estimated  model parameter estimates based on 500 realizations of the $\kp$-lognormal process with harmonic oscillator kernel~\eqref{eq:ldho-kernel}. The estimates are normalized by dividing with the true model parameter values, namely $\mu=1$, $\sigma=1$, $\kp=3$, $\tau_c=30$ and $\omega_d = 2\pi/50$.  The open circles in the middle of the violins mark the median of the distribution.  The estimates from 31 realizations that yield  $\hat{\omega}_{d}>1$ are not presented in this plot.}
\label{fig:Theta-LDHO-Simulations}
\end{figure}

\paragraph{Parameter estimation}
We apply MLE as described in Section~\ref{ssec:mle-joint} to estimate
$\bmphi, \bmthe$.
As initial estimates for $\mu$ and $\sigma$ we use the values obtained from the marginal NLL minimization, while $\hat{\tau}_{c;0}=T/2$ and $\hat{\omega}_{d;0}=2\pi/(T/10)$ are used for $\bmthe$.
The distribution of the ML parameter estimates for the ensemble of 500 realizations is summarized in the violin plots of Fig.~\ref{fig:Theta-LDHO-Simulations}. Each plot shows the distribution of the ratio $\hat{\theta}_{i;n}/\theta_{i}$, where $\hat{\theta}_{i;n}$ is the estimate of the $i$-th parameter from the set $(\mu, \sigma, \kp, \tau_{c}, \omega_{d})$ based on the $n$-th realization ($n=1, \ldots, 500$), and $\theta_i$ is the true value. We excluded 31 realizations that produce unrealistic cyclical frequency estimates  $\hat{\omega}_{d}>1$. As evidenced in these plots, the medians of $\hat{\sigma}$ and $\hat{\omega}_{d}$, respectively, underestimate the true standard deviation and relaxation time.  We suspect that this is due to the fact that the persistence of correlations, as expressed by the  ratios $N/\tau_{c} \approx 34$ and $N/T_{d}=N\omega_{d}/2\pi \approx 20.5$, does not allow fully resolving the variability of $X(t;\om)$.  For computational reasons, NLL optimization  is conducted using  local search with the interior-point algorithm. Hence, such large $\hat{\omega}_d$ are likely due to local optima of the NLL function.

\begin{figure}[!ht]
\centering
\includegraphics[width=0.5\linewidth]{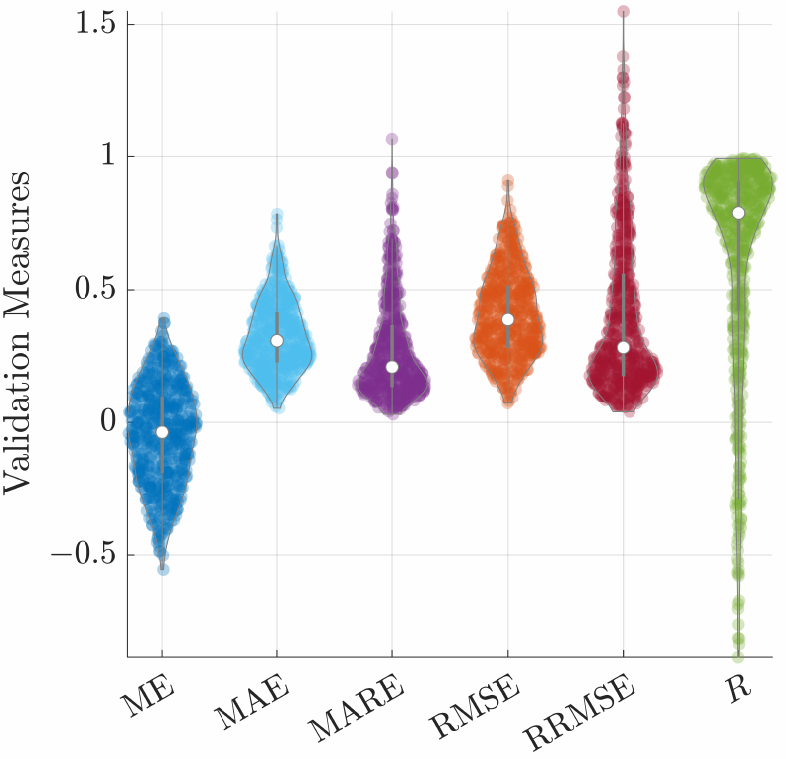}
\caption{Violin plots displaying the distribution of cross-validation measures based on 500 realizations of the $\kp$-lognormal process with harmonic oscillator kernel~\eqref{eq:ldho-kernel} and the following parameters: $\mu=1$, $\sigma=1$, $\kp=3$, $\tau_c=30$ and $\omega_d = 2\pi/50$.  The open circles in the middle of the violins mark the median.}
\label{fig:cross-val-ldho-histograms}
\end{figure}

\paragraph{Forecasting} Multi-step forecasts are generated using both the median and mode predictors. The median predictor is based on $\hat{\mu}_{0.5} \triangleq \expk(\mu_{\ast})$, where $\mu_{\ast}$ is the conditional mean of the latent Gaussian process at $\bfso$ given by~\eqref{eq:predictive-mean}.  The mode predictor is based on~\eqref{eq:mode-predictor}.
The \emph{cross-validation (CV)} measures  involve the following: Mean error (ME);  mean absolute error (MAE);  mean absolute relative error (MARE); root mean square error (RMSE); relative RMSE (RRMSE); Pearson correlation coefficient ($R$).  Violin plots of the CV measures obtained from $\Nsim=500$ realizations are shown in Fig.~\ref{fig:cross-val-ldho-histograms}. The  plots present kernel density estimates of the PDF for each CV measure. Two notable features are: (i) the long RRMSE   tail,  which is due to high error values for low  process values  and (ii) the long lower tail of $R$ that extends into negative values for a small number of realizations, indicating poor correlation of the forecasts with the true values.  We believe that such results are due to trapping of the NLL optimization in local minima, and they could be avoided if a global optimization method is used.

Average CV measures over the $\Nsim$-state ensemble are listed in Table~\ref{tab:cross-val-ldho}. The median-based predictor $\hat{\mu}_{0.5}$ outperforms  the mode-based predictor with respect to all the measures.  We also calculate CV measures by omitting 31 realizations of the process that lead to unreasonably high frequency estimates $\hat{\omega}_d >1$ (recall that $\omega_d \approx 0.126$).  Somewhat surprisingly, this does not seem to improve the forecasting performance.

\begin{table}[!ht]
\centering
\caption{Average cross-validation measures over the 51-step forecasting horizon. The reported values represent averages over 500 realizations of the LDHO stochastic process with parameters $\mu=1$, $\sigma=1$, $\kp=3$, $\tau_c=30$ and $\omega_d = 2\pi/50$.  The median predictor is based on $\expk(\mu_{\ast})$ where $\mu_{\ast}$ is given by~\eqref{eq:predictive-mean}, while the mode predictor is based on~\eqref{eq:mode-predictor}. The ``no-outliers'' forecasts exclude realizations with very high estimates of $\omega_d$. }
\begin{tabular}{c|cccccc}
\hline
Prediction Method & ME & MAE & MARE & RMSE & RRMSE & $R$ \\
 \hline
Median predictor  &   \bf{$-$0.0472}  &  \bf{0.3254}  &  \bf{0.2732}  &  \bf{0.4079}  &  \bf{0.3987}  &  \bf{0.6055}    \\
Mode predictor  & $-$0.1011  & 0.3475 & 0.2954 & 0.4569 & 0.4520  & 0.5628  \\
\hline
Median (no outliers) & $-$0.0508  &  0.3292 & 0.2785 & 0.4125 &  0.4074  &  0.5936  \\
Mode (no outliers) & $-$0.1078 & 0.3528  & 0.3023 & 0.4644 & 0.4640  & 0.5483 \\
\end{tabular}
\label{tab:cross-val-ldho}
\end{table}

\begin{figure}[!ht]
\centering
\includegraphics[width=0.5\linewidth]{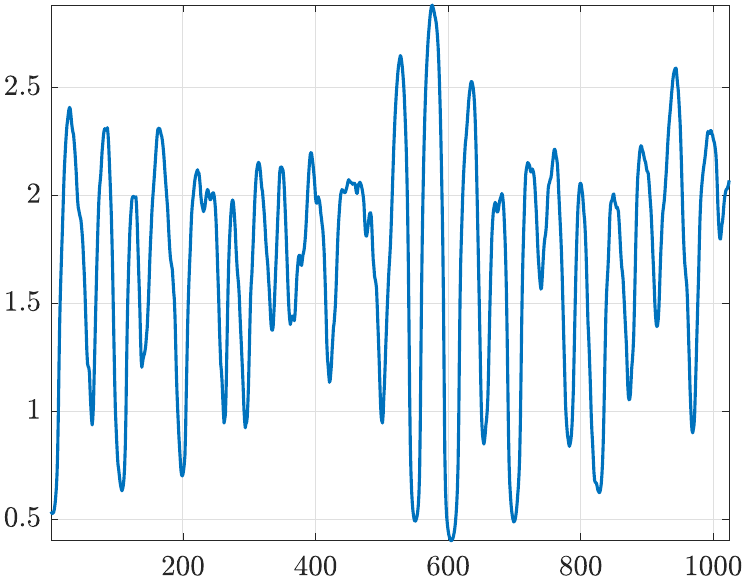}
\caption{Time series generated from the $\kp$-LN process with joint PDF~\eqref{eq:klogn-pdf-joint} and LDHO covariance kernel~\eqref{eq:ldho-kernel} with parameter set $\bmze = (\mu, \sigma, \kp,  \tau_{c}, \omega_{d})=(1, 1, 3, 30, 0.13)$.}
\label{fig:simulated-klogn-process}
\end{figure}

\begin{figure}[!ht]
\centering
\includegraphics[width=0.49\linewidth]{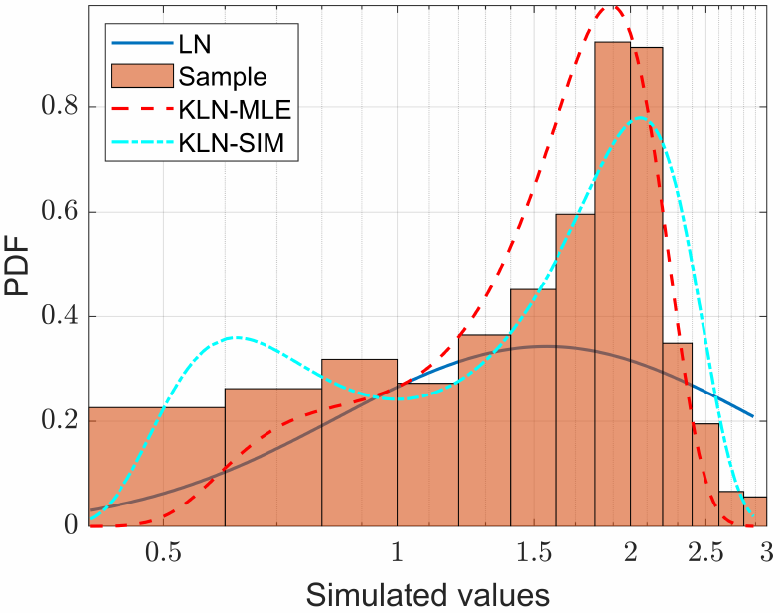}
\includegraphics[width=0.49\linewidth]{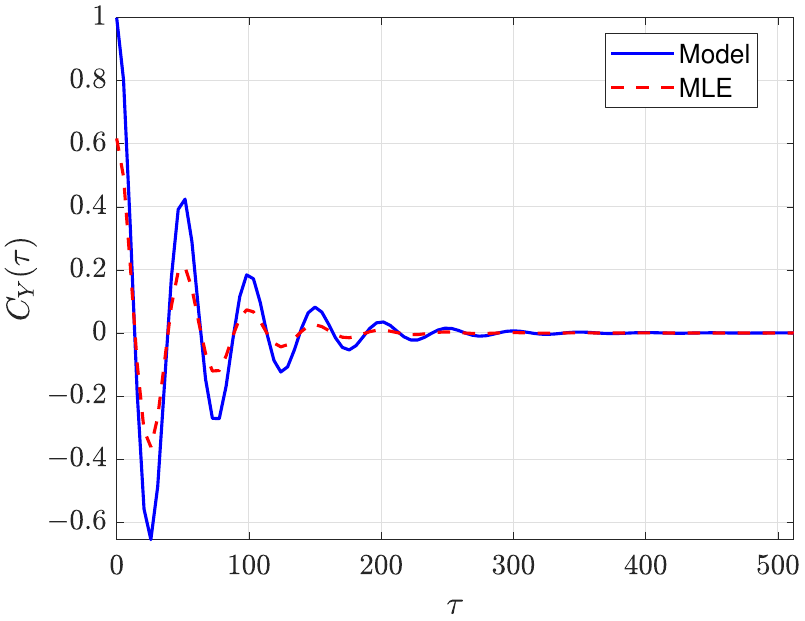}
\caption{\textbf{Left:} Sample histogram of $\kp$-lognormal process (bars),  $\kp$-lognormal PDF corresponding to the parameter set $\bmze = (\mu, \sigma, \kp,  \tau_{c}, \omega_{d})=(1, 1, 3, 30, 0.13)$ (dash-dot line),  $\kp$-lognormal PDF obtained for the MLE $\hat{\bmze}=(\hat{\mu}, \hat{\sigma}, \hat{\kp}, \hat{\tau}_{c}, \hat{\omega}_{d}) \approx (0.82, 0.79, 2.82, 23.63, 0.13)$ (broken line), and  best-fit lognormal PDF (continuous line).  \textbf{Right:}  Model LDHO kernel $C_{Y}(\tau)$ (continuous line) based on $\sigma, \tau_c$, and $\omega_d$ with the LDHO kernel based on  $\hat{\sigma}, \hat{\tau}_c$, and $\hat{\omega}_d$ (broken line).}
\label{fig:simulated-PDF-kernel}
\end{figure}

\begin{figure}
\centering
\includegraphics[width=0.49\linewidth]{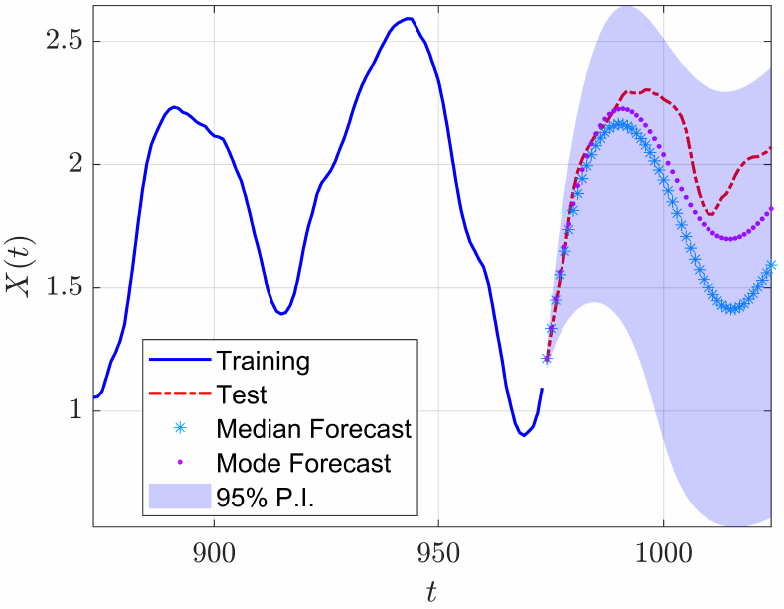}
\includegraphics[width=0.49\linewidth]{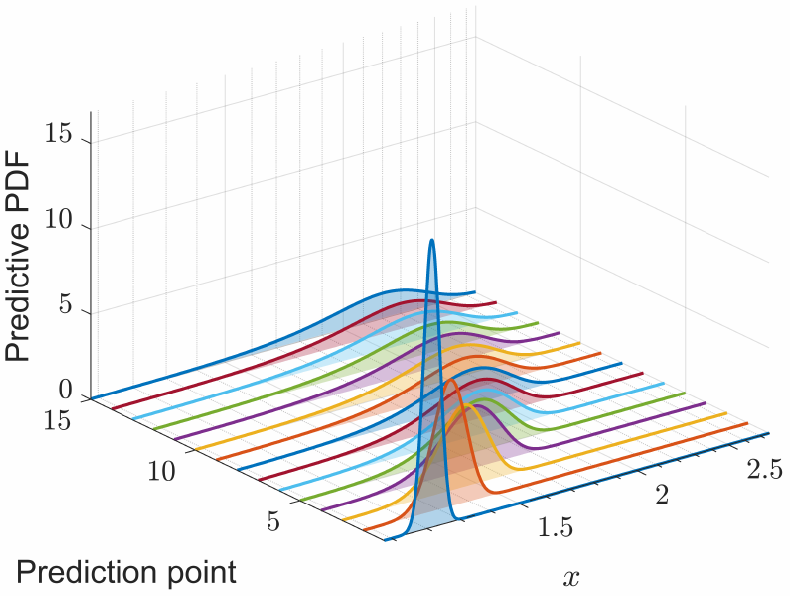}
\caption{Multi-step forecast for the time series shown in Fig.~\ref{fig:simulated-klogn-process}. \textbf{Left.}  Diagram showing the last 100 points of the training set (continuous line, blue online), the values of the test set (dash-dot line, red online), the median-based forecasts (stars, cyan online), and the mode-based  forecasts (dots, purple online). The shaded area denotes the 95\% prediction interval.  \textbf{Right.} Predictive PDFs for the first 15 points in the test set based on~\eqref{eq:predictive-marginal}. }
\label{fig:simulated-forecast}
\end{figure}

\paragraph{Single realization analysis} We focus on the first realization of the ensemble, namely the time series shown in Fig.~\ref{fig:simulated-klogn-process}. The ML estimated parameter vector is
$\hat{\bmze}=(\hat{\mu}, \hat{\sigma}, \hat{\kp}, \hat{\tau}_{c}, \hat{\omega}_{d}) \approx (0.82, 0.79, 2.82, 23.63, 0.13)$---using precision of two decimal places. The optimal NLL for this solution is ${\mathrm{NLL}} \approx -2.3\times 10^{-3}$. The left frame of Fig.~\ref{fig:simulated-PDF-kernel} displays the sample histogram (bars), the $\kp$-lognormal PDF for the population parameter vector $\bmze=(1, 1, 3,  30, 0.13)^\top$ (dash-dot line), the $\kp$-lognormal PDF obtained for the MLE $\hat{\bmze}$ (broken line), and the best-fit lognormal PDF, also derived by means of MLE (continuous line). The lognormal PDF is not a good fit for the bimodality of the underlying process, while the MLE $\kp$-lognormal PDF provides a better match with the sample histogram and the model PDF.  The right frame of Fig.~\ref{fig:simulated-PDF-kernel} confirms the agreement between  the model LDHO kernel $C_{Y}(\tau)$ (continuous line) based on $\sigma, \tau_c$, and $\omega_d$ with the LDHO kernel obtained by means of  $\hat{\sigma}, \hat{\tau}_c$, and $\hat{\omega}_d$ (broken line). Figure~\ref{fig:simulated-forecast} (left frame) shows the forecasts over the test set (51 points) using both the median-based and mode-based~\eqref{eq:mode-predictor} predictors as well as the 95\% prediction intervals~\eqref{eq:prediction-interval}. The mode-based predictor yields forecasts that track the true test values better  than the median-based predictor. This performance is confirmed by the CV measures reported in Table~\ref{tab:cross-val-ldho-single-Real}. However, the superior performance of the mode-based predictor is not universal as shown by Table~\ref{tab:cross-val-ldho},  which reveals better overall performance of the median predictor over the 500 realizations. The right frame of Fig.~\ref{fig:simulated-forecast} displays the evolution of the predictive marginal PDF~\eqref{eq:predictive-marginal} (increasing mean and standard deviation) for the first 15 test points.

For short-term forecasts, the one-step-ahead strategy is more suitable.  In this case,  the model  which is determined from the training data is used without updating as in multi-step forecasting. However, the  one-step-ahead forecasts are generated in a recursive manner, by extending  the training data through inclusion of the test values from earlier times. For one-step-ahead forecasting (not shown), all the predictions are in excellent agreement with the true test values.

\begin{table}[!ht]
\centering
\caption{Cross-validation measures over the 51-step forecasting horizon based on a single random realization (cf. Fig.~\ref{fig:simulated-klogn-process}) of the  stochastic process with LDHO kernel~\eqref{eq:ldho-kernel} and parameters $\mu=1$, $\sigma=1$, $\kp=3$, $\tau_c=30$ and $\omega_d = 2\pi/50$. The median predictor is based on $\expk(\mu_{\ast})$ where $\mu_{\ast}$ is given by~\eqref{eq:predictive-mean}, while the mode predictor is based on~\eqref{eq:mode-predictor}.  }
\begin{tabular}{c|cccccc}
\hline
Prediction Method & ME & MAE & MARE & RMSE & RRMSE & $R$ \\
 \hline
Median predictor  &   0.2612  &  0.2642    & 0.1296  &  0.3251  &  0.1606  &  0.7471    \\
Mode predictor  & 0.1278  &  0.1344    & 0.0651  &  0.1692   & 0.0814  &  0.9006  \\
\hline
\end{tabular}
\label{tab:cross-val-ldho-single-Real}
\end{table}

\rem[Using variable $\kp$] If we  optimize the  $\kp$-lognormal model allowing $\kp$ to vary,  global optimization yields essentially a lognormal solution with  $\hat{\bmze}=(\hat{\mu}, \hat{\sigma}, \hat{\kp}, \hat{\tau}_{c}, \hat{\omega}_{d}) \approx (0.44, 0.21, \,1.65\,10^{-4}, 169, 339\,)$. These estimates, however, differ significantly from the true parameter values and lead to  higher ${\mathrm{NLL}} \approx -1.82\times\,10^{-3}$ than the fixed-$\kp$ solution. Moreover, the lognormal solution yields poor representation of the temporal correlation structure and inferior CV measures. We believe that the failure of the global optimization in this case is due, at least partly, to the  non-convexity of the likelihood for Gaussian processes with quasi-periodic kernels~\cite{Li23}.

\subsection{Spatial Regression using $\kp$-lognormal Warping}
\label{ssec:applied-k-lognormal}

\begin{figure}[!ht]
\centering
\includegraphics[width=0.5\linewidth]{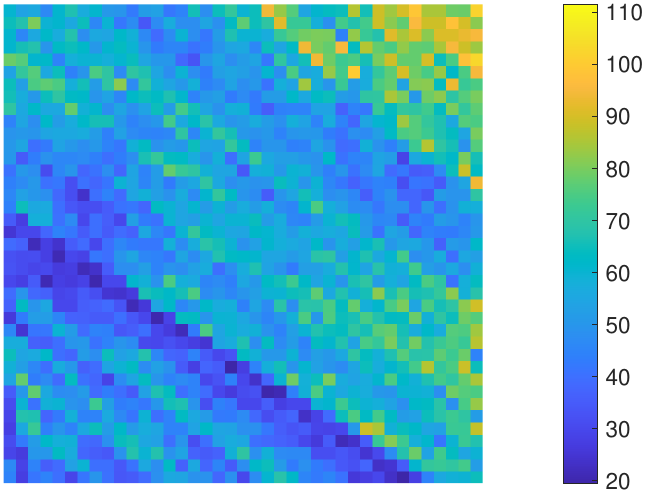}
\caption{Spatial configuration of the gridded Berea permeability dataset.  The colorbar measures permeability in mD.}
\label{fig:Berea-spatial}
\end{figure}

As an example of spatial regression, we apply $\kp$-lognormal processes to the Berea permeability dataset.  We use a randomly selected training set with $N_{\rm tr}=500, 1100$ points and predict the permeability values over the test set that includes $N_{\rm te}=1100, 500$ points.  The permeability field,  shown in Fig.~\ref{fig:Berea-spatial}, is assumed to follow the stochastic process $P(\bfs)=X(\bfs)+ \epsilon(\bfs)$, where $\epsilon(\bfs)$ is a Gaussian random noise field with variance $\sigma^{2}_{\epsilon}$. The prevailing spatial patterns indicate roughness and anisotropic dependence.  Accordingly, the permeability correlations are modeled in terms of a latent Gaussian process with  exponential kernel and elliptical anisotropy~\cite{dth08}. The anisotropy parameters are the ratio of the principal correlation lengths $\rho$ and the rotation angle $\vphi$ of the principal anisotropy axes with respect to the coordinate system.
The anisotropic exponential correlation kernel is given by
\begin{subequations}
\label{eq:Berea-kernel}
\beq
\label{eq:Berea-kernel-1}
\rho_{Y}(\bfr) = \exp\left(-\sqrt{\bfr^\top {\mathbf{M}}^{-1}\,\bfr} /\xi\right)\,,
\eeq
where $\bfr=\bfs-\bfs'$ is the spatial lag,  ${\mathbf{M}}^{-1}$ is the combined rotation and length rescaling matrix from the grid axes to the principal anisotropy system given by~\cite{dth08}
\beq
\label{eq:Berea-kernel-2}
{\mathbf{M}}^{-1} = \left[
\begin{array}{cc}
 \cos^{2}\vphi + \frac{1}{\rho^2}\sin^{2}\vphi    &  \cos\vphi\, \sin\vphi \left(1 - \frac{1}{\rho^2}\right) \\
 \cos\vphi\, \sin\vphi \left(1 - \frac{1}{\rho^2}\right)    & \sin^{2}\vphi + \frac{1}{\rho^2}\cos^{2}\vphi
\end{array}
\right]\,.
\eeq
\end{subequations}

The value of $\kp$ is set to the estimate $\hat{\kp}=0.556$ obtained in Section~\ref{ssec:Berea}. The remaining model parameters, which include $\mu, \sigma, \xi, \rho, \varphi$, are estimated by MLE as described in Section~\ref{ssec:mle-joint}.

\begin{table}[!ht]
\centering
\caption{ML parameter estimates and cross-validation measures for the Berea permeability dataset. The ML estimates are based on two randomly selected  training sets with $N_{\rm tr}=500, \, 1100$. The cross-validation measures are calculated over the respective test sets containing $N_{\rm tr}=1100, \, 500$ sites. The median predictor is based on $\expk(\mu_{\ast})$ where $\mu_{\ast}$ is given by~\eqref{eq:predictive-mean}, while the mode predictor is based on~\eqref{eq:mode-predictor}.  The exponential spatial covariance kernel~\eqref{eq:Berea-kernel} is used.  The rotation angle is given in both radians and degrees.}
\begin{tabular}{c|cccccc}
\hline
MLE estimates & $\kp$ & $\mu$ & $\sigma^{2}_{\epsilon}$ & $\sigma^2$ & $\xi$ & $\rho$,  $\varphi$ \\
\hline
$N_{\rm tr}=500$ & 0.88  & 20.81 &   2.60  & 40.31  &  5.42 &  16.53,     0.97 (55.58$\deg$)
 \\
 \hline
Prediction Method & ME & MAE & MARE & RMSE & RRMSE & $R$ \\
 \hline
Median predictor  &  0.55 & 6.65 &  0.13 &   8.67 &    0.16&  0.83 \\
Mode predictor  &   0.62 &  6.66  &  0.12 & 8.68 &    0.16 & 0.83  \\
\hline\hline
MLE estimates & $\kp$ & $\mu$ & $\sigma^{2}_{\epsilon}$ & $\sigma^2$ & $\xi$ & $\rho$,  $\varphi$ \\
\hline
$N_{\rm tr}=1100$ & 0.88  & 21.24 &   1.84  & 34.64  &  3.39 &  10.3, 0.95 (54.43$\deg$)
 \\
 \hline
Prediction Method & ME & MAE & MARE & RMSE & RRMSE & $R$ \\
 \hline
Median predictor  &   0.21  &  5.59  & 0.11 &  7.18 & 0.14  &  0.89 \\
Mode predictor  &   0.28 &  5.59 & 0.11    & 7.18  &  0.14  &  0.89   \\
\hline
\end{tabular}
\label{tab:cross-val-Berea}
\end{table}

The parameter estimates and the cross-validation measures for the two training data sets ($N_{\rm tr}=500, 1100$) are shown in Table~\ref{tab:cross-val-Berea}. For  both training sets, significant anisotropy is detected (in agreement with the visuall assessment of the permeability plot) indicating a principal axes system oriented at $\approx 55\deg$ with a correlation length ratio $>10$.

The differences between the median-based and mode-based predictors are minimal in the first case and non-existent in the latter  (except for the mean error which is in both cases smaller for the median-based predictor). The relative RMSE is 16\% for $N_{\rm tr}=500$ and $14\%$ for $N_{\rm te}=1100$ which underlines the difficulty in predicting the rough permeability profile shown in Fig.~\ref{fig:Berea-spatial}.
We also conducted the same numerical experiment as above using a Mat\'{e}rn covariance kernel for the latent Gaussian process. The cross-validation results are essentially identical to those obtained with the exponential kernel  (see Section~4 in the Supplement). This is not surprising since the Mat\'{e}rn kernel with $\nu=1/2$ is equivalent to the exponential kernel~\cite{dth20}.

\section{Conclusion}
\label{sec:Conclusions}
This paper investigates the statistical properties and the estimation of the $\kp$-lognormal probability distribution, where $\kp>0$ is a parameter that controls the shape and the asymptotic behavior of the distribution's right tail. This work also presents a $\kp$-lognormal stochastic process. We show that in the limit $\kp=0$ the standard  lognormal distribution is recovered. For $\kp>0$, we obtain probability distributions with flexible, lighter-than-lognormal right tails. Hence, the $\kp$-lognormal family provides  a unifying model which includes the standard lognormal as well as bimodal distributions, and it also approximates quite closely the normal distribution.
Such versatility can be useful in  practice. For example, simulation-based gene expression studies find both lognormal and normal protein number distributions, which switch  to bimodality by introducing a DNA looping mechanism~\cite{Krishna05}. The $\kp$-lognormal is a candidate model for such applications. On the other hand, bimodal distributions are characterized by the location, height and width of the two peaks as well as the overall asymmetry of the distribution. It is  unlikely that the three-parameter $\kp$-lognormal distribution can  adequately adapt to all of these features, but it provides a starting point for further investigation.

We present closed-form expressions for the main statistical functions (PDF, CDF, quantile function, hazard rate) of the $\kp$-lognormal distribution and for the median.  The presence of bimodality in regions of the parameter space $(\mu,\sigma,\kp)$ is established using theoretical analysis and numerical calculations. We obtain the hazard rate for the $\kp$-lognormal distribution, and we show that for ($\kp<1$) $\kp>1$ it is an asymptotically (decreasing) increasing function. This adaptability of the $\kp$-lognormal renders it more suitable than the lognormal for  failure-time analysis.  We also derive (i) lower and upper bounds for the statistical moments in terms of the error function and hypergeometric functions and (ii) power series expansions for the moments which involve a recursive equation for the coefficients.  Finally, we prove the scaling relation~\eqref{eq:moment-scaling} which implies that moments of any positive order are obtained  form the first-order moment, if the latter is known for all $(\mu,\sigma,\kp)$.

We define $\kp$-lognormal stochastic processes and focus on practical aspects such as model parameter estimation and simulation (generation of states).
We also show how to exploit the $\kp$-lognormal distribution in the framework of warped Gaussian process regression. We develop predictors based on both the median and the mode of the conditional $\kp$-lognormal distribution, and we present case studies that involve time series forecasting and interpolation of spatial data.

Even though we addressed the main goal set forth at the beginning of the paper, namely to develop a deformed lognormal probability distribution with flexible upper tail for asymmetric data,  the advances presented herein raise additional questions for  future exploration. Some of these challenges are briefly described below.

\paragraph{Lognormal deformations} Besides the PDF~\eqref{eq:klogn-pdf}, other deformations of the lognormal distribution are in principle possible; the resulting modified distributions do not necessarily  exhibit shorter-than-lognormal tails.  For example, the exponential term that appears in the standard lognormal PDF~\eqref{eq:klogn-pdf} could be replaced by $\left[ 1 - \frac{1}{\kp'}\left(\ln (x) -\mu\right)^2 / 2\sigma^2 \right]^{\kp'}$; this function recovers the natural exponential term in the lognormal PDF at the limit $\kp' \to \infty$.  Using suitable normalization, the above leads to the so-called log-kappa distribution~\cite{Leitner09}.  The latter tends to the standard lognormal for $\kp \to \infty$ but presents heavier-than-lognormal tails for finite $\kp$. The log-kappa model describes the observed distributions of magnetic field and plasma fluctuations in the
solar wind more accurately than the lognormal.

A PDF with two deformation parameters can be obtained by applying both the \kpe and \kpl deformations leading to
\beq
\label{eq:double-deformation}
f_{X}(x)= \frac{1}{Z}\, \exp_{\kp'}\left[-\frac{1}{2\sigma^2}\left( \lnk (x) - \mu\right)^2\right]\,
\eeq
where $\kp ,\kp' >0$ are two deformation parameters that satisfy $\kp' <2\kp$ and $Z>0$ is the normalization factor. The right tail of this distribution behaves as a power-law, that is,
\beq
\label{eq:double-asymptotic}
f_{X}(x) \sim \frac{(2\sigma^{2})^{1/\kp'}}{Z}\left(\frac{2\kp}{\kp'}\right)^{1/\kp'}\, x^{-2\kp/\kp'}\,,\;\mbox{for} \; x \to \infty\,.
\eeq
In the special case $\kp' \to 0$,  the deformed PDF~\eqref{eq:double-deformation} has some interesting properties: (1) For $\kp=0$, the PDF~\eqref{eq:double-deformation} becomes the lognormal. (2) For $\kp>0$ and $x \sim 0$, the PDF is approximately lognormal, because $\lnk (x) \approx \ln (x)$ for $x \approx 0$. (3) For $\kp>0$ and $\lnk (x) -\mu \gg \sigma$, it follows from~\eqref{eq:lnk-asympt} that $f_{X}(x) \sim \frac{1}{Z}\, \exp\left( -\frac{x^{2\kp}}{8\sigma^{2}\kp^2}\right)$ if $\kp>0$. (3a) If $0< 2\kp <1$,  the PDF exhibits stretched exponential upper tail;  (3b) for $\kp=1$, the upper tail of the PDF is the same as for the normal PDF with zero mean and variance $4\sigma^2$.

Note, however, that the above deformations are not obtained by means of a closed-form transformation function $g(\cdot)$ such that the transformed variable $Y(\om)=g(X(\om))$ is normally distributed.  Hence, in contrast with the $\kp$-lognormal distribution presented herein, PDFs such as~\eqref{eq:double-asymptotic} and the log-kappa PDF do not allow straightforward extension to respective joint densities by means of Jabobi's multivariate theorem (cf. Theorem~\ref{theorem:Jacobi-mult}). Such lognormal deformations are therefore more suitable for single-point characterization of the underlying variables but not easily extended to a stochastic process framework.

\paragraph{Joint $\kp$-lognormal densities} In addition to the joint lognormal PDF derived from the respective normal PDF and Jacobi's multivariate theorem,  several other  multivariate lognormal distributions have been proposed. Such distributions  do not possess a lognormal marginal but have applications in the analysis of insurance data~\cite{Nadarajah22}. Could the deformed exponential and logarithm functions be used to  derive generalizations of such distributions?

\paragraph{Relation with Central limit theorem?} It is well known that by virtue of the Central limit theorem (applied in the domain of logarithms), the lognormal model represents the asymptotic distribution for multiplicative averaging, namely for the $N$-th root of the product of $N$ independent, identically distributed random variables.  One can then ask whether the $\kp$-lognormal distribution appears as  a pre-asymptotic form of multiplicative averaging for  finite $N$  or as a fixed point if the aggregated processes have  long-range correlations.

\appendices

\section{Determining the Modes of $\kp$-Lognormal PDF}
\label{app:modes}

This section presents the proof of Theorem~\ref{theorem:peaks}.

\begin{IEEEproof}
If $\kp=0$ the corresponding PDF is the lognormal which is known to be umimodal with the mode given by $x_{\rm{mode}}=\exp\left(\mu-\sigma^2\right)$. Therefore, the rest of the proof focuses on $\kp>0$.  

The PDF~\eqref{eq:klogn-pdf} can be expressed as $f_{X}(x)=c\,f_{1}(x)\,f_{2}(x)$, where $c=1/\left(\sqrt{2\pi}\sigma \right)$, and
\beq
\label{eq:fx-summands}
f_{1}(x)=\exp\left[ -\left( \lnk (x) - \mu\right)^2 /(2\sigma^2)\right]\,,\quad f_{2}(x)= \frac{1}{2}\, \left( x^{\kp-1}+ x^{-\kp-1} \right)\,,
\eeq
are \emph{continuously differentiable}, positive-valued functions for $x > 0$.
Let us denote the first derivative by a prime for brevity. Then, the first derivative of the PDF is
\beq
\label{eq:fx-1-2}
f'_{X}(x) = c \left[ f'_{1}(x)f_{2}(x) + f_{1}(x) f'_{2}(x)\right]\,.
\eeq

The function $f_{1}(x)$ has a maximum at $\lnk (x)=\mu$ where the exponent becomes zero. Hence, $f'_{1}(\xmed)=0$. Moreover,  it follows from~\eqref{eq:lnk-first-deriv} that $f_{2}(x)= \dd \lnk(x)/\dd x$.  The following lemmas will be useful in the proof.

\begin{lemma}[Derivative of $f_{2}$ for $\kp \le 1$]
\label{lemma:f2-kp-leq-1}
If $\kp \le 1$, according to Proposition~\ref{propo:kpl-convexity}   the \kpl is an increasing concave function. Therefore,  $f_{2}(x)$, the \kpl derivative,  is a positive, monotonically decreasing function, and thus  $f'_{2}(x)<0$ for all $x\in\R_{0,+}$.
\end{lemma}

\begin{lemma}[Derivative of $f_{2}$ for $\kp > 1$]
\label{lemma:f2-kp-gt-1}
For $\kp >1$, according to Proposition~\ref{propo:kpl-convexity} the \kpl is a concave function for $0<x\le x_{+}$ and convex for $x>x_{+}$, where according to the analysis following~\eqref{eq:lnk-second-deriv},
\beq
\label{eq:inflection-point}
x_{+} \triangleq \left( \frac{\kp +1}{\kp -1}\right)^{1/2\kp}
\eeq
is the \emph{inflection point}. Therefore, $f_{2}(x)$ is monotonically decreasing for $x< x_{+}$ and monotonically increasing for $x>x_{+}$.
\end{lemma}

\begin{lemma}
\label{lemma_signs-fder}
Taking into account~\eqref{eq:fx-1-2} the following statements are true:
(i) For all $\kp>0$,   it holds that $f'_{X}(\xmed)=f'_{2}(\xmed)f_{1}(\xmed)$ since $f'_{1}(\xmed)=0$; thus the sign of $f'_{X}(\xmed)$ is the same as the sign of $f'_{2}(\xmed)$.  (ii) For $\kp>1$, since $f'_{2}(x_{+})=0$, it holds that $f'_{X}(x_{+})=f'_{2}(x_{+})f_{1}(x_{+})$; thus the sign of $f'_{X}(x_{+})$ is the same as the sign of $f'_{1}(x_{+})$.
\end{lemma}

We   consider the cases $0 <\kp \le 1$ and $\kp>1$ separately.   Recall that $\xmed = \expk(\mu)$ is the \emph{median} of the $\kp$-lognormal distribution.

\paragraph{$0< \kp \le 1$} We will establish the existence of at least one mode of $f_{X}(\cdot)$ in $(0, \xmed)$ by showing that the continuous function $f'_{X}(x)$ changes sign at least once in $(0, \xmed)$. We also show that the PDF has no modes for $x>\xmed$.
\begin{enumerate}
\item Since $f_{2}(x)$ decreases monotonically (according to Lemma~\ref{lemma:f2-kp-leq-1}) and $f_{1}(x)$ is positive everywhere, $f_{1}(x) f'_{2}(x)<0$ for all $x >0$.  On the other hand, $f'_{1}(x)f_{2}(x) \lessgtr 0$ for $x \gtrless \xmed$.  Hence   critical points $x_\ast$ of $f_{X}(x)$  where $f'_{X}(x_\ast)=0$, if they exist,  are found inside  the interval $(0, \xmed )$ where the two summands of $f'_{X}(x)$ have opposite signs.

\item In addition,  $f'_{2}(\xmed)<0$, $f_{1}(\xmed)=1$ and  $f'_{1}(\xmed)=0$. Hence, it follows from Lemmma~\ref{lemma_signs-fder} that $f'_{X}(\xmed) =f'_{2}(\xmed)<0$.

\item Since $f_{X}(x)$ is a continuous, non-negative function and $f_{X}(0)=0$,  there exists at least some range of values between zero and $\xmed$ where $f'_{X}(x)>0$.

\item Based on statements (2)-(3) above,  $f'_{X}$ changes sign at least once in $(0, \xmed)$.  Since $f'_{X}$ is differentiable, by Bolzano's theorem  it has at least one root $x_{\ast} \in (0, \xmed)$ corresponding to (at least) one peak  of $f_{X}(x)$. Furthermore, for $x>\xmed$ both $f'_{1}f_{2}<0$ and $f'_{2}f_{1}<0$ (the latter according to Lemma~\ref{lemma:f2-kp-leq-1}) leading to $f'_{X}<0$; therefore, the PDF does not have critical points for $x>\xmed$.
\end{enumerate}

\medskip
\paragraph{$\kp >1$} We  establish the existence of at least one mode of $f_{X}(\cdot)$ by determining intervals where the continuous function $f'_{X}(x)$ changes sign.

\begin{enumerate}
\item The function $f'_{1}(x)f_{2}(x)$ is negative for $x>\xmed$ and positive for $x<\xmed$ (due to the maximum of $f_{1}$  at $\xmed)$.
\item The function $f_{1}(x) f'_{2}(x)$ changes sign around  $x_{+}$ according to Lemma~\ref{lemma:f2-kp-gt-1}; thus, $f_{1}(x) f'_{2}(x)< 0$ for $x < x_{+}$ while $f_{1}(x) f'_{2}(x) > 0$ for $x > x_{+}$.
\item First we consider the case $\xmed<x_{+}$. If $\mu<0$ the condition is automatically satisfied since $x_{+}>0$.  If $\mu>0$, the condition requires  $\mu < \left(\kp\sqrt{\kp^{2} -1}\right)^{-1}$. According to Table~\ref{tab:signs-pdf-derivative-1} (Rows 1-4), the critical points $x_\ast$ of $f_{X}(x)$ should be sought within the intervals $(0, \xmed )$ and $(x_{+},\infty)$ where the summands  of $f'_{X}(x)$ have opposite signs.

\noindent
(i) According to Lemma~\ref{lemma_signs-fder}, $f'_{X}(\xmed) =f'_{2}(\xmed)<0$. Since $f_{X}(x)$ is a continuous nonnegative function and $f_{X}(0)=0$,  there exists at least some range of values between zero and $\xmed$ where $f'_{X}(x)>0$. Based on the above,
$f'_{X}$ changes sign at least once in $(0, \xmed)$. Therefore, by Bolzano's theorem, $f'_{X}$ has at least one root $x_{\ast} \in (0, \xmed)$ corresponding to (at least) one peak of $f_{X}(x)$. (ii)  It also holds that $f'_{X}<0$ for $x\in (\xmed, x_{+})$, since both summands of $f'_{X}$ are negative (see 1 and 2 above and Table~\ref{tab:signs-pdf-derivative-1}).  Furthermore, using Lemma~\ref{lemma_signs-fder} again,  $f'_{X}(x_{+}) =f'_{1}(x_{+})\,f_{2}(x_{+})<0$.  If there  exist points in  $(x_{+},\infty)$ where $f'_{X}$  changes sign (since  in this interval the two summands have opposite signs  according to 1 and 2),  one (or more) peak(s) are possible for $x>x_{+}$.

\item Next, we consider the case $\xmed>x_{+}$.  The critical points $x_\ast$ are now found inside the intervals $(0, x_{+})$ and $(\xmed ,\infty)$.

\noindent In this case, $f'_{1}(x_{+})>0$ because $f_{1}$ has not reached its maximum at $x_{+} < \xmed$. Therefore, from Lemma~\ref{lemma_signs-fder} it follows   that $f'_{X}(x_{+})>0$. In addition, $f'_{X}(\xmed)>0$ as well because $f'_{2}(\xmed)>0$ (cf. Lemma~\ref{lemma:f2-kp-gt-1}). According to Table~\ref{tab:signs-pdf-derivative-1} (Rows 5-8), $f'_{X}$ is positive inside $(x_{+},\xmed)$, and it may change sign within the intervals $(0,x_{+})$ and $(\xmed,\infty)$ where the summands have opposite signs. Thus, possible critical points of $f_X(x)$  lie inside these intervals.

\item Finally, for the degenerate case $\xmed = x_{+}$ there is a critical point inside $(0, \xmed)$ (based on the arguments used for $\xmed < x_{+}$), and possibly more  for higher values.

\end{enumerate}

\begin{table}[!ht]
\centering
\caption{Signs of the PDF-derivative summands~\eqref{eq:fx-summands} as well as of the $\kp$-lognormal PDF derivative~\eqref{eq:fx-1-2} per interval for $\kp>1$. The case $x_{+}>\xmed$  corresponds to Rows 1-4, while the case $x_{+} < \xmed$ to Rows 5-8.  $\xmed$ is the $\kp$-lognormal median defined in~\eqref{eq:klogn-median} and $x_{+}$ is the $\kp$-lognormal PDF inflection point~\eqref{eq:inflection-point}. The symbol $+ \to -$ denotes a transition from positive to negative sign, while the symbol $+/-$ implies that the sign is undetermined ($f'=0$ is also possible).  For $x_{+}>\xmed$ the following boundary inequalities hold: $f'_{X}(\xmed)<0$ and $f'_{X}(x_{+})<0$.  For $x_{+}<\xmed$ the following boundary inequalities hold: $f'_{X}(\xmed)>0$ and $f'_{X}(x_{+})>0$.}
\label{tab:signs-pdf-derivative-1}
\begin{tabular}{c|c|c|c}
Intervals & $(0, \xmed)$ & $(\xmed, x_{+})$ & $(x_{+}, \infty)$
 \\[1ex]
 \hline
$f'_{1}\,f_{2}$ & $+$ & $-$
 & $-$ \\[1ex]
 \hline
$f_{1}\,f'_{2}$ & $-$ & $-$  & $+$
\\[1ex]
\hline
$f'_{X}$  & $+ \to -$ & $-$ & $ -/+$
\\[1ex]
\hline\hline
Intervals & $(0, x_{+} )$ & $(\,x_{+}, \xmed \,)$ & $(\xmed, \infty)$
 \\[1ex]
 \hline
$f'_{1}\,f_{2}$ & $+$ & $+$
 & $-$ \\[1ex]
 \hline
$f_{1}\,f'_{2}$ & $-$ & $+$  & $+$
\\[1ex]
\hline
$f'_{X}$  & $+/ -$ & $+ $ & $+ /-$
\\[1ex]
\hline
\end{tabular}
\end{table}

\medskip

To determine the number  of critical points of the $\kp$-lognormal PDF~\eqref{eq:klogn-pdf}, we investigate the behavior of $f'_{X}(x)$ with respect to $x$. For $\kp>0$ it holds that

\begin{align}
\label{eq:logknormal-pdf-deriv}
f'_{X}(x) = &\frac{-1}{8\, \sqrt{2\pi}\sigma^{3}\kp\,x^{2}} \, \E^{- \left(\lnk (x) - \mu\right)^2/2\sigma^2} \, g_{0}(x) \,,
\nonumber \\[1ex]
g_{0}(x) & = x^{3\kp} - x^{-3\kp} + x^{\kp}-x^{-\kp}-2\mu\kp \,x^{2\kp} -4\mu \kp -2\mu\kp \,x^{-2\kp}
\nonumber \\
& \quad -4 x^{\kp}\kp^{2}{\sigma}^2 +4 x^{\kp}\sigma^{2}\kp + 4 x^{-\kp}\kp^{2}\sigma^{2} + 4 x^{-\kp}\sigma^{2}\kp \,.
\end{align}
At $x=0$ both the PDF and its first derivative vanish due to the  exponential term.  Hence, we will seek extrema of the PDF by searching for the roots of $g_{0}(x)$ in $(0, \infty)$.  Since $x>0$ the roots of $g_{0}(x)$ coincide with those of
$g_{1}(x)= x^{3\kp}\,g_{0}(x)$. Furthermore, we define  $p_{1}(z) \triangleq g_{1}(x=z^{1/\kp})$ which leads to the sixth-degree  polynomial~\eqref{eq:polynom-df-dx}.
According to the fundamental theorem of algebra, the polynomial $p_{1}(z)$ has exactly six, in general complex, roots which correspond to roots of the $\kp$-lognormal PDF. The real roots of $p_{1}(z)$ determine the behavior (number and location of peaks) of $f_{X}(x)$.

Note that for all $\kp>0$, $p_{1}(z)$ is a univariate polynomial with real-valued coefficients. Thus, we can use \emph{Descartes' rule of signs}: if the nonzero terms of $p_{1}(z)$ are ordered by descending value of the exponent,  the number of positive roots of $p_{1}(z)$ is equal to the number $N_{\pm}$ of sign changes between consecutive (nonzero) coefficients or is less than $N_{\pm}$ by an even number. Roots of multiplicity $n$ are counted as $n$ distinct roots.

The  coefficients $a, b, c$ in $p_{1}(z)$ can each be positive, negative, or zero depending on $\kp, \mu, \sigma$---recall that $a=2\mu\kp$,  $b=1 -4\kp\sigma^{2}(\kp-1)$ and $c=4\kp\sigma^{2} (\kp+1) -1$.  Hence, there are $3^3=27$ combinations of $a, b, c$  that lead to different sign sequences. We exclude from consideration the case $\kp=0$ which leads to the unimodal lognormal distribution. The ``critical'' values of $\kp, \mu, \sigma$ where the coefficients $a, b, c$ change sign are given respectively by $\mu_{\rm crit}=0$, $\sigma^{2}_{\rm crit;1}=\frac{1}{4\kp(\kp-1)}$ (if $\kp>1$), and $\sigma^{2}_{\rm crit;2}= \frac{1}{4\kp(\kp+1)}$.

Table~\ref{tab:number-of-roots} lists the possible combinations,  sign changes of the $p_{1}(z)$ polynomial coefficients, and  number of positive roots according to Descartes' rule. For $\kp\le 1$ it holds that $b>0$; thus, $b$ can become negative only if $\kp>1$.  In addition, since $c+b = 8\,\kp\,\sigma^{2}$, combinations $b \le 0 \wedge c \le 0$ are not allowed (even for $\kp>1$). This means that 12 out of the 27 combinations are not feasible (as shown in Table~\ref{tab:number-of-roots}). For the remaining sign combinations, there are at most $n_r$ roots  of $p_{1}(z)$ where $n_r \in \{1, 3, 5\}$.  If $z_{r}$  corresponds to a root of $p_{1}(z)$, the respective peak of the $\kp$-lognormal PDF is given by
$x_{\rm{mode}}=z_{r}^{1/\kappa}$.

\begin{table}[!ht]
\centering
\caption{Number of positive roots of the polynomial $p_{1}(z)$ given by~\eqref{eq:polynom-df-dx} for all possible sign combinations of the coefficients $a=2\mu\kp$,  $b=1 -4\kp^{2}\sigma^{2} + 4\sigma^{2}\kp$ and $c=4\kp\sigma^{2} (\kp+1) -1$. The second column lists the sign combinations for $a, b, c$. The third column shows the number of sign changes for the coefficients of $p_{1}(z)$ arranged in descending order of the exponent. The fourth column represents the possible number of positive roots according to Descartes' rule. One root corresponds to a unimodal $\kp$-lognormal PDF, three roots to a bimodal  $\kp$-lognormal PDF.  Rows in grey correspond to non-permissible combinations of $b$ and $c$ that fail to satisfy the constraint $b+c>0$.}
\begin{tabular}{c|c|c|c}
Number & Signs of $a,b,c$  &  Sign  changes of $p_{1}(z)$ coefficients  &  Number of positive roots\\
\hline\hline
\rowcolor{lgray}1& $0, 0, 0$    &    $+\; -$      &  1 \\ \hline
\rowcolor{lgray}2 & $0, 0, -$    &    $+\; -\; -\;$    &  1 \\ \hline
3 & $0, 0, +$    &    $+\; +\; -\;$    &  1 \\ \hline
\rowcolor{lgray}4 & $0, -, 0$    &    $+\; -\; -\;$    &  1 \\ \hline
5 & $0, +, 0$    &    $+\; +\; -\;$    &  1 \\ \hline
\rowcolor{lgray}6 & $-, 0, 0$    &    $+\; +\; +\; +\;-\;$    &  1 \\ \hline
\rowcolor{lgray}7 & $+, 0, 0$    &    $+\; -\; -\; -\;-\;$    &  1 \\ \hline
\rowcolor{lgray} 8 & $0, -, -$    &    $+\; -\; -\; -$    &  1 \\ \hline
\rowcolor{lgray}9 & $-, 0, -$    &    $+\; +\; +\;-\;+\; -$    &  1, 3 \\ \hline
\rowcolor{lgray}10 & $-, -, 0$    &    $+\; +\; -\; +\;+\; -$    &  1, 3 \\ \hline
11 & $0, +, +$    &    $+\; +\; +\; -$    &  1 \\ \hline
12 & $+, 0, +$    &    $+\; +\; +\;-\;+\; -$    &  1, 3 \\ \hline
13 & $+, +, 0$    &    $+\; -\; +\; -\;-\; -$    &  1, 3 \\ \hline
14 & $0, +, -$    &    $+\; +\; -\; -$    &  1 \\ \hline
15 & $0, -, +$    &    $+\; -\; +\;-$    &  1, 3 \\ \hline
\rowcolor{lgray}16 & $+, 0, -$    &    $+\; -\; -\; -\;-\; -$    &  1 \\ \hline
17 & $-, 0, +$    &    $+\; +\; +\; +\; +\; -$    &  1 \\ \hline
\rowcolor{lgray}18 & $+, -, 0$    &    $+\; -\; -\;-\;-\; -$    &  1 \\ \hline
19 & $-, +, 0$    &    $+\; +\; +\; +\;+\; -$    &  1 \\ \hline\hline
20 & $-, -, +$    &    $+\; +\; -\;+\;+\;+\; -$    &  1, 3 \\ \hline
21 & $-, +, -$    &    $+\; +\; +\;+\;-\; +\;-$    &  1, 3 \\ \hline
\rowcolor{lgray}22 & $+, -, -$    &    $+\; -\; -\;-\;-\; -\; -$    &  1 \\ \hline
23 & $+, +, -$    &    $+\; -\; +\;-\;-\; -\; -$    &  1, 3 \\ \hline
24 & $+, -, +$    &    $+\; -\; -\;-\;+\; -\; -$    &  1, 3 \\ \hline
25 & $-, +, +$    &    $+\; +\; +\;+\;+\; +\; -$    &  1 \\ \hline
\rowcolor{lgray} 26 & $-, -, -$    &    $+\; +\; -\;+\;-\;+\; -$    &  1, 3, 5 \\ \hline
 27 & $+, +, +$    &    $+\; -\; +\;-\;+\; -\; -$    &  1, 3, 5 \\ \hline\end{tabular}
\label{tab:number-of-roots}
\end{table}


\end{IEEEproof}


\section{Proof of the Moment Scaling Relation}
\label{app:scaling}
This section presents the proof of Theorem~\ref{theorem:scaling}.

\begin{IEEEproof}
The $\ell$-th order non-centered moment of a $\kp$-logarithmic distribution is defined by means of~\eqref{eq:kpl-moment-ell}
\beq
\label{eq:kpl-moment-ell}
m_{X;\ell}(\kp;\mu,\sigma)= \int_{0}^{\infty} \,dx\, x^{\ell}\, f_{X}(x)\,,
\eeq
with  $f_X(x)$ the PDF of the $\kp$-lognormal distribution  given in \eqref{eq:klogn-pdf}.
Using the fact that $Y=\lnk X \overset{d}{=} \N(\mu, \sigma^2)$   and the conservation of probability under the  nonlinear, monotone   \kpe  transform $X=\expk(Y)$, the integral~\eqref{eq:kpl-moment-ell} becomes
\beq
\label{eq:kpl-moment-ell-2}
m_{X;\ell}(\kp;\mu,\sigma)= \int_{-\infty}^{\infty} \,dy\, g^{\ell}(y)\, f_{Y}(y;\mu,\sigma^2)\,= \EE_{Y}\left[ \left(\expk Y \right)^\ell\right],
\eeq
where $f_{Y}(y;\mu,\sigma^2)=\frac{1}{\sigma}\phi(\frac{y-\mu}{\sigma})$ is the PDF of the normally distributed $Y$.
Hence, the moment of order $\ell \in \mathbb{N}$ is given by
\begin{align}
\label{eq:moment-expk}
m_{X;\ell}(\kp;\mu,\sigma) \triangleq & \, \EE_{Y}\left[ \left(\expk Y \right)^\ell\right]= \EE_{Y}\left[ \exp_{\kp'}(\ell \,Y)\,\right]\,.
\end{align}
The second equation is a result of the fact that $g^{\ell}(y)= \left[\expk(y) \right]^\ell$ and
\beq
\label{eq:expk-power}
\left[\expk(y) \right]^\ell=\exp_{\kp'}(\ell \,y)\,, \; \kp'=\kp/\ell.
\eeq
Based on~\eqref{eq:moment-expk}, the $\kp$-lognormal moments are obtained from the  expectation integral~\eqref{eq:moments-integral}  which can be easily evaluated by means of numerical integration.

To prove the scaling relation~\eqref{eq:moment-scaling}, we use~\eqref{eq:expk-power} and the variable transformation $\ell y \to \tilde{y}$  in order to transform the moment integral~\eqref{eq:moments-integral}   as follows
\begin{align*}
m_{X;\ell}(\kp;\mu,\sigma)= & \int_{-\infty}^\infty \,dy \exp_{\kp/\ell}(\ell y)\,f_{Y}(y;\mu,\sigma^2) = \int_{-\infty}^\infty \,d\tilde{y}\ \exp_{\kp/\ell}(\tilde{y}) \,f_Y(\tilde{y};\ell\mu,\ell^{2}\sigma^{2})\,= m_{X;1}(\kp/\ell; \ell\mu, \ell\sigma)\,,
\end{align*}
where in the above $\tilde{Y}\overset{d}{=}\N(\ell\mu,\ell^{2}\sigma^{2})$.  This concludes the proof.
\end{IEEEproof}

\section{Proof of the Moment Bounds Relations}
\label{app:bounds}

This section presents the proof of Theorem~\ref{theorem:bounds}, that is, the derivation of the $\kp$-lognormal moment bounds.

\begin{IEEEproof}
We use the integral~\eqref{eq:moments-integral} for the moments and decompose it into two branches over $R_{0,-}$ and $\R_{0,+}$ respectively
\begin{align}
\label{eq:moments-proof}
m_{X;\ell}(\kp;\mu,\sigma)  = & \frac{1}{\sqrt{2\pi}\sigma}\,
\int_{0}^{\infty} d y\, \E^{-\frac{1}{2\sigma^2}(y-\mu)^2}\, \exp_{\kp/\ell}(\ell y)
\nonumber \\
+ &  \frac{1}{\sqrt{2\pi}\sigma}\,
\int_{-\infty}^{0} d y\, \E^{-\frac{1}{2\sigma^2}(y-\mu)^2}\, \exp_{\kp/\ell}(\ell y) \,.
\end{align}
The bounds will be derived by considering each branch separately.
The \kpe function satisfies the following bounds for any $\kp'>0$ (we will replace $\kp'$ with $\kp/\ell$ below)
\begin{subequations}
\label{eq:expk-bounds}
\begin{align}
\label{eq:bounds-ygt0}
2^{1/\kp'}\, \left(1+ \kp^{2}y^{2} \right)^{1/2\kp'} & \ge \exp_{\kp'}(\ell y)  \ge \left( 2\kp y\right)^{1/\kp'}\,, \;  \mbox{for}\; y\ge 0\,.
\\[1ex]
\label{eq:bounds-ylt0}
1 & \ge \exp_{\kp'}(\ell y)  \ge  \exp(\ell y) \,, \;  \mbox{for}\; y<0\,.
\end{align}
\end{subequations}
Both the upper and lower bounds of inequality~\eqref{eq:bounds-ygt0} are derived by using $\sqrt{1+\kp^2 y^2} > \kp y$ (the inequality is valid for $y\ge 0$): for  the lower bound,  $\sqrt{1+\kp^2 y^2}$ in $\exp_{\kp'}(\ell y)$ is  bounded from below  by $\kp y$, while for the upper bound, $\kp y$ in $\exp_{\kp'}(\ell y)$ is bounded from above by $\sqrt{1+\kp^2 y^2}$. The upper bound in~\eqref{eq:bounds-ylt0} is due to the monotonic increase  of $\exp_{\kp'}(\ell y)$ with respect to $y$ which ensures that it is bounded from above by $\exp_{\kp'}(0)=1$. On the other hand, the lower bound is due to the fact that $\exp_{\kp'}(\ell y)$ is a decreasing function of $\kp'$ bounded from below by $\exp(\ell y)$ for $y<0$ according to Proposition~\ref{propo:bounds-kpe}.

\paragraph{Lower moment bound} Based on~\eqref{eq:moments-proof} and~\eqref{eq:expk-bounds}, the lower bound $LB_{X;\ell}(\kp;\mu,\sigma)$ for $m_{X;\ell}(\kp;\mu,\sigma)$ is given by
\begin{align}
\label{eq:lb-1}
LB_{X;\ell}(\kp;\mu,\sigma)  = & \frac{1}{\sqrt{2\pi}\sigma}\, \left( \frac{2\kp}{\ell} \right)^{\ell/\kp}
\int_{0}^{\infty} d y\, \E^{-\frac{1}{2\sigma^2}(y-\mu)^2}\,  (\ell y)^{\ell/\kp}
\nonumber \\
& + \frac{1}{\sqrt{2\pi}\sigma}\,
\int_{-\infty}^{0} d y\, \E^{-\frac{1}{2\sigma^2}(y-\mu)^2 +\ell y} \,.
\end{align}

The integral over $\R_{0,-}$ (second term in the right-hand side) in the above is given by
\[
\frac{1}{\sqrt{2\pi}\sigma}\,
\int_{-\infty}^{0} d y\, \E^{-\frac{1}{2\sigma^2}(y-\mu)^2 + \ell y} = \E^{\ell \mu +\ell^{2}\sigma^{2}/2}
\left( \frac{1 - \erf\left( \frac{\mu + \ell \sigma^2}{\sqrt{2}\sigma}\right)}{2} \right) \,.
\]
The integral over the positive real line is defined  by
\begin{align*}
I_{+} \triangleq \left(\frac{2\kp}{\ell}\right)^{\ell/\kp}\,\int_{0}^{\infty} \frac{d y}{\sqrt{2\pi}\sigma}\, \E^{-\frac{1}{2\sigma^2}(y-\mu)^2}\,  (\ell y)^{\ell/\kp} \,.
\end{align*}
Evaluating the integral leads to
\begin{align}
\label{eq:ilb}
I_{+}   = &
\frac{(2\kp)^{\ell/\kp} \, 2^{\frac{\ell-\kp}{2\kp}} \, \sigma^{\ell/\kp} \, }{\sqrt{2\pi}}\, \E^{-\frac{\mu^2}{2\sigma^2}}\, \left[ \Gamma\left(\frac{\ell+ \kp}{2\kp}\right) \hyperg \left(\frac{\ell+\kp}{2\kp}, \, \frac{1}{2}\,; \, \frac{\mu^2}{2\sigma^2}\right) \right.
\nonumber
\\
& \left. \quad \quad \quad \quad + \,
\Gamma\left(\frac{\ell+ 2\kp}{2\kp}\right)\frac{\mu\sqrt{2}}{\sigma}\,  \hyperg\left(\frac{\ell+ 2\kp}{2\kp}, \, \frac{3}{2}\,; \, \frac{\mu^2}{2\sigma^2}\right) \right]\,,
\end{align}
where $\hyperg(\al,\gamma;z)$ is the \emph{confluent hypergeometric function}~\cite[9.210]{Gradshteyn07}.

To prove~\eqref{eq:ilb}, we express the integral $I_{+}$ as follows
\begin{subequations}
\label{eq:Iplus}
\begin{align}
I_{+}= & c_{+} \, \E^{-\mu^{2}/(2\sigma^2)}\,\int_{0}^{\infty} d y \,\E^{-\beta y^{2} - \gamma y}\,  y^{\ell/\kp} \triangleq  c_{+} \, \E^{-\mu^{2}/(2\sigma^2)}\, \tilde{I}_{+}\,,
\\
\tilde{I}_{+} = & \int_{0}^{\infty} d y \,\E^{-\beta y^{2} - \gamma y}\,  y^{\ell/\kp}\,,
\\
 c_{+} = & \frac{(2\kp)^{\ell/\kp}}{\sqrt{2\pi}\sigma}\,, \; \beta=\frac{1}{2\sigma^2}\,,  \; \gamma=-\frac{\mu}{\sigma^2}\,.
\end{align}
\end{subequations}
Based on~\cite[Eq. (3.462.1)]{Gradshteyn07} the integral $\tilde{I}_{+}$ becomes
\begin{align}
\label{eq:tIplus-Dnu}
\tilde{I}_{+}= & (2\beta)^{-\nu/2}\, \Gamma(\nu)\, \E^{\gamma^{2}/8\beta}\,D_{-\nu}\left(\frac{\gamma}{\sqrt{2\beta}}\right)\,,
\nonumber \\
= & \sigma^{\nu}\,\Gamma(\nu)\,\E^{\mu^2/(4\sigma^2)}\, D_{-\nu}\left(z\right)
\end{align}
where $D_{-\nu}(z)$ is the \emph{parabolic cylinder function}, $z \triangleq {\gamma}/{\sqrt{2\beta}}=-\mu/\sigma$, and $\nu \triangleq \ell/\kp+1=(\ell+\kp)/\kp$. $D_{-\nu}(z)$ can be expressed in terms of the confluent hypergeometric function as follows~\cite[9.240]{Gradshteyn07}
\beq
\label{eq:Dnu}
D_{-\nu}(z)= 2^{-\nu/2}\,\E^{-z^{2}/4}\, \left\{ \frac{\sqrt{\pi}}{\Gamma(\frac{\nu+1}{2})}\, \hyperg\left( \frac{\nu}{2},\frac{1}{2}\,;\, \frac{z^2}{2}\right) - \frac{\sqrt{2\pi}z}{\Gamma(\frac{\nu}{2})}\, \hyperg\left( \frac{\nu+1}{2},\frac{3}{2}\,;\, \frac{z^2}{2}\right) \right\}\,.
\eeq
Taking into account that $z=-\mu/\sigma$,  and plugging~\eqref{eq:Dnu} in~\eqref{eq:tIplus-Dnu}, after cancelation of the exponential terms, $\tilde{I}_{+}$ becomes
\beq
\label{eq:tIplus}
\tilde{I}_{+} = \left( \frac{\sigma^{2}}{2}\right)^{\nu/2}\, \left\{ \frac{\sqrt{\pi}\Gamma(\nu)}{\Gamma(\frac{\nu+1}{2})}\, \hyperg\left( \frac{\nu}{2},\frac{1}{2}\,;\, \frac{\mu^2}{2\sigma^2}\right) +\frac{\mu}{\sigma} \frac{\sqrt{2\pi}\Gamma(\nu)}{\Gamma(\frac{\nu}{2})}\, \hyperg\left( \frac{\nu+1}{2},\frac{3}{2}\,;\, \frac{\mu^2}{2\sigma^2}\right) \right\}\,.
\eeq
Next, we use Legendre's duplication formula for the Gamma function~\cite[6.1.17]{Abramowitz72}
\[
\frac{\Gamma(\nu)\sqrt{\pi}}{\Gamma(\frac{\nu+1}{2})}=2^{\nu-1} \, \Gamma(\nu/2)\,,
\]
based on which $\tilde{I}_{+}$ becomes
\beq
\label{eq:tIplus-1F1}
\tilde{I}_{+} =  \sigma^{\nu}\, 2^{\nu/2-1}\, \left\{  \Gamma\left(\frac{\nu}{2}\right) \, \hyperg\left( \frac{\nu}{2},\frac{1}{2}\,;\, \frac{\mu^2}{2\sigma^2}\right) +\frac{\mu}{\sigma} \, \sqrt{2}\,\Gamma\left(\frac{\nu+1}{2}\right)\, \hyperg\left( \frac{\nu+1}{2},\frac{3}{2}\,;\, \frac{\mu^2}{2\sigma^2}\right) \right\}\,.
\eeq
Finally, recalling  $c_{+}$ from~\eqref{eq:Iplus} and replacing $\nu$ with $(\ell+\kp)/\kp$  in~\eqref{eq:tIplus-1F1}, we obtain~\eqref{eq:ilb}. This concludes the proof for the lower bound.

\paragraph{Upper moment bound} The upper bound for the moments is  based on the respective bound of the \kpe in~\eqref{eq:expk-bounds}; the latter imply that $m_{X;\ell}(\kp;\mu,\sigma)$ is bound from above by the expression
\begin{align}
\label{eq:ub-1}
\widetilde{UB}_{X;\ell}(\kp;\mu,\sigma)  = & \widetilde{I}_b + \frac{1}{2}\left[ 1 - \erf\left( \frac{\mu}{\sqrt{2}\sigma}\right) \right] \,,
\end{align}
where  the term containing  $\erf(\cdot)$ is obtained by integrating the normal density over $\R_{0,-}$ and $\widetilde{I}_b$ is the integral
\[
\widetilde{I}_b = \frac{1}{\sqrt{2\pi}\sigma}\, \,2^{\ell/\kp}\,
\int_{0}^{\infty} d y\, \E^{-\frac{1}{2\sigma^2}(y-\mu)^2}\,  \left(1+ \kp^{2}y^{2} \right)^{\ell/2\kp} \,.
\]

Taking into account that $(1+y^2)^{\ell/2\kp} \le (1+y^2)^{n_\ast}$, where $n_\ast = \lceil \ell/2\kp \rceil$, and the fact that the integrand in $\widetilde{I}_b$ is non-negative for all $y \in \R$, it holds that
\begin{align}
\label{eq:I-b}
I_b \triangleq \,\frac{2^{\ell/\kp}\,}{\sqrt{2\pi}\sigma}\, \,
\int_{0}^{\infty} d y\, \E^{-\frac{1}{2\sigma^2}(y-\mu)^2}\,  \left(1+ \kp^{2}y^{2} \right)^{n_\ast}  \ge \widetilde{I}_b \,,
\end{align}
Therefore, we obtain the upper bound ${UB}_{X;\ell}(\kp;\mu,\sigma)$ such that
\begin{align}
\label{eq:ub-2}
  {UB}_{X;\ell}(\kp;\mu,\sigma) & \triangleq {I}_b + \frac{1}{2}\left[ 1 - \erf\left( \frac{\mu}{\sqrt{2}\sigma}\right) \right] \ge \widetilde{UB}_{X;\ell}(\kp;\mu,\sigma)  \,.
\end{align}

To evaluate $I_b$ we use the binomial expansion, and we calculate the exponential integral in terms of the confluent hypergeometric function (as is done above for the lower limit) leading to
\begin{align*}
I_b = &
2^{\ell/\kp} \, \sum_{m=0}^{n_\ast}\,\,  \frac{n_\ast !}{m! \, (n_\ast - m)!}\,\kp^{2m}
\int_{0}^{\infty} \frac{d y}{\sqrt{2\pi}\sigma}\,\, \E^{-\frac{1}{2\sigma^2}(y-\mu)^2}\,  y^{2m}
\\
= & \frac{2^{\ell/\kp}\, \E^{-\mu^2/2\sigma^2}}{\sqrt{\pi}}\, \, \sum_{m=0}^{n_\ast}\,\,  \frac{n_\ast ! \, 2^{m-1}}{m! \, (n_\ast - m)!}\, \kp^{2m}\sigma^{2m}\, \left[\Gamma\left(m+\frac{1}{2}\right)\, \hyperg\left( m+\frac{1}{2}\,, \frac{1}{2}\,;\, \frac{\mu^2}{2\sigma^2}\right) \right.
\\
& \left. \quad\quad\quad\quad\quad\quad\quad\quad + \Gamma(m+1)\, \frac{\mu\sqrt{2}}{\sigma}\,\hyperg\left( m+1\,, \frac{3}{2}\,;\, \frac{\mu^2}{2\sigma^2}\right)  \right]\,.
\end{align*}

\end{IEEEproof}

\section{Proof of Power-Series Moment Expansions}
\label{app:power-series}
This section presents the proof of Theorem~\ref{theorem:power-series-moments}.
\begin{IEEEproof}  We use the moment integral relation~\eqref{eq:moments-integral} in combination with the Taylor series expansion of the $\kp$-exponential around $\mu$.  Let us denote $Q(y;\kp,\ell) \triangleq \exp_{\kp/\ell}(\ell y)$, so that $Q(y;\kp,\ell)=\left( \sqrt{1+ (\kp y)^{2}}+ \kp\,y\right)^{\ell/\kp}$.
Then the moment integral~\eqref{eq:moments-integral} becomes
\beq
\label{eq:moments-integral-2}
m_{X;\ell}(\kp;\mu,\sigma) =\int_{-\infty}^\infty Q(y;\kp,\ell)\,f_Y(y;\mu,\sigma)~dy\,.
\eeq
The Taylor expansion of $Q(y;\kp,\ell)$ around $\mu$  is given by
\beq
\label{eq:expk-taylor-mu}
Q(y;\kp,\ell) = Q(\mu;\kp,\ell) +
\sum_{n=1}^{\infty} Q^{(n)}(\mu;\kp,\ell)\,\frac{(y-\mu)^{n}}{n\,!}\,,
\eeq
where $Q^{(n)}(\mu;\kp,\ell) = \left. {\dd^{n}Q(\mu;\kp,\ell)}/{\dd y^{n}}
\right|_{y=\mu}\,$.

Inserting the Taylor expansion in~\eqref{eq:moments-integral-2} and interchanging the order between the summation and the integral we obtain the following power series
\beq
\label{eq:moments-integral-3}
m_{X;\ell}(\kp;\mu,\sigma) =Q(\mu;\kp,\ell) +
\sum_{n=1}^{\infty} \frac{1}{n!}\,Q^{(n)}(\mu;\kp,\ell) \,\EE[(Y-\mu)^{n}]\,.
\eeq
The centered moments $\EE[(Y-\mu)^{n}]$ are calculated using the Wick-Isserlis theorem which states that while the odd powers of $\EE\left[ (Y-\mu)^{n}\right]$ are equal to zero, the even powers equal $\delta_{n,2q} \, \frac{\sigma^{2q} (2q)!}{q!\, 2^{q}}$, where $q \in \mathbb{N}$~\cite{dth20}.

Next we turn to  the derivatives $Q^{(n)}(\mu;\kp,\ell)$.
For $n\ge 1$, the derivatives can be computed using the expression
\beq
\label{eq:Qn-deriv}
Q^{(n)}(\mu;\kp,\ell) = \frac{\ell\, g_{n}(\mu;\kp,\ell)}{\left({1+ \kp^{2} \mu^{2}}\right)^{n-1/2}}\, Q(\mu;\kp,\ell)\,, \; \mbox{for} \; n \ge 1\,,
\eeq
with the $g_{n}(\mu;\kp,\ell)$ defined by the recursive formula~\eqref{eq:gn}.
It is straightforward to see by direct differentiation of $Q(\mu;\kp,\ell)$ that
$Q^{(1)}(\mu;\kp,\ell)=\ell\,Q(\mu;\kp,\ell)/\sqrt{1+\kp^{2}\mu^{2}} $ is given by~\eqref{eq:Qn-deriv} with $g_{1}(\mu;\kp,\ell)=1$.

The general validity of~\eqref{eq:Qn-deriv} for $n>1$ is shown by induction. Assuming that~\eqref{eq:Qn-deriv} holds for $n$, using the product (Leibniz) rule for differentiation  we obtain
\begin{align}
\label{eq:Qnplusone-deriv}
Q^{(n+1)}(\mu;\kp,\ell) = & \frac{\ell^{2}\, g_{n}(\mu;\kp,\ell)}{\left({1+ \kp^{2} \mu^{2}}\right)^{n}}\, Q(\mu;\kp,\ell) - \frac{\ell\,\kp^{2}\,\mu\,(2n-1) g_{n}(\mu;\kp,\ell)}{\left({1+ \kp^{2} \mu^{2}}\right)^{n+1/2}}\, Q(\mu;\kp,\ell)
\\
&  + \frac{\dd g_{n}(\mu;\kp,\ell)}{\dd \mu}\, \frac{\ell\,  Q(\mu;\kp,\ell)}{\left({1+ \kp^{2} \mu^{2}}\right)^{n-1/2}}\,.
\end{align}
Substituting in~\eqref{eq:Qnplusone-deriv}, the expression derived from~\eqref{eq:gn}:
\[
\frac{\dd g_{n}(\mu;\kp,\ell)}{\dd \mu}\,=\frac{g_{n+1}(\mu;\kp,\ell)+\left((2n-1)\kp^2\mu-\ell \sqrt{1+\kp^2\mu^2}\right)g_{n}(\mu;\kp,\ell)}{1+\kp^2\mu^2},
\]
we take
\[
Q^{(n+1)}(\mu;\kp,\ell) =\frac{\ell\, g_{n+1}(\mu;\kp,\ell)}{\left({1+ \kp^{2} \mu^{2}}\right)^{n+1/2}}\, Q(\mu;\kp,\ell)\,.
\]
This concludes the proof.
\end{IEEEproof}

\section{Hessian of the Negative Log-Likelihood for the $\kp$-lognormal Distribution}
\label{app:hessian}

This section presents the Hessian matrix of the negative log-likelihood for the $\kp$-lognormal distribution. The Hessian is defined by means of the following partial derivatives with respect to $\bmphi=(\mu, \sigma, \kp)$:
\beq
\mathbf{H} \triangleq  \begin{pmatrix}
    \frac{\pa^{2}\nll(\bmphi)}{\pa\mu^2} & \frac{\pa^{2}\nll(\bmphi)}{\pa\mu\pa\sigma} & \frac{\pa^{2}\nll(\bmphi)}{\pa\mu\pa\kp}
    \\[1ex]
    \frac{\pa^{2}\nll(\bmphi)}{\pa\sigma\pa\mu} & \frac{\pa^{2}\nll(\bmphi)}{\pa\sigma^2} &  \frac{\pa^{2}\nll(\bmphi)}{\pa\sigma\pa\kp} \\[1ex]
     \frac{\pa^{2}\nll(\bmphi)}{\pa\kp\pa\mu} & \frac{\pa^{2}\nll(\bmphi)}{\pa\kp\pa\sigma} & \frac{\pa^{2}\nll(\bmphi)}{\pa\kp^2}
  \end{pmatrix}\,.
\eeq
Using~\eqref{eq:nll-gradient} for the NLL gradient $\nabla_{\bmphi}\nll(\bmphi)$, we calculate $\mathbf{H}=\nabla_{\bmphi}\nabla_{\bmphi}\nll(\bmphi)$. After some algebra  the Hessian $\mathbf{H}$ is given  by the following symmetric matrix
\beq
\mathbf{H}= \begin{pmatrix}
    \frac{1}{N\,\sigma^2} & \frac{2\sum_{i=1}^{N} \left( \lnk (x_{i}) - \mu\right)}{N\,\sigma^3} & \frac{-\sum_{i=1}^{N}\ln(x_i)\, s_{i;k} +\frac{\lnk(x_{i})}{\kp}}{N\sigma^2} \\[1ex]
    \times &  \frac{3\sum_{i=1}^{N} \left( \lnk (x_{i})-\mu \right)^2 - N\sigma^2}{N\,\sigma^4}  &  \frac{-2 \sum_{i=1}^{N}\left( \lnk (x_{i})-\mu \right)\left( \ln (x_{i})\,s_{i,k} - \frac{\lnk (x_{i})}{\kappa}\right)}{N\,\sigma^3} \\[1ex]
     \times  & \times & \frac{\pa^{2}\nll(\bmphi)}{\pa\kp^2}
  \end{pmatrix}\,,
\eeq
where $s_{i;k} \triangleq \frac{\left( x_{i}^\kappa + x_{i}^{-\kappa}\right) }{2\kappa}$. The $H_{i,j}$ entries below the diagonal ($i>j$) are marked by $\times$. Due to the symmetry of the Hessian, it holds that $H_{i,j}=H_{j,i}$. Finally, the third diagonal entry of $\mathbf{H}$ is given by
\begin{align}
\label{eq:nll-kk}
\frac{\pa^{2}\nll(\bmphi)}{\pa\kp^2} & = \frac{1}{N\sigma^2}\sum_{i=1}^{N}\left(s_{i;k}\ln (x_i)-\frac{\lnk (x_i)}{\kp} \right)^2
\nonumber \\
& + \frac{1}{N\sigma^2}\sum_{i=1}^{N}\left(\lnk( x_i) - \mu \right)
\left( \ln^{2}(x_{i})\,\lnk (x_{i}) -\frac{2\ln (x_{i})}{\kp}s_{i;k} +\frac{2\lnk (x_{i})}{\kp^2} \right)
\nonumber \\
& - \frac{1}{N}\sum_{i=1}^{N}\left[ -\ln^{2}(x_{i})  + \ln^{2}(x_{i}) \frac{(\lnk (x_{i}))^2}{s_{i;k}^2} \right]\,.
\end{align}

\section*{Acknowledgment}
We would like to thank Prof. Jaime G\'{o}mez-Hern\'{a}ndez, School of Civil Engineering, Technical University of Valencia (Spain), for providing the Berea permeability data used in this study, and Prof.  Athanasios Liavas (School of Electrical and Computer Engineering, Technical University of Crete) for his helpful input.

\Urlmuskip=0mu plus 1mu\relax


\providecommand{\noopsort}[1]{}\providecommand{\singleletter}[1]{#1}%

\onecolumn
\renewcommand\thesection{S.\arabic{section}}
\setcounter{subsection}{0}
		\renewcommand\thesubsection{\thesection.\arabic{subsection}}

\begin{center}\Large\textbf{Supplementary Information}\end{center}

\section*{S1. Properties of $\kp$-deformed exponential and logarithm}
\label{sec:additional-properties}
This section focuses on mathematical properties of the \kpe and \kpl.

\subsection*{Taylor series expansion of \kpe}
\label{sec:expk-taylor}
The power-series expansion of $\expk(y)$ around $y=0$ is given by means of the following expression~\cite{Kaniadakis13}:
\beq
\label{eq:taylor-kpe}
\expk(y) =  \sum_{n=0}^{\infty} \xi_{n}(\kp) \, \frac{y^{n}}{n!}\,, \; \mbox{for} \; (\kp y)^2 < 1,
\eeq
where the $\kp$-correction functions $ \{ \xi_{n}(\kp)\}_{n=0}^{\infty}$ are polynomials of $\kp$ defined by the following recurrence relations
\begin{subequations}
\label{eq:kpe-polynomials}
\begin{align}
& \xi_{0}(\kp) = \xi_{1}(\kp)=1,
\\
& \xi_{n}(\kp) = \prod_{j=1}^{n-1} \left[ \, 1 - (2j -n) \kp  \, \right], \; n>1\,.
\end{align}
\end{subequations}

\noindent The polynomials $\xi_{n}(\kp)$ for the first seven orders $n=0, \ldots, 6$ are given by
$ \xi_{0}(\kp) = \xi_{1}(\kp)= \xi_{2}(\kp) = 1$, $\xi_{3}(\kp) = 1-\kp^2$, $\xi_{4}(\kp) = 1- 4\kp^2$, $\xi_{5}(\kp) = ( 1-\kp^2) (1- 9\kp^2)$, $\xi_{6}(\kp) = ( 1- 4\kp^2) (1- 16\kp^2)$.
The first three terms of the Taylor expansion~\eqref{eq:taylor-kpe} coincide with those of the ordinary exponential, i.e.,
\[
\expk(y) = 1 + y + \frac{y^2}{2} + (1 - \kp^2) \,\frac{y^3}{3!} + \Or(\kp^{2} y^{4} ).
\]
The Taylor  expansion implies that for $y \to 0$ and  $\kp \to 0$  the
\kpe function  $\expk(y)$ converges to the ordinary exponential.  More precisely,
\beq
\label{eq:expk-taylor}
\expk(y) = \exp(y) + \sum_{n=3}^{\infty} \left[ \,\xi_{n}(\kp) -1 \right)]\, \frac{y^{n}}{n!} = \exp(y) + \Or({\kp^2\, y^3}).
\eeq
The above  is useful in perturbation analysis to express analytically the deviation between the \kpe and $\exp(x)$. 
In addition,  based on the Taylor expansion~\eqref{eq:taylor-kpe},  the sign of the leading corrections of $\expk(y)$ with respect to $\exp(y)$ is negative.

\medskip

\begin{propo}[Recurrence relations for \kpe corrections]
\label{propo:expk-corrections-recurrence}
Let  ${\xi'}_{n}(\kp) \triangleq 1 - \xi_{n}(\kp)$ for $n \in \mathbb{N}$ and $\kp>0$ represent  the difference between the power-series coefficients of the \kpe and those of the natural exponential.  These corrections satisfy the recurrence relations
\begin{subequations}
\label{eq:recurrence-xi-prime}
\begin{align}
\label{eq:recurrence-xi-prime-a}
\xi'_{0}(\kp)=& \xi'_{1}(\kp)=\xi'_{2}(\kp)=0 \,,
\\[1ex]
\label{eq:recurrence-xi-prime-b}
\xi'_{n+2}(\kp)=& \xi'_{n}(\kp) + n^{2}\kp^{2}\left[ 1 -\xi'_{n}(\kp) \right]\,, \; \mbox{for} \; n \ge 0\,. 
\end{align}
Furthermore, the leading-order in $\kp^{2}$ term of $\xi'_{n+2}(\kp)$ is given by
\begin{align}
\label{eq:xik-xorrections-leading}
L(\xi'_{n+2}(\kp)) =   \left\{
\begin{array}{cc}
 \kp^{2}\left[\,2^2 + 4^2 + \ldots (2k)^2 \right]\,,    &  \mbox{if} \; n=2k\,,
\\[1ex]
   \kp^{2}\left[\,1^2 + 3^2 + \ldots (2k+1)^2 \right]\,,   & \mbox{if} \; n=2k+1\,.
\end{array} \right.
\end{align}  
\end{subequations}
\end{propo}

\begin{IEEEproof}
Based on~\eqref{eq:kpe-polynomials} it is straightforward  to obtain the recursive relation 
\beq
\label{eq:xik-ratio}
\frac{\xi_{n+2}(\kp)}{\xi_{n}(\kp)} = 1 - n^{2}\kp^2 \,, \; \mbox{for} \; n \ge 1\,.
\eeq
The values of $\xi'_{n}(\kp)$ for $n=0, 1, 2$ in~\eqref{eq:recurrence-xi-prime-a} follow  from the respective $\xi_{n}(\kp)$. The recursive relation~\eqref{eq:recurrence-xi-prime-b} is obtained directly from~\eqref{eq:xik-ratio}.   The above  implies that $\xi'_{n}(\kp)=c_{n} \kp^2 + \Or(\kp^4)$ for all $n\ge 3$.  By defining  $L(\xi'_{n+2}(\kp)) \triangleq c_{n} \kp^2$, the recursive relation $L(\xi'_{n+2}(\kp)) = L(\xi'_{n}(\kp)) + n^{2}\kp^{2}$ follows  from the above. Expanding the recursive terms, we obtain the series~\eqref{eq:xik-xorrections-leading} for the leading-order corrections.  

\emph{Note:}
Similarly, second-order terms which contain contributions $\propto \kp^{4}$ satisfy the recursive relation  
\[
L_{2}(\xi'_{n+2}(\kp)) = L_{2}(\xi'_{n}(\kp)) \, \left( 1 + n^{2}\kp^{2} \right)
\]
with $L_{2}(\xi'_{n}(\kp))=0$ for $n=0, 1, \ldots, 4$,  $L_{2}(\xi'_{5}(\kp))=- 1^{2} \cdot 3^{2}$, $L_{2}(\xi'_{4}(\kp))=- 2^{2} \cdot 4^{2}$, and so on. 
\end{IEEEproof}

\subsection*{Properties of \kpl}

\paragraph{Integral relation} The \kpl satisfies the following equation~\cite{Kaniadakis13}
\beq
\label{eq:kpl-integral}
\lnk (x) = \frac{1}{2}\int_{1/x}^{x}\, \frac{\dd t}{t^{\kp+1}}\,, 
\eeq
which generalizes the symmetric integral representation of $\ln x$. In the standard integral representation of $\ln x$ the integration limits are $1$ and $x$ and the constant coefficient is 1 (instead of $1/2$).

\medskip \paragraph{Representation involving logarithm and hyperbolic sine} 
The \kpl can be expressed using the natural logarithm and the hyperbolic sine  as follows~\cite{Kaniadakis13}:
\beq
\label{eq:kpl-sinh}
\lnk x= \frac{1}{\kp}\, \sinh(\kp\,\ln x)\,.
\eeq

\section*{S2. Comparing the tail of the normal and lognormal distributions}
\label{sec:compare-tails-normal-lognormal}

In order to understand the upper tail difference between the lognormal and the normal distributions, we calculate extreme values assuming that both distributions have the same mean, $\mx$, and variance, $\sigma^{2}_X$. Thus, we select $\mu=0$, $\sigma=1.5$, we evaluate the lognormal mean, $\mx$ and variance, $\sigma^{2}_X$, based on~\eqref{eq:ln-mean-mode-median} and~\eqref{eq:logn-var-skew} respectively, and we use them in the normal distribution as well. Next, using the quantile function~\eqref{eq:qf-ln} for the lognormal and $Q_{N}(p)=\mx+\sigma_{X}\sqrt{2} \,\erf^{-1}(2p-1)$ for the normal distributions and  the lognormal median $\mu_{0.5} = \exp(\mu)$ based on~\eqref{eq:ln-mean-mode-median}, the ratios $x_{\max}/\mu_{0.5}$ are respectively given by 
\begin{subequations}
\label{eq:nhev}
\begin{align}
\frac{x_{\max}}{\mu_{0.5}} & \triangleq \frac{Q_{\mathrm{LN}}\left(p^\ast(N) \right)}{Q_{\mathrm{LN}}\left(0.5\right)} = \exp\left[  \sqrt{2\sigma^2}\, \erf^{-1}(1-2/N)\right]\,, \; \textrm{lognormal} 
\\
\frac{x_{\max}}{\mu_{0.5}} & \triangleq \frac{Q_{\mathrm{N}}\left(p^\ast(N) \right)}{Q_{\mathrm{N}}\left(0.5\right)}  = 
\frac{\mx + \sigma_{X}\sqrt{2} \,\erf^{-1}(1-2/N) }{\mx}\,, \; \textrm{normal}\,,
\end{align}
\end{subequations}
 where $Q_{\mathrm{LN}}$ and $Q_{\mathrm{N}}$ are  the quantile functions of the lognormal and the normal distributions respectively. 

\begin{figure}[!ht]
\centering
\includegraphics[width=0.5\linewidth]{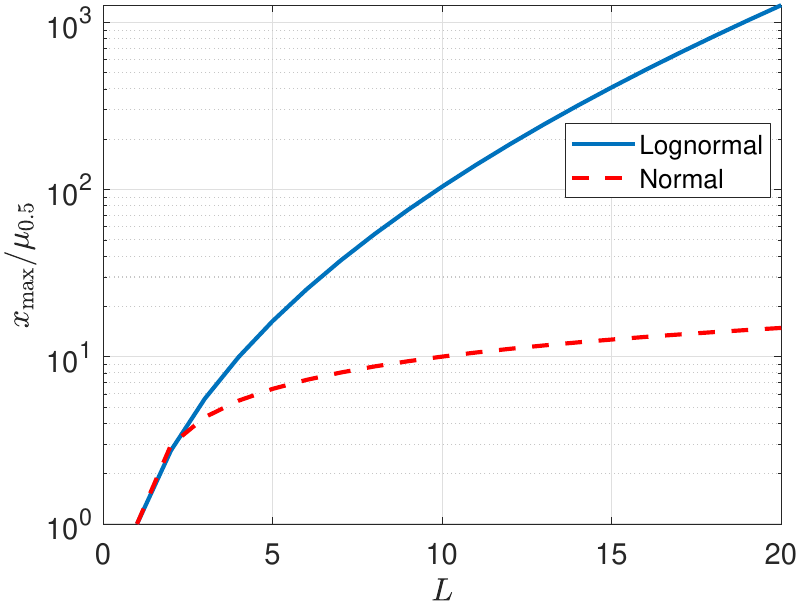}
\caption{Ratios $x_{\max}/\mu_{0.5}$ representing normalized typical extreme values for   $N=2^L$ independent random variables versus $L$ for the lognormal (continuous lines) and the normal (broken line) distributions--- $x_{\max}=Q(1-\frac{1}{N})$  where $N=2^L$  and $\mu_{0.5}=Q(0.5)$.  Different $L$ values correspond to  probability levels between $p^\ast(2)=0.5$  and $p^\ast(20) \approx 0.999999046$.  The lognormal distribution refers to the exponential of a normal variable with $\mu=0$ and $\sigma=1.5$ (continuous line);  the normal distribution has the same mean and variance as the lognormal, that is, $\mu_{X}=\E^{2.25/2} \approx 3.08$ and ${\sigma}^{2}_{X}=(\E^{2.25}-1)\E^{2.25} \approx 80.53$ (broken line) based on~\eqref{eq:ln-mean-mode-median} and~\eqref{eq:logn-var-skew}. }
\label{fig:ratio_val_lognormal}
\end{figure} 
We compare the ratios $x_{\max}/\mu_{0.5}$ versus $L$ where $N=2^{L}$ is the system size for a random variable $X$ which follows (a) the normal and (b) the lognormal  distribution with the same  mean and variance. 
For the lognormal we assume  $\mu=0$ and $\sigma=1.5$ leading to $\mu_{X} =\exp(\sigma^2 /2)\approx 3.08$ and $\varx =\exp(\sigma^2)\left[ \exp(\sigma^2) -1\right] \approx 80.53$.  Hence, $\mu_{0.5}=1$ for the lognormal (due to $\mu=0$) and $\mu_{0.5}=\exp(\sigma^2 /2)$ for the normal. 
The curves of the normalized extreme values~\eqref{eq:nhev} shown in Fig.~\ref{fig:ratio_val_lognormal}  reveal a  steep increase of $x_{\max}/\mu_{0.5}$ with $L$ for the lognormal. For $L=20$, corresponding to a system  size $N \approx 10^6$ (e.g., for a square grid with $10^3$ nodes per side),  $x_{\max}/\mu_{0.5} \approx 1.27\,\times 10^{3}$, while the respective value for the normal distribution $x_{\max}/\mu_{0.5} \approx 14.88$, is about two orders of magnitude smaller.   However, very large values of the ratio $x_{\max}/\mu_{0.5}$  are not suitable for all physical processes with skewed probability distributions.

\section*{S3. Application to Jura Mountain Heavy Metal Dataset}
\label{sec:Jura}
This section investigates the application of the $\kp$-lognormal distribution to the  Jura dataset. The  latter contains heavy metal concentrations in the topsoil (units are mg/kg) measured at 259 locations in the Swiss Jura mountains~\cite{Atteia94,Goovaerts97}. The seven metals include Cadmium (Cd), Copper (Cu), Lead (Pb), Cobalt (Co), Chromium (Cr), Nickel (Ni), and Zinc (Zn). 
The statistics of the data are shown in Table~\ref{tab:Jura_stats}. 

\begin{table}[!ht]
\centering
\caption{Summary statistics for the seven metal concentrations of the Jura dataset.}
\begin{tabular}{lrrrrrrr}
Element & {Mean} & {Median} & {Min} & {Max} & {Standard deviation} & {Skewness} & {Kurtosis}\\ \hline\hline
{Cd} & 1.31 & 1.07 & 0.14 & 5.13 & 0.92 & 1.51 & 5.58\\ \hline
{Cu} & 23.73 & 17.60 & 3.96 & 166.40 & 20.71 & 2.88 & 15.12\\ \hline
{Pb} & 53.92 & 46.40 & 18.96 & 229.56 & 29.79 & 2.91 & 14.60\\ \hline
{Co} & 9.30 & 9.76 & 1.55 & 17.72 & 3.58 & $-0.18$ & 2.20\\ \hline
{Cr} & 35.07 & 34.84 & 8.72 & 67.60 & 10.96 & 0.29 & 3.02\\ \hline
{Ni} & 19.73 & 20.56 & 4.20 & 53.20 & 8.23 & 0.16 & 3.24\\ \hline
{Zn} & 75.08 & 73.56 & 25.20 & 219.32 & 29.02 & 1.03 & 5.16\\ \hline
\end{tabular}
\label{tab:Jura_stats}
\end{table}

Most concentrations exhibit non-zero skewness and non-Gaussian ($\neq 3$) kurtosis coefficients (cf. Columns 7--8 in Table~\ref{tab:Jura_stats}). The logarithmic transformation is typically used for such data to reduce skewness  before applying kriging or Gaussian process regression~\cite{Goovaerts97,Wilson11}. This procedure is known as Gaussian anamorphosis in geostatistics~\cite{Chiles12,dth20} and data warping in Gaussian process regression~\cite{Snelson03,dth22_warp}.

In Table~\ref{tab:Jura} we list the $\kp$-lognormal parameter estimates $\mu, \sigma, \kp$ (derived by MLE) for the seven different concentrations. These are complemented by the NLL per site as well as the AIC and BIC values. The table also includes estimates of the lognormal parameters $\mu_0$ and $\sigma_0$ as well as the respective NLL, AIC, and BIC values for comparison. Both AIC and  BIC  favor the $\kp$-lognormal for Co, Cr, and Ni while the lognormal is preferred for Cd, Cu, Pb, Zn.  For Cu and Pb the estimated  $\kp$  is close to zero, indicating the proximity to the lognormal. For Cd the estimated $\kp$ is 0.43; nonetheless BIC favors the lognormal distribution, indicating that the deformed logarithm does not impart a strong advantage over the lognormal. For Co the estimated $\kp$ is slightly higher than one.  The estimated distributions are all unimodal---confirmed by numerical evaluation of the roots of $p_{1}(z)$. Note that the $\kp$-lognormal BIC values for Co, Cr, and Ni  are clearly lower than the respective lognormal values. On the other hand, the difference between the BIC values of lognormal and $\kp$-lognormal is considerably smaller for Cd, Cu, Pb and Zn (cases in which the lognormal is preferred).

\begin{table}[!ht]
\centering
\caption{Statistics of the  Jura heavy metal dataset (259 sampling points). Columns 2-4 contain the optimal $\kp$-lognormal parameters. Columns 5-7 display respectively the negative log-likelihood (NLL) per sampling location, as well as the AIC and BIC selection criteria. Columns 8-9 contain the optimal lognormal parameters, while Columns 10-12 the NLL per site, the AIC and BIC respectively. Bold font indicates the model with the lowest BIC (AIC follows the same pattern).}
\begin{tabular}{l|cccccc||ccccc}
\hline\hline
 & \multicolumn{6}{c||}{$\kp$-lognormal} & \multicolumn{5}{c}{Lognormal}\\
 \hline
Element & {$\mu$} & {$\sigma$} & {$\kappa$} & {NLL} & {AIC} & {BIC} & {$\mu_0$} & {$\sigma_0$} & {NLL$_0$} & {AIC$_0$} & {BIC$_0$}\\\hline
{Cd} & 0.03 & 0.74 & 0.43 & 1.11 & 579.37 & 590.04 & 0.04 & 2.90 & 1.11 & 578.38 & \textbf{585.49}\\ \hline
{Cu} & 2.90 & 0.70 & $6.5\cdot10^{-4}$ & 3.97 & 2061.38 & 2072.05 & 3.89 & 2.14 & 3.97 & 2059.38 & \textbf{2066.50}\\ \hline
{Pb} & 3.89 & 0.42 & $4.2\cdot10^{-4}$ & 4.44 & 2308.36 & 2319.03 & 3.50 & 2.87 & 4.44 & 2306.37 & \textbf{2313.48}\\ \hline
{Co} & 4.85 & 1.98 & 1.04 & 2.69 & 1397.81 & \textbf{1408.48} & 4.25 & 0.71 & 2.80 & 1454.77 & 1461.88\\ \hline
{Cr} & 8.39 & 1.88 & 0.70 & 3.80 & 1975.91 & \textbf{1986.58} & 0.70 & 0.42 & 3.84 & 1995.60 & 2002.72\\ \hline
{Ni} & 6.90 & 2.48 & 0.82 & 3.52 & 1828.56 & \textbf{1839.23} & 0.47 & 0.34 & 3.62 & 1877.68 & 1884.79\\ \hline
{Zn} & 4.97 & 0.58 & 0.23 & 4.71 & 2445.59 & 2456.27 & 0.51 & 0.39 & 4.71 & 2445.25 & \textbf{2452.37}\\ \hline
\end{tabular}
\label{tab:Jura}
\end{table}

\begin{figure}
\centering
\includegraphics[width=0.9\linewidth]{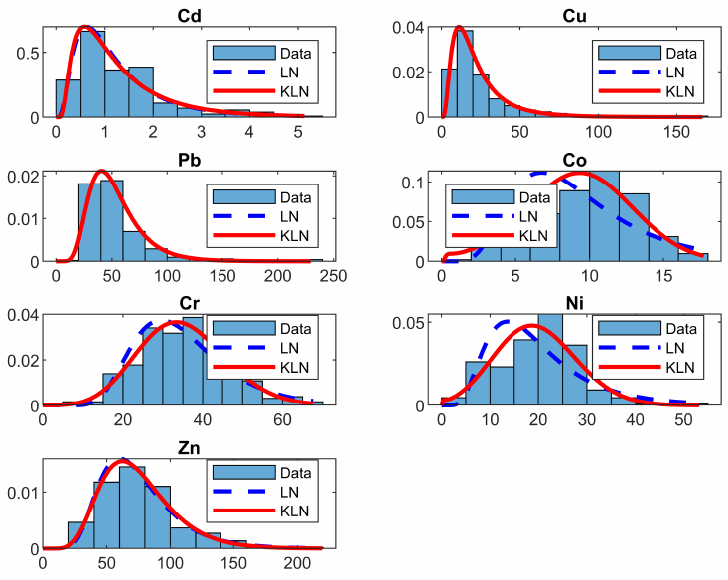}
\caption{Histograms of Jura heavy metal datasets (bins) with the optimal $\kp$-lognormal (continuous red line) and lognormal (blue broken line) PDFs. The PDF curves are obtained using the MLE optimal parameters for the two distributions listed in Table~\ref{tab:Jura}.}
\label{fig:Jura_histo}
\end{figure}

The histograms of the datasets are shown along with the fits to the optimal  $\kp$-lognormal and lognormal distributions (continuous lines) in Fig.~\ref{fig:Jura_histo}. The tails of the $\kp$-lognormal fit are evidently shorter than those of the lognormal for Co, Cr,  and Ni concentrations. At the same time, the  overall fit of the $\kp$-lognormal to the histograms is better. On the other hand, for Cd, Cu, Pb and Zn there is no visible difference between the lognormal and $\kp$-lognormal distributions. For Cu and Pb this is expected, since $\hkp \approx 0$.  For Cr the estimated $\kp$  is $\hkp \approx 0.43$, but the NLL per site is practically the same for the lognormal and the $\kp$-lognormal, indicating that both distributions provide  an equally good fit to the data.  

\begin{figure}
\centering
\includegraphics[width=0.85\linewidth]{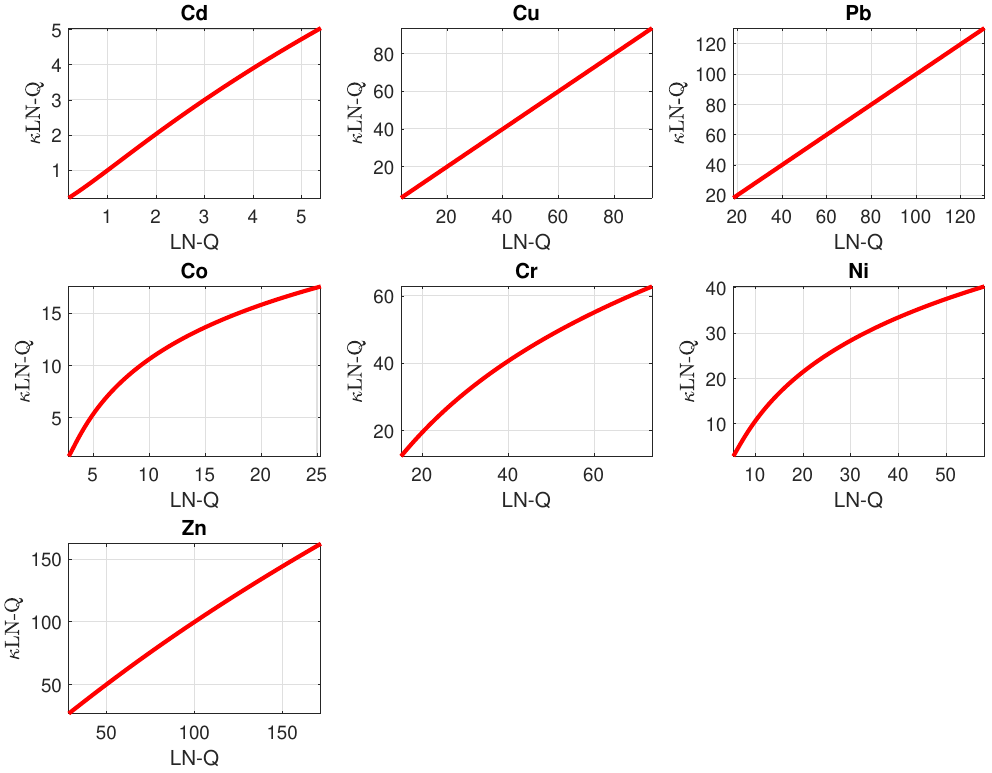}
\caption{Quantile-quantile plots of the lognormal distribution (horizontal axis) versus the $\kp$-lognormal (vertical axis) based on the MLE parameters (determined by global optimization) for the seven metal concentrations of the Jura dataset.   }
\label{fig:Jura_QQ}
\end{figure}

Certain comments are in order at this point: First, the equally good (visually) fit of both distributions to the Jura Cd data does not refute the fact that the $\kp$-lognormal with  $\kp =0.43$ has a shorter right tail than the lognormal. Thus, the $\kp$-lognormal will assign  lower probability to higher Cd concentrations than the lognormal.  
This pattern is also observed for Co and Ni and to a lesser extent for Zn and Cr as is evident in the respective Q-Q plots of Fig.~\ref{fig:Jura_QQ}: the horizontal axis, corresponding to the lognormal quantiles, has a considerably larger range than the vertical which corresponds to the $\kp$-lognormal. The Q-Q plots  in this case employ $N_Q=500$ points.  

Secondly,  while the BIC and AIC  selection criteria favor the lognormal if the latter is viewed as a two-parameter distribution, they cannot distinguish between the two distributions if the lognormal is viewed as a special case of the $\kp$-lognormal for $\kp=0$. Also note that for Co, Cr and Ni the long tail of the lognormal forces the mode of the PDF to lower values than the $\kp$-lognormal. The latter, however, provides a better fit to the histogram peaks for these data, as evidenced in Fig.~\ref{fig:Jura_histo}.   

\section*{S4. Application to Berea Permeability Dataset}
\label{sec:sup-Berea}

This section focuses on the application of warped-GPR to the Berea permeability dataset using a different covariance kernel for the latent Gaussian process.

We assume the latent Gaussian process has an anisotropic Mat\'{e}rn  covariance kernel which is given by 
\beq
\label{eq:Matern}
C_{Y}({h})=\frac{\sigma^2 \, 2^{1-\nu}}{\Gamma(\nu)} \Bigg(\frac{\sqrt{2\nu} \, h}{\xi}\Bigg)^{\nu} K_{\nu}\Bigg(\frac{\sqrt{2\nu} \, h}{\xi}\Bigg)\,,
\eeq
where $h \triangleq \sqrt{\bfr^\top {\mathbf{M}}^{-1}\,\bfr}$, $\nu$ is the smoothness index, $\Gamma(\cdot)$ is the Gamma function, and $K_{\nu}(\cdot)$ is the modified Bessel function of the second kind of order $\nu$.  Two training sets that comprise randomly selected 500 and 1100 points respectively are used. The initial value of the smoothness index is $\nu_0=1.5$. 

The ML estimated parameters are shown in Table~\ref{tab:theta-Berea-Matern}. For both training sets the ML estimated smoothness index is $\hat{\nu}=0.5$, which resonates with the ``rough''  spatial profile of the permeability. The larger training set leads to smaller nugget variance $\sigma^{2}_{\epsilon}$, correlation scale $\xi$ and anisotropy ratio $\rho$. 

The cross-validation measures for both training sets are given in Table~\ref{tab:cross-val-Berea-Matern}.  Only the statistics for the median-based predictor are listed since the mode-based predictor yields almost identical statistics (except for higher ME). 

\begin{table}[!ht]
\centering
\caption{ML parameter estimates for the Berea permeability dataset. The ML estimates are based on two randomly selected  training sets with $N_{\rm tr}=500, \, 1100$. The Mat\'{e}rn spatial covariance kernel~\eqref{eq:Matern} is used for the latent Gaussian process.}
\begin{tabular}{c|cccccccc}
\hline
MLE estimates & $\kp$ & $\mu$ & $\sigma^{2}_{\epsilon}$ & $\sigma^2$ & $\xi$ & $\nu$ & $\rho$ &  $\varphi$ \\
$N_{\rm tr}=500$ & 0.88 & 20.81  &  2.60 &   40.31  &  5.42 &    0.5   &    16.53  &   0.97\\\hline\hline
MLE estimates & $\kp$ & $\mu$ & $\sigma^{2}_{\epsilon}$ & $\sigma^2$ & $\xi$ & $\nu$ & $\rho$ &  $\varphi$ \\
$N_{\rm tr}=1100$ & 0.88 & 21.24  &  1.84 &   34.64  &  3.39 &      0.5   &    10.32  &   0.95\\
\hline
\end{tabular}
\label{tab:theta-Berea-Matern}
\end{table}

\begin{table}[!ht]
\centering
\caption{Cross-validation measures for the Berea permeability dataset obtained for two randomly selected  training sets with $N_{\rm tr}=500, \, 1100$. The median predictor is based on $\expk(\mu_{\ast})$ where $\mu_{\ast}$ is given by~\eqref{eq:predictive-mean}, while the mode predictor is based on~\eqref{eq:mode-predictor}.  The Matern spatial covariance kernel~\eqref{eq:Matern} is used for the latent Gaussian process.  }
\begin{tabular}{c|cccccc}
\hline
$N_{\rm tr}=500$ & ME & MAE & MARE & RMSE & RRMSE & $R$ \\
 \hline
Median predictor  &  0.55 & 6.65 &  0.12 &   8.67 &    0.16&  0.83 \\
 \hline
$N_{\rm tr}=1100$ & ME & MAE & MARE & RMSE & RRMSE & $R$ \\
 \hline
Median predictor  &   0.21  &  5.59  & 0.11 &  7.18 & 0.14  &  0.89 \\
\hline
\end{tabular}
\label{tab:cross-val-Berea-Matern}
\end{table}

\begin{figure}[!ht]
\centering
\includegraphics[width=0.49\linewidth]{Berea_Perm_spatial.pdf}
\includegraphics[width=0.49\linewidth]{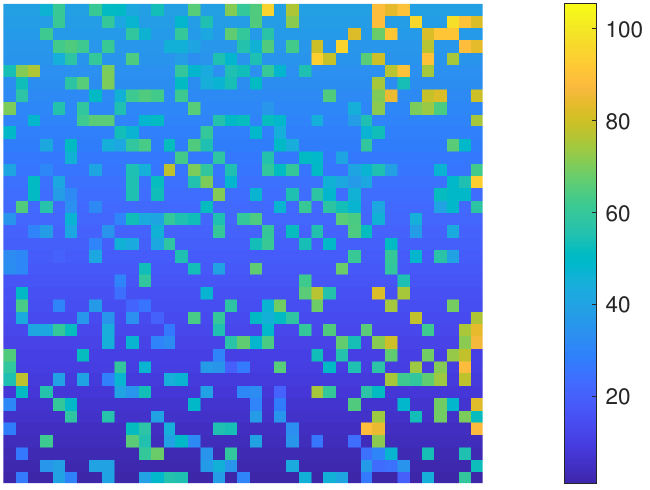}
\caption{\textbf{Left:} Berea permeability data.  \textbf{Right:}  Reconstructed permeability field using Mat\'{e}rn covariance kernel and a training set comprising 500 randomly selected values.}
\label{fig:Berea-Matern-500-spatial}
\end{figure}

\begin{figure}[!ht]
\centering
\includegraphics[width=0.49\linewidth]{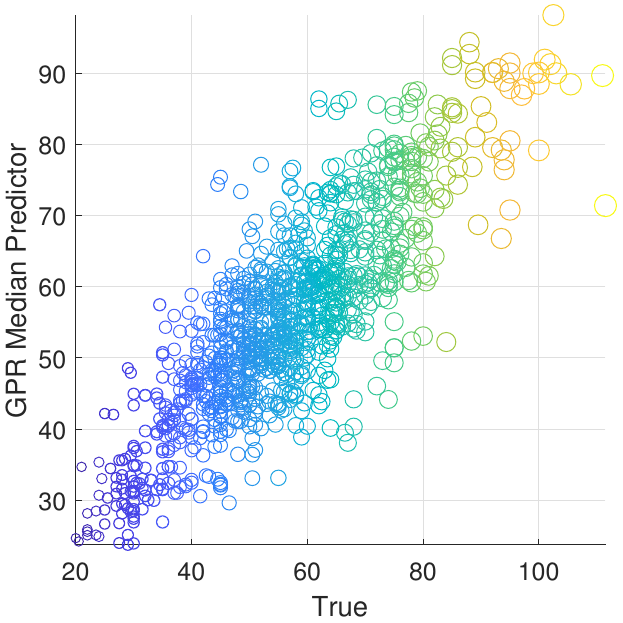}
\includegraphics[width=0.49\linewidth]{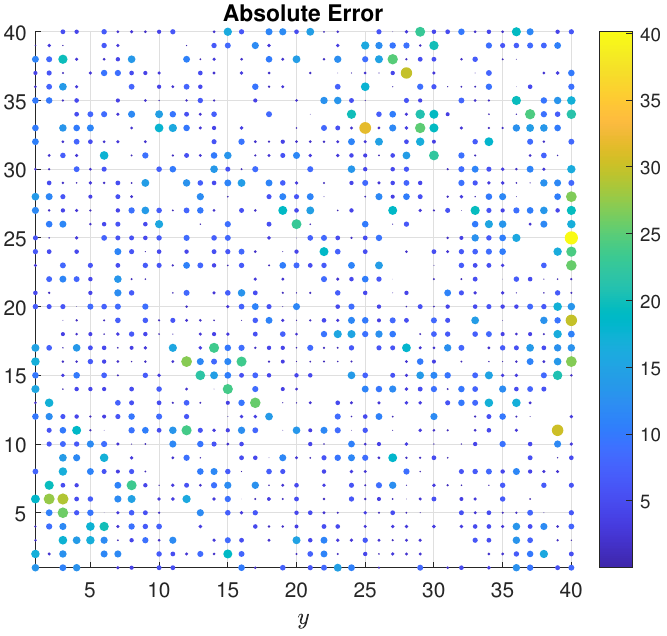}
\caption{\textbf{Left:} Scatter plot of test set permeability values versus the median-based warped-GPR predictions using the Mat\'{e}rn covariance kernel.  \textbf{Right:}  Spatial distribution of absolute values of prediction errors. Empty sites correspond to training set points. The training set comprises 500 randomly selected values.}
\label{fig:Berea-Matern-500-crosval}
\end{figure}

\begin{figure}[!ht]
\centering
\includegraphics[width=0.49\linewidth]{Berea_Perm_spatial.pdf}
\includegraphics[width=0.49\linewidth]{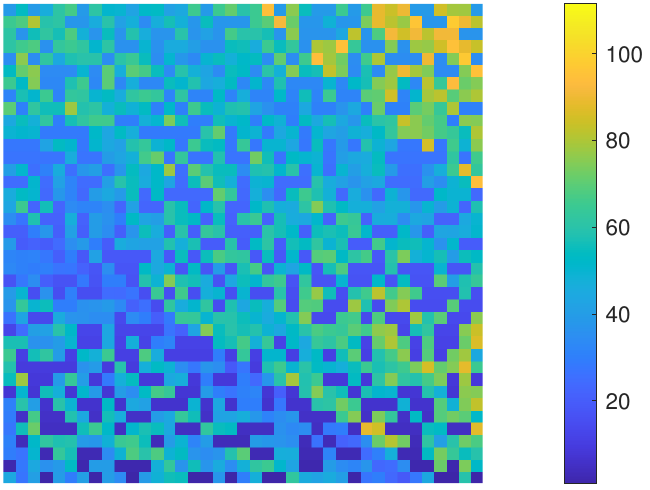}
\caption{\textbf{Left:} Berea permeability data.  \textbf{Right:}  Reconstructed permeability field using Mat\'{e}rn covariance kernel and a training set comprising 1100 randomly selected values.}
\label{fig:Berea-Matern-1100-spatial}
\end{figure}

\begin{figure}[!ht]
\centering
\includegraphics[width=0.49\linewidth]{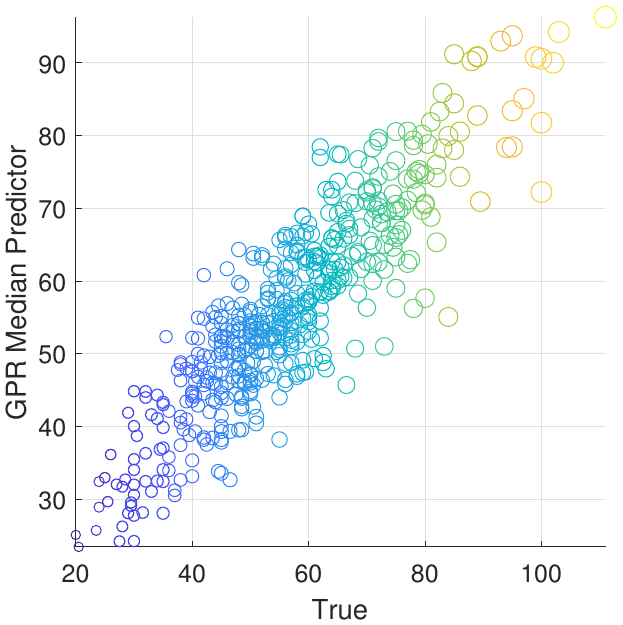}
\includegraphics[width=0.49\linewidth]{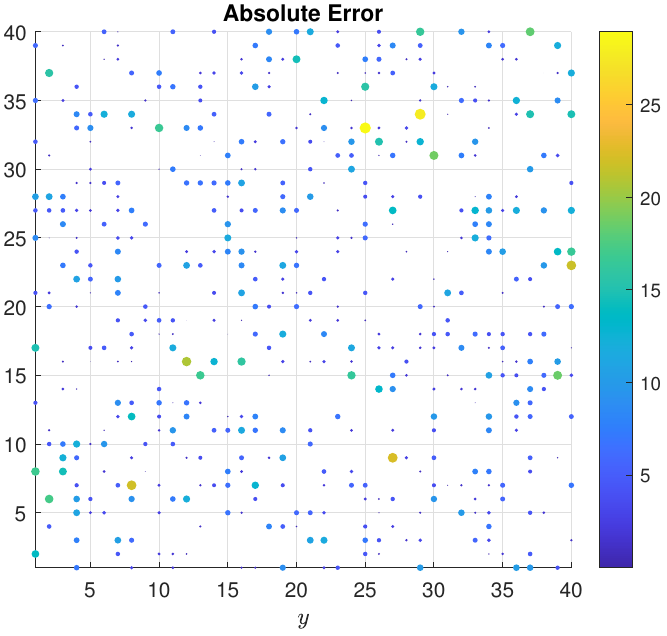}
\caption{\textbf{Left:} Scatter plot of test set permeability values versus the median-based warped-GPR predictions using the Mat\'{e}rn covariance kernel.  \textbf{Right:}  Spatial distribution of absolute values of prediction errors. Empty sites correspond to training set points. The training set comprises 1100 randomly selected values.}
\label{fig:Berea-Matern-1100-crosval}
\end{figure}

\clearpage

\providecommand{\noopsort}[1]{}\providecommand{\singleletter}[1]{#1}%

 \end{document}